\documentclass[twocolumn,iop,apj]{aastex631} 
\usepackage{times}
\usepackage[sort&compress]{natbib}

\usepackage{amssymb,amsmath,array}
\usepackage{graphicx} 
\usepackage{color}
\usepackage{ulem} 
\usepackage{float}
\usepackage{listings}
\usepackage{xparse}
\usepackage{forest}

\newcolumntype{F}[1]{>{\minipage{\dimexpr#1}\centering\arraybackslash}c<{\vspace\tabcolsep\endminipage}}
\newcommand{\logp}[2]{\mathcal{L}^{#1}_{#2}}

\DeclareMathOperator{\sech}{sech}

\newcommand{\be}{\begin{equation}}
\newcommand{\ee}{\end{equation}}

\newcommand{\gray}{$\gamma$-ray}

\newcommand{\hi}{H~{\sc i}}
\newcommand{\htwo}{H$_2$}
\newcommand{\hii}{H~{\sc ii}}
\newcommand{\e}[2]{$^{#1}_{#2}$}

\newcommand{\Xco}{$X_{\rm CO}$}

\newcommand{\fermilat}{{\it Fermi}--LAT}

\newcommand{\gardian}{{\sc GaRDiAn}}

\newcommand{\GP}{{\sc GalProp}}
\newcommand{\WR}{{\sc WebRun}}

\newcommand{\frankie}{{\sc FRaNKIE}}

\newcommand{\galgas}{{\sc GalGas}}
\newcommand{\galprop}{{\sc GalProp}}
\newcommand{\galtoolslib}{{\sc GalToolsLib}}
\newcommand{\helmod}{{\sc HelMod}}

\shorttitle{The v57 \GP{} Release}
\shortauthors{Porter~et~al.}

\begin{document}



\title{The \GP{} cosmic-ray propagation and non-thermal emissions framework: Release v57}
\author{T.~.A.~Porter}
\affiliation{W. W. Hansen Experimental Physics Laboratory and Kavli Institute for Particle Astrophysics and Cosmology, Stanford University, Stanford, CA 94305, USA}
\author{G.~J\'ohannesson}
\affiliation{Science Institute, University of Iceland, IS-107 Reykjavik, Iceland}
\author{I.~V.~Moskalenko}
\affiliation{W. W. Hansen Experimental Physics Laboratory and Kavli Institute for Particle Astrophysics and Cosmology, Stanford University, Stanford, CA 94305, USA}

\begin{abstract}

The past decade has brought impressive advances in the astrophysics of cosmic rays (CRs) and multiwavelength astronomy, thanks to the new instrumentation launched into space and built on the ground.
Modern technologies employed by those instruments provide measurements with unmatched precision, enabling searches for subtle signatures of dark matter (DM) and new physics.
Understanding the astrophysical backgrounds to better precision than the observed data is vital in moving to this new territory.
The state-of-the-art CR propagation code called \galprop{} is designed to address exactly this challenge.
Having 25 years of development behind it, the \galprop{} framework has become a de-facto standard in the astrophysics of CRs, diffuse photon emissions (radio- to \gray{s}), and searches for new physics.
\galprop{} uses information from astronomy, particle physics, and nuclear physics to predict CRs and their associated emissions self-consistently, providing a unifying modelling framework.
The range of its physical validity covers 18 orders of magnitude in energy, from sub-keV to PeV energies for particles and from $\mu$eV to PeV energies for photons.
The framework and the datasets are public and are extensively used by many experimental collaborations and by thousands of individual researchers worldwide for interpretation of their data and for making predictions.
This paper details the latest release of the \galprop{} framework and updated cross sections, further developments of its initially auxiliary datasets for models of the interstellar medium that grew into independent studies of the Galactic structure--distributions of gas, dust, radiation and magnetic fields--as well as the extension of its modelling capabilities.
Example applications included with the distribution illustrating usage of the new features are also described.
\end{abstract}



\section{Introduction} \label{sec:intro}

\setcounter{footnote}{0}


The last decade has brought spectacular advances in the astrophysics of CRs and space- and ground-based astronomy.
Launches of missions that employ forefront detector technologies have enabled measurements with large effective areas, wide fields of view, and precision that until recently we could not even dream of.
Among those missions are the Alpha Magnetic Spectrometer--02 (AMS-02), the {\it Fermi} Large Area Telescope ({\it Fermi}-LAT), the Payload for Antimatter Matter Exploration and Light-nuclei Astrophysics (PAMELA), the NUCLEON experiment, the CALorimetric Electron Telescope (CALET), the DArk Matter Particle Explorer mission (DAMPE), and the Cosmic-Ray Energetics and Mass investigation (ISS-CREAM).
Outstanding results have been also delivered by mature missions, such as the Cosmic Ray Isotope Spectrometer on board of the Advanced Composition Explorer (ACE-CRIS), and Voyager~1 and~2 spacecrafts, currently at 151 au/126 au from the Sun, respectively.
Indirect observations of high-energy processes in the Galaxy and beyond are made by X-ray and \gray{} telescopes: the International Gamma-Ray Astrophysics Laboratory (INTEGRAL), \fermilat{}, the High-Altitude Water Cherenkov \gray{} observatory (HAWC), and by atmospheric Cherenkov telescopes, the High Energy Stereoscopic System (H.E.S.S.), Major Atmospheric Gamma-ray Imaging Cherenkov Telescopes (MAGIC), the Very Energetic Radiation Imaging Telescope Array System (VERITAS), and the Large High Altitude Air Shower Observatory (LHASSO).
High-resolution data relevant to studies of the cosmic microwave background (CMB) are provided by the {\it Wilkinson} Microwave Anisotropy Probe (WMAP), and {\it Planck} mission.
These significant improvement in the precision of observations may enable major discoveries only if our modelling efforts attain similar or better precision.

Coherent interpretation of the individual slices of information about the internal working of the Milky Way (MW) provided by such experiments requires a self-consistent approach.
The research tool that we have developed over a number of years is the state-of-the-art \galprop{} code that does exactly that: it provides a self-consistent interpretation tool, and combines, in a single framework, the results of individual past, current, and future experiments in astrophysics and astronomy spanning in energy coverage, types of instrumentation, and the nature of detected species. 

The public \galprop{} framework is a numerical package that describes the propagation of Galactic CRs and the production of diffuse emissions, which can also be used in conjunction with other software packages, such as SuperBayes, \helmod{}, etc.
The project now has 25 years of development behind it.
The original FORTRAN90 code has been public since 1998 \citep{1998ApJ...493..694M, 1998ApJ...509..212S}, and a rewritten C++ version was produced in 2001.
Subsequent public releases have been made as computational capabilities have progressed and more precise data have become available requiring increasing detail of modelling.
The last major release (v56) followed improvements over a number of years \citep{2016ApJ...824...16J,Moskalenko:2015ptr}, along with enhanced capabities for full 3D modelling for both the CR sources, the interstellar radiation field \citep[ISRF;][]{2017ApJ...846...67P}, and interstellar gas distributions \citep{2018ApJ...856...45J}.
The next version of the \GP{} code (v57) is made available with this paper.
Substantial new features are included, with emphasis to making realistic time-dependent 3D modelling of CR propagation through the interstellar medium (ISM) and the production of the associated diffusion emissions computationally tractable.
These developments are of particular relevance for interpretation of the new data into the very high energy (VHE; $\gtrsim$100~GeV) range coming from different instruments both space- and ground-based, e.g., CALET and DAMPE for the former, and HAWC, LHASSO, and others for the latter.
The releases of the \GP\ framework and supporting data products are available at the dedicated website, which also provides the \WR\ facility to run \GP{} via a web browser interface\footnote{http://galprop.stanford.edu/}. 

\section{\GP{} Framework} \label{sec:gp_code}

Theoretical understanding of CR propagation in the ISM is the framework that the \GP{} code is built around.
The key idea is that all CR-related data, including direct measurements, \gray{s}, sychrotron radiation, etc., are subject to the same physics and must therefore be modelled self-consistently \citep{1998A&A...338L..75M}.
The goal for the \galprop{}-based models is to be as realistic as possible, making use of the available astrophysical information, nuclear and particle data \citep{2007ARNPS..57..285S}.
Below, we provide a summary of the \GP{} framework and its features up to the v56 release.

The \galprop{} code solves a system of about 90 time-dependent transport equations (partial differential equations in 3D or 4D: spatial variables plus energy) with a given source distribution and boundary conditions to give the intensity distributions for all CR species through the ISM: $^1$H--\,$^{64}$Ni, $\bar{p}$, $e^\pm$ \citep{1998ApJ...509..212S, 2007ARNPS..57..285S, 2009arXiv0907.0559S}.
The propagation equations include terms for convection, distributed reacceleration, energy losses, nuclear fragmentation, radioactive decay, and production of secondary particles and isotopes.
The spatial boundary conditions assume free particle escape.
For a given halo size the diffusion coefficient, as a function of rigidity, and other propagation parameters can be determined from secondary-to-primary nuclei ratios, typically B/C, [Sc+Ti+V]/Fe, and/or $\bar{p}/p$.
If reacceleration is included, the momentum-space diffusion coefficient $D_{pp}$ is related to the spatial coefficient $D_{xx} = \beta D_0 \rho^\delta$ \citep{1994ApJ...431..705S}, where $\beta=v/c$ is the particle velocity, $\rho$ is the magnetic rigidity, and $\delta = 1/3$ for a Kolmogorov spectrum of interstellar turbulence \citep{1941DoSSR..30..301K}, or $\delta = 1/2$ for an Iroshnikov--Kraichnan cascade \citep{1964SvA.....7..566I, 1965PhFl....8.1385K}, but can also be arbitrary.
The spatial diffusion coefficient can also depend on position.
This option was developed and utilised by \citet{2015ApJ...799...86A} where the positional dependence of the spatial diffusion coefficient was linked to the distribution of the Galactic magnetic field (GMF) strength.

The composition scheme introduced for v56 \GP\ for specifying the source model allows the spatial density distribution, spectral characteristics, and respective contributions to be customised at the user's discretion.
  This enables the modelling of multiple populations, e.g., CR injection by supernova remnants (SNRs) and pulsar wind nebulae (PWNe).
  Possible components for the spatial density model include an axisymmetric disc, spiral arms, various central bulges, and other structures.
  Each basic component can be further split up and fine-tuned with different radial profiles, so that different classes of sources have their own population spatial distribution, injection spectra, and isotopic abundances, allowing for a very flexible description of a galaxy.
  
  The injection spectra of CR species for a source density distribution are parametrised by a multiple broken power law in rigidity:
\begin{equation}  \label{eq:src_spec}
q(\rho) \propto (\rho/\rho_0)^{-\gamma_0}\prod_{i=0}^N\left[1 + (\rho/\rho_i)^\frac{\gamma_i - \gamma_{i+1}}{s_i}\right]^{s_i},
\end{equation}
where $\gamma_{i =0,\dots,N+1}$ are the spectral indices, $\rho_{i = 0,\dots,N}$ are the break rigidities, and $s_i$ are the smoothing parameters ($s_i$ is negative/positive for $|\gamma_i |\lessgtr |\gamma_{i+1} |$).
Each primary isotope can have unique spectral parameters. 

The nuclear reaction network is built using the Nuclear Data Sheets, where a detailed description of its method of construction is given by \citet{2020ApJS..250...27B}.
Included are multistage chains of $p$, $n$, $d$, $t$, $^3$He, $\alpha$, and $\beta^\pm$-decays, and K-electron capture, as well as, in several cases, more complicated reactions.
It is a core part of \galprop{}, but has also used been used for other studies.
Examples include investigating the accuracy of the isotopic production cross sections employed in astrophysical applications \citep[e.g.,][]{2015arXiv151009212T,2018PhRvC..98c4611G, 2019PhRvD..99j3023E}, and in other propagation codes \citep[e.g.,][]{2008JCAP...10..018E, 2016JCAP...04E.001E, 2017JCAP...02..015E}. 

The \galprop{} code computes a complete network of primary, secondary, and tertiary isotope production starting from input CR source abundances.
Because the decay branching ratios and half-lifes of fully stripped and hydrogen-like ions may differ, \galprop{} includes the processes of K-electron capture, electron pickup from the neutral ISM gas, and formation of hydrogen-like ions as well as the inverse process of electron stripping \citep{1973RvMP...45..273P, 1978PhDT........12W, 1979PhDT........67C}.
Meanwhile, the fully stripped and hydrogen-like ions are treated as separate species.
Also included are knock-on electrons \citep{1966PhRv..150.1088A, 2003ApJ...594..709B} that may significantly contribute to hard X-ray--soft \gray{} diffuse emission through inverse Compton (IC) scattering and bremsstrahlung \citep{2008ApJ...682..400P}. 

The production of secondary particles in \galprop{} is calculated taking into account $pp$-, $pA$-, $Ap$-, and $AA$ reactions.
Calculations of $\bar{p}$ production and propagation are detailed by \citet{2002ApJ...565..280M, 2003ApJ...586.1050M}, \citet{2015ApJ...803...54K}, and \citet{2019CoPhC.24506846K}, where inelastically scattered (tertiary) $\bar{p}$ and (secondary) $p$ are treated as separate species.
Production of neutral mesons ($\pi^0$, $K^0$, $\bar{K}^0$, etc.), and secondary $e^\pm$ is calculated using the formalism by \citet{1986A&A...157..223D} and \citet{1986ApJ...307...47D} as described by \citet{1998ApJ...493..694M}, or more recent parameterisations \citep{2006ApJ...647..692K, 2012PhRvD..86d3004K, 2014ApJ...789..136K, 2019CoPhC.24506846K}. 

The \gray{} emissivities are calculated using the propagated CR distributions, including primary $e^-$, secondary $e^\pm$, and knock-on $e^-$, as well as inelastically scattered (secondary) protons \citep{2004ApJ...613..962S, 2008ApJ...682..400P}.
Gas-related \gray{} intensities ($\pi^0$ decay, bremsstrahlung) are computed from the emissivities using the column densities of \htwo{} + \hi{} (+ \hii{}, ionised hydrogen) gas for galactocentric annuli based on 2.6-mm carbon monoxide CO (a tracer of \htwo{}) and 21-cm \hi{} surveys \citep{2012ApJ...750....3A}, with corrections for gas not traced by these data \citep[e.g.,][]{2016ApJ...819...44A}.
The IC emissivities use the formalism for an anisotropic background photon distribution \citep{2000ApJ...528..357M}.
Synchrotron emissivities are calculated for total and polarised components.
The line-of-sight (LOS) integration, including absorption effects, of the corresponding emissivities with the ISM distributions of gas, ISRF, and GMF yields \gray{} and synchrotron intensity sky maps. 

The \galprop{} framework also has well-developed options to propagate particles produced by {\it exotic sources and processes}, such as annihilation or decay of dark matter (DM) particles, and calculate the associated emissions (DM sky maps).
\galprop{} can be used alone or run in conjunction with dedicated packages for modelling the production via these mechanisms \citep[e.g., DarkSUSY;][]{2004JCAP...07..008G,2018JCAP...07..033B}.

For the CR interactions with the interstellar gas, \galprop{} runs can use different density models. 
The ISM gas consists mostly of H and He with a ratio of 10:1 by number \citep{2001RvMP...73.1031F}.
Hydrogen can be found in the different states, atomic (\hi), molecular (H$_2$), or ionised (\hii), while He is mostly neutral.
\hi{} is $\sim$60\% of the mass, while H$_2$ and \hii{} contain 25\% and 15\%, respectively \citep{2001RvMP...73.1031F}.
The \hii{} gas has a low number density and scale height $\sim$few~100~pc.
The H$_2$ gas is clumpy and forms high-density molecular clouds.

For 2D calculations, analytical models for the gas density distribution are available \citep{2002ApJ...565..280M}.
The radial distribution for \hi{} is taken from \citet{1976ApJ...208..346G} while the vertical distribution is from \citet{1990ARA&A..28..215D} for galactocentric radial distances $0\le R\le 8$~kpc and \citet{1986A&A...155..380C} for $R\ge10$~kpc with linear interpolation in between.
The CO gas distribution is taken from \citet{1988ApJ...324..248B} for 1.5~kpc$<$$R$$<$10~kpc, and from \citet{1990A&A...230...21W} for $R$$\ge$10~kpc, and is augmented with the \cite{2007A&A...467..611F} model for $R$$\le$1.5 kpc.
Both of the 2D atomic and molecular gas distributions are rescaled to the common solar radius of $R_\odot = 8.5$~kpc.
The \hii{} gas distribution is given by the NE2001 model \citep{2002astro.ph..7156C, 2003astro.ph..1598C, 2004ASPC..317..211C} with the updates from \citet{2008PASA...25..184G}.

For 3D simulations, the \hi\ and $^{12}$CO distributions from \cite{2018ApJ...856...45J} are available.
These were developed using a maximum-likelihood forward-folding optimisation applied to the LAB-\hi\ \citep{2005A&A...440..775K} and CfA composite CO data \citep{2001ApJ...547..792D, 2004ASPC..317...66D}.
Compared to the 2D models, the added degrees of freedom (spiral arms, bar) allow the optimised distributions to better reproduce the features observed in the line-emission surveys.

For the CR electron/positron interactions with the ISRF there are 2D and 3D models available within the \GP\ framework.
The ISRF is the distribution of the low-energy photon populations that are the result of emission by stars, and the scattering, absorption, and reemission of absorbed starlight by dust in the ISM.
The ISRF models have been calculated using a full radiation transfer treatment accounting for the scattering/absorption/reemission processes in the ISM.

The 2D ISRF models are described by \cite{2005ICRC....4...77P} and \cite{2008ApJ...682..400P}.
The former model provides only the energy density distribution, while the latter
has a spatially varying UV-to-far-IR distribution with all-sky intensity maps.
The angular distribution with position is necessary to account for the important directional amplification/suppression effects due to the anisotropic IC scattering cross section \citep{2000ApJ...528..357M, 2007AIPC..921..490M}.

The 3D ISRF models for the MW were developed by \citet{2017ApJ...846...67P} based on spatially smooth stellar and dust models.
The ISRF models have designations R12 and F98 that correspond to the respective references supplying the stellar/dust distributions \citep{2012A&A...545A..39R,1998ApJ...492..495F}.
Both the R12 and F98 models provide equivalent solutions for the ISRF intensity distribution throughout the MW, but neither gives an overall best match with the data.
Toward the inner Galaxy, where the ISRF intensity is most uncertain, they are giving lower and upper bounds for its distribution, as determined by pair-absorption effects on sources toward the Galactic centre (GC) \citep[][]{2018PhRvD..98d1302P}.
\galprop{} simulations made with both can be used to estimate the bounding systematic modelling uncertainty from lack of knowledge of the precise distribution of the ISRF across the MW.

The GMF consists of the large-scale regular \citep{1985A&A...153...17B} and small-scale random \citep[e.g.,][]{2008A&A...477..573S} components that are about equal in intensity.
The random fields are mostly produced by the supernovae and other outflows,  which result in randomly oriented fields with a typical spatial scale of $\lesssim$100~pc \citep{1995MNRAS.277.1243G, 2008ApJ...680..362H}.
Also, there may be the anisotropic random (``striated'') fields, which refer to a large-scale ordering originating from stretching or compression of the random field \citep{2001SSRv...99..243B}.
This component is expected to be aligned to the large-scale regular field, with frequent reversal of its direction on small scales.
\galprop{} includes multiple large-scale MW GMF models \citep{2008A&A...477..573S, 2010MNRAS.401.1013J, 2010RAA....10.1287S, 2011ApJ...738..192P, 2012ApJ...757...14J}. 

\subsection{Heliospheric Transport and Comparison with CR data}\label{helmod}

Comparison of the interstellar CR spectra calculated by \GP{} with the direct measurements taken inside the heliosphere is treated using the \citet{1965P&SS...13....9P} equation, where the numerical solutions are provided by the \helmod{}\footnote{http://www.helmod.org/} code \citep{2017ApJ...840..115B, 2018ApJ...854...94B, 2018ApJ...858...61B, 2018AdSpR..62.2859B, 2019AdSpR..64.2459B}.
The solar modulation affects all CRs with rigidities below $\sim$30~GV, and thus must be accounted for when using the CR data for determining the propagation model parameters for the ISM: the spatial diffusion coefficient, the halo size, etc.

\helmod{} is a Monte Carlo code developed to model the heliospheric transport of Galactic CRs including terms for the
diffusion, adiabatic energy changes, effective convection resulting from the convection with solar wind, and drift (charge-sign) effects.
The heliospheric propagation parameters are tuned to data of many spacecraft, including Voyager 1~and~2,  ACE-CRIS, AMS-02, and Ulysses (providing data for locations outside of the ecliptic).
\helmod{} can model the spectra of CR species for an arbitrary level of solar modulation and the polarity of the solar magnetic field.
The combination of \galprop{} with \helmod{} forms a self-consistent framework for treatment of CR propagation from the CR sources down to the inner heliosphere.
It is run iteratively to optimise the local interstellar spectra (LIS) of CR species.
Examples are the LIS of $\bar{p}$ and $e^{-}$ \citep{2017ApJ...840..115B, 2018ApJ...854...94B}, and a complete set of $_{1}$H--\,$_{28}$Ni nuclei LIS for the rigidity range from MV--100 TV \citep{2020ApJS..250...27B, 2021ApJ...913....5B, 2021arXiv210601626B}.

The ultimate goal of this development is to produce well-defined LIS for all CR species (H--Ni nuclei, $\bar{p}$, and $e^\pm$) to disentange the interstellar and heliospheric propagation.
The derived, and in some cases predicted, LIS \citep{2020ApJS..250...27B, 2021ApJ...913....5B, 2021arXiv210601626B} could be used as a substitute for CR measurements in interstellar space.
       This potentially eliminates the need to account for the solar modulation entirely, which requires expert knowledge and well-developed modelling.
 
\section{Features of the New Release}\label{sec:features}

The main new features in the v57 release of \GP\ are the following:
\begin{itemize}
\item A new installer to ease the configuration and compilation of the required support libraries and \GP\ code.
\item New run modes that to enable robust completion for the time-dependent runs.  
  Restarting is now possible, if the calculation is interrupted, for CR propagation/non-thermal emissions production.
  The latter can also be post-processed for both steady-state and time-dependent runs using the calculated CR distributions.
\item New solvers for the propagation equation with revised differencing scheme to make treatment of edge cases more robust, and to support the nonuniform spatial grids.
\item Nonuniform grids are now supported for improved resolution where it is most needed.
\item New source distributions, including a sampler for producing spatial distributions of time-dependent discrete CR sources.
\item Improved parameterisations for calculations of the total inelastic cross sections for $p+A$ and He $+$ $A$ reactions have been made.
  Included also are new routines for the production cross sections for isotopes of hydrogen, $^2$H and $^3$H, and helium, $^3$He, in $p+A$ and He $+$ $A$ reactions, as well as $^2$H production in the $pp$ reaction. 
\end{itemize}
The features are explained with more details below.

\subsection{Dependencies and Installation}\label{sec:install}

Architecturally, \GP\ versions prior to the v56 release were monolithic, with the code and configuration required to detect external libraries and build the \GP\ library and executable contained in the source distribution.
Because of the reuse of core functionality across our different code bases--\GP\ as well as \frankie\ \citep{2015ICRC...34..908P,2017ApJ...846...67P}, \galgas\ \citep{2018ApJ...856...45J}, \gardian\ \citep{2012ApJ...750....3A}--we introduced with v56 the \galtoolslib\ support library, which separates the installation of the common utility code and supporting dependencies from that for the individual packages. \galtoolslib\ includes utility code for parameter parsing (e.g., reading the {\it galdef} configuration file of a \GP\ run), specifying spatial distributions (e.g., for CR source densities), libraries for the representation of results (e.g., sky maps with HEALPix), core physics routines for the nuclear reaction network, energy losses, and emission processes, and other commonly reused code. The higher-level packages (e.g., \GP) link with \galtoolslib, retrieving all build information directly from it. 

\galtoolslib\ has a number of external package dependencies: Boost\footnote{https://www.boost.org/}, CFitsIO\footnote{https://heasarc.gsfc.nasa.gov/fitsio/fitsio.html} and CCfits\footnote{https://heasarc.gsfc.nasa.gov/fitsio/ccfits/}, CLHep\footnote{http://proj-clhep.web.cern.ch/proj-clhep/}, the Gnu Scientific Library\footnote{https://www.gnu.org/software/gsl/}, HEALPix\footnote{http://healpix.jpl.nasa.gov/}, WCSLIB\footnote{http://www.atnf.csiro.au/people/mcalabre/WCS/}, and Xerces-C\footnote{https://xerces.apache.org/xerces-c/}.
Most of these are available for the operating systems that we support via external package managers (e.g., repositories for various Linux distributions, or Macports for OSX).
However, targeting libraries available via these installation methods can be problematic because of variation of versions and their features available across distributions.
To reduce these issues and provide a consistent installation process, with the v57 release we directly include all support libraries at the tested versions necessary for a successful \GP\ installation, and provide an installer (requiring minimal set-up) that takes care of all configuration and build steps.
The entire installation is fairly self-contained so that the only dependencies are on the system and tool chain libraries.
In our use and testing this has provided a more streamlined process than the v56 release.
    
Appendix~\ref{app:A} shows the results for a successful installation and execution for the \GP{} binary produced by the process.
Appendix~\ref{app:dir_struct} gives the directory structure of the installation, including where the respective libraries and binaries are built.

\subsection{Run Modes}\label{sec:run_modes}

A \GP\ solution can be made for either steady-state or time-dependent propagation models.
  The default is to obtain the steady-state model solution, which has been the standard for CR modelling over the years.
  Previous versions of \GP\ were optimised for these runs and had very limited functionality for restarting and reprocessing.
  Calculating time-dependent solutions is generally much more time consuming, and having robust check-pointing and post-processing enhances the user experience.
  Different run modes were therefore introduced to facilitate these new features.
  
  The default run mode produces, at minimum, the solution for the propagated CR intensities over the spatial/energy grid specified by the user-supplied configuration.
  Other outputs that are typically produced (set by the user) may include the secondary emissions (\gray{s}, synchrotron), ionisation rate distribution, and so forth.
  With this mode, and appropriate configuration, everything is computed in one pass.
  This operates similarly to previous versions of \GP\ and is suitable for steady-state models.
  
  Additional run modes are supported that enable restarting/post-processing.
  The restarting functionality can be used for the time-dependent runs.
  Because these can be time consuming, we introduced the facility to resume from an interrupted calculation that has had check-pointing enabled.
  The post-processing mode can be used with a fully complete steady-state or time-dependent solution (including for each of the intermediate check-pointed distributions) to compute additional observables.
  As described above for the standard processing, these include secondary emissions, etc.
  This mode is useful for situations where the CR intensity solution is obtained first, possibly on a cluster batch system, and transferred to another to produce other data products for subsequent analysis/interpretation.
  The check-pointing also allows for the calculations of secondary emission at regular intervals, which is not possible with the default run mode.

\subsection{Solvers}\label{sec:solvers}

Earlier \GP\ versions employed solvers for the propagation equations based on the so-called operator splitting method.
With this release we now include two classes of solvers.
The operator splitting class remains, but with improvements for vectorisation to speed up the methods.
The other class uses the full set of equations defining the finite differences.
The classes behave somewhat differently and will be described in the following subsections.
The differencing formalism has changed from that employed by previous versions of the code to accommodate the nonuniform grids (Sec.~\ref{sec:grids}); details are given in Appendix~\ref{app:finite_differences}.
We give in Appendix~\ref{app:tests} operation details for the solvers and tests of the solutions obtained with them.

\subsubsection{Operator splitting}

The first set of solvers is based on the operator splitting method that has been employed in previous versions of the code. For these, each dimension is solved for independently, assuming the others are fixed in the evaluation. There are three different types of solvers, depending on the handling for the time derivative.

\begin{itemize}
  \item {\it Explicit.} In this scheme, the time-updating step is
    \[
      f_{i+1} = f_{i} + \left(\frac{\partial f}{\partial t}\right)_{i} \Delta t,
    \]
    where the derivative is evaluated based on the current value of the gradient.   This method is first-order accurate in $\Delta t$ and only stable for small time steps. Combining the operator splitting scheme with this updating method makes the solution accurate, because the updating scheme depends on known quantities only.
  \item {\it Implicit.} In this scheme, the time-updating step is
    \[
      f_{i+1} = f_{i} + \left( \frac{\partial f}{\partial t} \right)_{i+1} \Delta t
    \]
    where the derivative is evaluated based on the next value of the gradient, resulting in a set of equations that has to be solved.   Because of operator splitting, the resulting matrix is trilinear and can be solved directly.   This method is first-order accurate in $\Delta t$ and stable for all $\Delta t$.   However, the operator splitting scheme may not be accurate, because we solve for only one dimension at a time.   Extensive testing has shown, though, that other factors such as the grid resolution play a bigger role.
  \item {\it Crank-Nicholson.} In this scheme, the time-updating step is
    \[
      f_{i+1} = f_{i} + \frac{1}{2}\left[\left(\frac{\partial f}{\partial t}\right)_{i} + \left( \frac{\partial f}{\partial t} \right)_{i+1}\right] \Delta t
    \]
    where the derivative is the average of the current value and that of the next value, again resulting in a set of equations that has to be solved.   These are again trilinear when operator splitting is applied.   This is a second-order accurate updating method that is again stable for all $\Delta t$.   It is thus the preferred updating step.   As for the implicit updating step, this may not be accurate with the operator splitting. 

\end{itemize}

As for previous \GP\ versions, the Crank-Nicholson and implicit solvers employ an iterative scheme to obtain steady-state solutions.
  Starting at the largest timescale of the problem (user defined, usually $\sim$$10^9$ years for a MW-like galaxy) the solution is found by evolving this time step a specific number of iterations (user defined, 20 iterations has been found to be sufficient for most cases).
  The time step is then reduced by a user-defined fraction (e.g., 0.7 works well) and the solution again iterated the same number of iterations.
  This is repeated until the step size reaches a user-defined minimum and the steady-state solution has been reached.

The operator splitting solvers were parallelised using OpenMP for \GP\ v56, and with this release the trilinear matrix inversion for them has been fully vectorised.
To ensure the auto-vectorisation, the grid step for the energy grid and spatial grid, either (radial) $R$- or $X$-axis (depending if the modelling configuration is 2D or 3D spatially), is adjusted to have a whole multiple of 8 in the number of grid cells.

\subsubsection{Direct solvers}

New for the v57 release is the possibility to obtain the propagation equation solution using iterative solvers for sparse linear systems.
Currently implemented is the BiGCStab solver from the Eigen project,\footnote{\url{https://eigen.tuxfamily.org/}. It is installed along with the other support libraries, see Appendix~\ref{app:A}.} using either the diagonal preconditioner or the IncompleteLUT preconditioner.
For multicore systems, the diagonal preconditioner may be the best choice because it allows for wider utilisation of the computational resources.
The IncompleteLUT preconditioner is not parallel, but can increase the single-core speed and may therefore be a better choice if more limited resources are available.

One of the benefits of the direct solvers is avoiding the iterative procedure of the operator splitting methods when solving steady-state models.
When employed for the time-dependent mode they are used together with the Crank-Nicholson time-updating scheme for an implicitly stable solution.
However, the solution may not be accurate.
Care must be taken to select a time step suitable for the smallest relevant timescale for the system being solved.

The solutions from the direct solvers have been extensively tested by comparing to those found using the operator splitting solvers.
The direct solvers find in all cases an equivalent solution, provided the start and stop time steps of the iterative method for the operator splitting solver have been properly adjusted for the configuration being calculated.

\subsection{Non-uniform Grids}\label{sec:grids}

\GP{} versions v56 and earlier allowed only uniform 2D ($R,Z$) or 3D ($X,Y,Z$) spatial grids for solutions of the propagation equations.
However, uniform grids, particularly in 3D, can be inefficient both in terms of calculation speed and in-memory image size.
With the new release we introduce an option to use nonequidistant grids that allow for increased spatial resolution over user-specified regions of the calculation volume.
This \GP\ enhancement is inspired by the Pencil Code\footnote{See 
  \url{http://pencil-code.nordita.org/doc/manual.pdf}, Section 5.4.}
\citep{2002CoPhC.147..471B} that specifies the nonuniform spacing based on analytic functions.
The so-called ``grid functions'' are an easy way to have adjustable (in space but not time) resolution for the solutions of differential equations using finite differences, and offer substantial speed and memory computational efficiencies.
The functions are solved on a uniform grid $\zeta$ with a unit step size $\Delta \zeta=1$, where the uniform grid is mapped onto the physical coordinate with a function $Q(\zeta)$.
Correspondingly, the differential equations to be solved have to be adjusted to account for this change of coordinates.
The first and second derivatives are adjusted as follows:
  \begin{align}
    \frac{\partial f}{\partial Q} &= \frac{d \zeta}{d Q}\frac{\partial f}{\partial \zeta},\\
    \frac{\partial^2 f}{\partial Q^2} &= \left( \frac{d \zeta}{d Q} \right)^2 \frac{\partial^2 f}{\partial \zeta^2} - \left( \frac{d \zeta}{d Q} \right)^3 \frac{d^2 Q}{d \zeta^2} \frac{\partial f}{\partial \zeta}.
    \label{eq:grid-second-derivative}
  \end{align}

It is useful to think about the grid functions in terms of their derivative $dQ/d\zeta$, which gives the physical step size change along the uniform ($\zeta$) grid.
The physical step size is at the true spatial resolution of the solution of the propagation equations.
For stability of the solution it is best that the second derivative $d^2Q/d\zeta^2$ is smaller than the first derivative, so that the correction (second term) for the second derivative in Eq.~(\ref{eq:grid-second-derivative}) is small.

Our implementation for the grid functions includes the trivial linear grid (\verb+linear+), and two others: the tangent grid (\verb+tan+) and the step grid (\verb+step+).
The linear grid is used for testing.
The tangent grid has strong utility across different modelling scenarios, e.g., Galaxy-wide CR propagation, down to about localised regions surrounding individual sources. 
The step grid is mainly intended to provide high resolution about individual localised regions while transitioning to a constant coarser resolution outside, potentially suitable for two-zone scenarios for individual so-called ``TeV halos'' \citep[e.g.,][]{2019PhRvD.100d3016S}. 

The example applications below (Sec.~\ref{sec:examples}) show the usage of the grid functions for different scenarios.
Appendix~\ref{app:grid_fns} gives the formulae for the transformations and the user-configurable parameters for each.

\subsection{Discretised Source Sampler}\label{sec:sampler}

\GP\ has included the capability to simulate the time-dependent CR injection and propagation from individual sources since its earliest versions.
Application to modelling the CR distributions through the ISM and associated non-thermal diffuse emissions from ensembles of sources indicated that the stochastic effect on the CR and high-energy \gray{s} intensities were important for correctly interpreting the data \citep{2001AIPC..587..533S,2001ICRC....5.1964S,2003ICRC....4.1989S}.
However, the computational limitations of the initial implementation meant it was a little-used feature.

To enable more efficient time-dependent CR propagation and interstellar emissions modelling, we have made many enhancements to the \GP\ code.
Among these, we introduce a new method for specifying the spatial distributions for the individual CR sources.
This allows for a fully reproducible and more versatile 3D discretisation scheme compared to the earlier implementation.

Prior versions allowed for the SN or CR source birth rate to be specified, and generated the locations of the individual sources according to the volumetric discretisation for the CR source density distribution over the propagation volume.
The allowed CR density models were from different hard-coded forms for describing those for SNRs, pulsars, etc., across the Galaxy, and were mostly 2D in galactocentric $R$ and $Z$.
The random number generation for this method relied on the system-supplied function via the standard C library. 

With the v57 release, we allow for much higher flexibility regarding the spatial density of the CR sources employing the distribution specification method introduced with v56 \citep{2017ApJ...846...67P}.
The smooth spatial density distribution is given via composition using predefined primitives (e.g., disc, arms) or user-defined profiles whose parameters can be configured via XML or adjusted at run-time via interfaces into the \GP\ code. 

The list of individual sources active over a given epoch is then obtained using the so-called ``discrete sampler'' from the given smooth source density distribution.
The discrete sampler uses an acceptance/rejection method and employs the pseudo-random number generator (PRNG) included in the \galtoolslib\ library.
For the same seed, the smooth density model discretisation is fully reproducible across different installations because the PRNG is part of the supporting library included with the \GP\ distribution.
The discretisation reproducibility also allows for direct testing of the effects of parameter variations for the CR injection regions in the time-dependent solutions, e.g., scenarios modifying the source luminosity time evolution.

\subsection{Updated formalism for calculation of the total inelastic cross sections}\label{sec:xs}

\GP\ now has three different parameterisations for the total inelastic cross sections for $p,d$, $^3$He, and $^4$He projectiles and arbitrary targets.
Two of them, by \citet{1996PhRvC..54.1329W} and \citet{BarPol1994}, have been included since the early versions but were not documented properly.
For v56 and earlier, the $^4$He$+$$A$ cross sections were scaled from $p+A$ cross sections using an empirical formula by \citet{1988PhRvC..37.1490F}.
Meanwhile, there is a better parameterisation by \citet{BarPol1994} that was not used often, but is now employed.

We have also included with this release an additional parameterisation by \citet{1996NIMPB.117..347T}.
The original formalism \citep{1996NIMPB.117..347T, 1997lrc..reptQ....T, 1999NIMPB.155..349T, 1999STIN...0004259T} contains numerous typos and confusing statements.
This caused different authors to invent their own modifications, which are scattered across many publications \citep[see][and references therein]{2021NJPh...23j1201L}. 
We could not find a publication that would provide a coherent set of parameters and expressions and, therefore, the results of the papers where this formalism is used are hard to reproduce.
We corrected inconsistencies and now provide this formalism also as an option for \GP\ runs.
  
Appendix \ref{app:xs} gives details of the above enhancements, including a full description of the corrected and tested formalism by \citet{1996NIMPB.117..347T}.

\subsection{Fragmentation of $^{3,4}$He to $^3$He, $^3$H, $^2$H, and $pp\to\pi^++^2$H reaction}\label{sec:light}

This release of \GP\ includes a significant upgrade of the formalism for calculation of the cross sections for $^4$He fragmentation into $^3$He, $t$, $d$, and for $d$ production in the $pp\to\pi^++d$ reaction. 

The following channels are distinguished in the literature: $^4$He$(p,pnX)^3$He, $^4$He$(p,dX)^3$He, $^4$He$(p,ppX)^3$H, $^4$He$(p,ppnX)d$, $^4$He$(p,dX)d$. They are considered separately as the physics involved is different. As a basis for our parameterisation we use formulations provided by \citet{1993STIN...9417666C}, but with the parameters derived from our own fits to the data.

We also provide our own parameterisations for the reactions $^3$He$(p,ppX)d$ and $pp\to\pi^++d$.
For the latter, we use all available data for direct $pp\to\pi^++d$ and inverse $\pi^++d\to pp$ reactions.
A simple approximation to the differential cross section of the deuteron production in the $pp\to\pi^++d$ reaction is also provided.

For heavier targets $p+A$ $(A>4)$ and/or projectiles $^4$He $+A$ $(A\ge4)$ we follow a formalism described in \citet{2012A&A...539A..88C}.

Appendix \ref{app:frag} gives the formulations of the above processes and operational details for using them in a \GP\ run.

\section{Example Applications}\label{sec:examples}

New with the v57 release are a collection of physical modelling applications that show usage of the \GP\ framework features.
The \verb+examples+ directory that contains them is accessible at the top level of the installation after the archive is extracted (see Appendix~\ref{app:dir_struct}).
All applications are nontrivial, showing the uses of \GP\ for 2D and 3D geometries for steady-state and time-dependent scenarios.
The different examples can be used as templates and easily extended.
For simplicity the heliospheric modulation is done using the so-called force-field approximation \citep{1968ApJ...154.1011G}.
However, substitution with a modern specialised code, such as \helmod{} (see Sec. \ref{helmod}) can be done for more realistic modelling of the heliospheric CR propagation.

The usual ``observables'' that are the output of a \GP\ run are the spatial distributions for the CR spectral intensities (FITS data cube) and intensity sky maps for different processes (HEALPix or Mapcube FITS files).
Standard methods for reading from these file formats can be used to extract results from the \GP\ run outputs.
For each example there is a documented usage and run sequence in the subdirectory. 
All configuration files necessary for reproducing the runs are provided in the individual example subdirectories.
We show expected results that can be obtained for the individual examples and discuss their relevance to interpretation of data.

\subsection{Propagation Model Parameter Optimisation}\label{sec:prop_model_opt}

Determination of model parameters using, e.g., secondary-to-primary ratios is a staple of CR propagation studies for the MW.
The v57 release includes an example of applying \GP\ for determining optimised propagation model parameters using a limited set of CR nuclei measurements.
We have coupled the \GP\ library with an external driver routine (a ``fitter'') to make the tuning procedure as automatic as possible.\footnote{The fitter is built as part of the installation procedure, and its source code is available in the {\tt utils/CRfitter} subdirectory from the top level of the uncompressed archive. The fitter relies on the standalone version of the {\sc Minuit2} library (available from \url{https://github.com/GooFit/Minuit2}) that is built as part of the installation.}
With the configuration as distributed, this example should yield results relatively quickly on a modestly provisioned modern laptop.

We use two CR source density and two gas density distributions, and optimise the same model parameters for each combination for a diffusion-reacceleration propagation scenario with a fixed CR confinement volume.
Covariance of propagation model parameters with the distribution of the CR sources and target gas density has long been recognised \citep[e.g.,][]{2012ApJ...750....3A}.
The four sets of parameters that are obtained give an indication of the variation possible by changing inputs even for a relatively simple \GP\ model configuration.

For minimal memory image and fast execution, we assume a 2D galactocentric cylindrically symmetric geometry where the IAU-recommended $R_S = 8.5$~kpc~\citep{1986MNRAS.221.1023K} is used for the distance from the Sun to the Galactic centre (GC).
The maximum radial size of the CR propagation region is $R_{\rm max} = 20$~kpc and the halo size is set to $Z_{\rm max} = 6$~kpc, consistent with that obtained by \citet{2016ApJ...824...16J}.
The spatial grid spacing uses the tangent grid function (Sec.~\ref{sec:grids}) with $\Delta_R=1.1$~kpc and $\Delta_Z=0.1$~kpc at the solar system's reference location.
The kinetic energy grid is logarithmic from 3~MeV to 10~TeV with 48~planes. 
The operator splitting solver with the Crank-Nicholson updating scheme is selected for this problem.

\begin{figure}[tb!]
  \centerline{
  \includegraphics[scale=0.8]{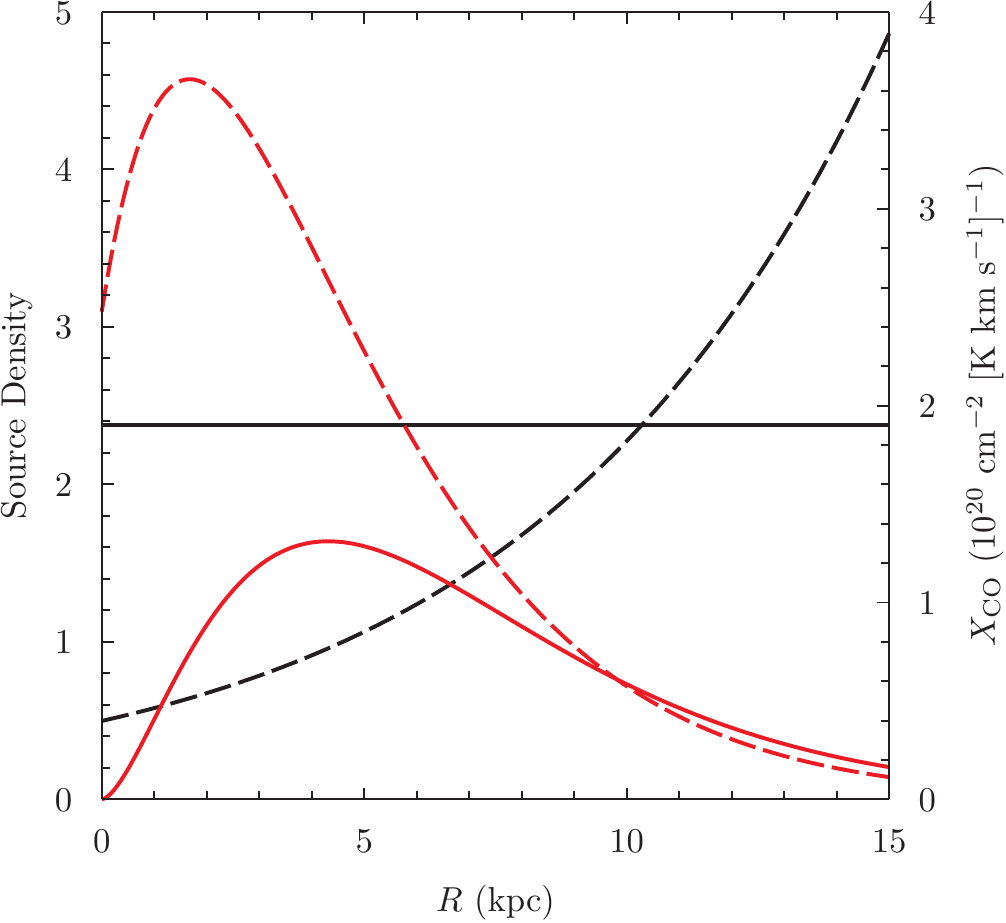}
  }
  \caption{
    CR source density and \Xco\ galactocentric radial distributions used for the 2D CR propagation model parameter optimisation example. CR source distributions: SNR (solid, red) and Pulsar (long dashed, red).   \Xco\ distributions: constant (solid, black) and $R$-dependent (long dashed, black).   The CR source density is normalised at the solar system's location.   Note the different scales for the left and right axes.
    \label{fig:param_opt}
  }
\end{figure}

For the the CR source density models we use the \citet{1998ApJ...504..761C} SNR distribution and the \citet{2004A&A...422..545Y} pulsar distribution. 
The former peaks around $R$$\sim$4~kpc while the latter is peaked at $R$$\sim$1--2~kpc (see Fig.~\ref{fig:param_opt}).
For both source densities the functional dependence perpendicular to the plane has a $\sech^2$ profile with a scale height of 200~pc.
The primary CR source spectra are modelled as broken power laws in rigidity, where the location of the break and the two indices are common but the normalisation for each species is independent. 

For the ISM, we use the 2D neutral gas (\hi{} and \htwo{}) distribution model (Sec.~\ref{sec:gp_code}) with 90\% hydrogen and 10\% helium by number, and the ionised gas distribution described by the hybrid \hii{} model included in the \GP\ code that is based on the NE2001 model of \citet{2002astro.ph..7156C} and the work of \citet{2008PASA...25..184G}, assuming a 7500~K electron temperature.
For the molecular gas, conversion of the $^{12}$CO observation derived distribution to H$_2$ is parameterised via the so-called \Xco=$N$H$_2$/I$_{\rm CO}$ conversion factor \citep[see, e.g., the review by][]{2013ARA&A..51..207B}.
We use two distributions for \Xco\ (both in units $10^{20}$~cm$^{-2}$~[K~km~s$^{-1}$]$^{-1}$): a constant value of 1.9 everywhere, and a galactocentric radial variation $10^{-0.4 + 0.066 R}$ for $R$$<$15~kpc with constant outside (see Fig.~\ref{fig:param_opt}).
This gives two models for the gas density distribution.

\begin{deluxetable*}{lcccc}[tb!]
\tablecolumns{5}
\tablewidth{0pc}
\tablecaption{Propagation model parameter optimisation example final values. \label{tab:param_opt} }
\tablehead{
   & \multicolumn{4}{c}{Input} \\
   \cline{2-5}
\multicolumn{1}{l}{Parameter} & 
\colhead{Pulsar/\Xco\ const.} & 
\colhead{Pulsar/\Xco\ $R$-dep.} &
\colhead{SNR/\Xco\ const.} &
\colhead{SNR/\Xco\ $R$-dep.}
}
\startdata
\tablenotemark{a}$D_{0,xx}$ [$10^{28}$cm$^2$\,s$^{-1}$] & $7.61\pm0.05$ & $7.1\pm0.1$ & $6.32\pm0.05$ & $5.99\pm0.05$ \\
\tablenotemark{a}$\delta$ & $0.350\pm0.002$ & $0.346\pm0.005$ & $0.361\pm0.002$ & $0.355\pm0.002$ \\
$v_{A}$ [km s$^{-1}$] & $36.3\pm0.4$ & $33.5\pm0.09$ & $31.7\pm0.5$ & $29.6\pm0.5$ \\
\tablenotemark{b}$\gamma_0$ & $1.863\pm0.002$ & $1.875\pm0.006$ & $1.858\pm0.005$ & $1.866\pm0.006$ \\
\tablenotemark{b}$\gamma_1$ & $2.37722\pm0.00001$ & $2.377\pm0.003$ & $2.370\pm0.003$ & $2.373\pm0.003$ \\
\tablenotemark{c}$q_{0,^{12}\rm C} \hfill [10^{-6}]\qquad\qquad$ & $ 3293\pm12$ & $  3278\pm18$ & $ 3313\pm16$ & $3294\pm17$\\
\tablenotemark{c}$q_{0,^{16}\rm O} \hfill [10^{-6}]\qquad\qquad$ & $ 4421\pm17$ & $  4403\pm23$ & $ 4444\pm23$ & $4420\pm25$ \\
\tablenotemark{d}$\Phi_{\rm AMS,I}$ [MV] & $ 533\pm7$ & $ 538\pm9$ & $ 500\pm9$ & $497\pm9$ \\
$\chi^2$ & $ 751$ & $ 758$ & $ 689$ & $700$ \\
\enddata
\tablenotetext{a}{$D(\rho) \propto \beta \rho^{\delta}$, $D_0$ is the normalisation at $\rho_0=4$~GV. 
}
\tablenotetext{b}{$\gamma_0$ and $\gamma_1$ are the power-law indices for the injection below and above the fixed rigidity break at $10^4$~GV.}
\tablenotetext{c}{The injection spectra for isotopes are normalised relative to the proton injection spectrum at 100~GeV nuc$^{-1}$.  
The normalisation constants for isotopes not listed here are the same as given in \citet{2016ApJ...824...16J}.}
\tablenotetext{d}{The force-field approximation is used for calculations of the solar modulation and is determined independently for each configuration. $\Phi_{\rm AMS,I}$ corresponds to the 2011--2016 observing period for the AMS--02 instrument.}
\end{deluxetable*}

For each configuration we tune the CR intensities, together with the spatial diffusion coefficient and its rigidity dependence, using a limited set of B/C data from Voyager~I and AMS-02.
The driver routine first reads in the relevant FITS data products, the CR data to be optimised against, the base \GP\ configuration (geometry, etc.), sets parameters to be fit along with their initial values, and initialises \GP.
It then iteratively calls \GP\ to obtain the CR intensities for the current parameter configuration and evaluates the quality of the fit with the data using a $\chi^2$ goodness-of-fit estimator, until the convergence criterion is met.
From the procedure we obtain the optimised parameters for each of the four input configurations along with fitter-estimated uncertainties.
The results are listed in Table~\ref{tab:param_opt}.
These were made on OSX 10.15 using the tools for the installation as outlined in Appendix~\ref{app:A}.
Typical wall clock execution time for each configuration fit to converge was 30--40~minutes on the 8-core laptop that was used.
(Repeating all configurations across other OSX versions and hardware configurations, along with Centos~7/8 workstation installations with different tools, gave similar numbers within the reported 1$\sigma$ uncertainties.)

From the table, the variance is clear over the four configurations for the optimised values.
The configurations weighted closer to the inner Galaxy for the CR source and gas densities require faster diffusion and stronger reacceleration to fit the local measurements.
The model parameters are generally not invariant for fixed confinement volumes when the CR source or gas distributions are also modified.
Such variation has been known from analysis of \gray{} data for some time \citep[see, e.g.,][]{2004A&A...422L..47S,2012ApJ...750....3A}.
However, it appears that it is often overlooked as a contributing uncertainty for interpreting CR data \citep[e.g.,][]{2019A&A...627A.158D,2019PhRvD..99l3028G}.

While not shown here, the fitter also provides estimates of parameter correlations.
The strongest correlations are between $D_{0,xx}$ and $v_A$, and the strongest anticorrelations are between the modulation potential and $\gamma_1$.
There is also correlation for $\gamma_1$ with $D_{0,xx}$ and $v_A$, while it is anticorrelated with $\delta$.
Meanwhile, the high-energy index $\gamma_2$ is also anticorrelated with the C and O abundances, which are themselves correlated.
The strength of the correlations varies between the different configurations, but they are all qualitatively similar for each one.
Dealing with the correlations is somewhat difficult for the simplified fitter used for this example, and some tuning of the initial parameter value was necessary to make sure that it converged properly.

Extension to more sophisticated Bayesian sampling frameworks, as employed by \citet{2011ApJ...729..106T} and \citet{2016ApJ...824...16J}, can be readily accomplished because the interface mechanism is the same.
These methods enable better treatments for model parameters and their correlations.
And, as described above already, coupling with \helmod\ \citep{2020ApJS..250...27B} can be done for a more physically accurate treatment of the CR propagation through the heliosphere, e.g., to compare with data taken at different epochs/heliocentric distances.

\subsection{Steady-state Interstellar Emission Models}\label{sec:steady-state}

The \GP\ framework has an extensive history for modelling the nonthermal interstellar emissions from the MW across the electromagnetic spectrum \citep[see, e.g.,][]{2000ApJ...537..763S, 2004ApJ...613..962S, 2008ApJ...688.1078A, 2008ApJ...682..400P, 2011MNRAS.416.1152J}.
The standard approach employs a CR source spatial density described as a smoothly varying function of position that does not evolve with time, and solves for the steady-state CR distribution throughout the MW.
For the CR nuclei, from kinetic energies $\sim$100~MeV~nucleon$^{-1}$ to $\gtrsim$100~TeV~nucleon$^{-1}$ this is generally adequate.
The slow energy losses coupled with the long residence times for these particles are thought to provide sufficient mixing to effectively erase the individual contributions of the CR sources, leading to a ``sea'' of CR particles through the ISM.
For lower energies, the steady-state assumption is less valid due to the fast ionisation losses and fragmentation.
For CR electrons/positrons, their much more rapid energy losses mean that the approximation may be physically inaccurate for $\gtrsim$100~GeV energies.
However, the steady-state approach remains widely used because of its modelling simplicity and because the majority of data is covered by its applicable energy range.

We include with this release an example for steady-state interstellar emission models that enables straightforward intercomparison between predicted observables, e.g., nonthermal intensity sky maps, over a grid of CR source models.
The \verb+steady_state+ subdirectory gives a set of 3D modelling configurations differing only by their CR source spatial density distributions, which are consistently normalised at the solar system's location.
The normalisation/optimisation method used for this example (see below) enables the models to be used to make predictions for broadband nonthermal emissions from radio to the $\gtrsim$100~TeV \gray{s}.

The three CR source distributions that we employ are the pulsar distribution in the disc from Sec.~\ref{sec:prop_model_opt} (which we term here SA0), and the SA50 and SA100 models from \citet{2017ApJ...846...67P}.
The latter two correspond to a 50/50\% split of the injected CR luminosity between disc-like and spiral arms (SA50), and pure spiral arms (SA100).
The primary CR source spectra and other parameters are determined for each model by the two-part optimisation procedure described below.

For the ISM components, we use the 3D neutral gas (atomic and molecular) distribution model described by \citet{2018ApJ...856...45J} with 90\% hydrogen and 10\% helium by number, and the ionised gas distribution from the parameter optimisation example (Sec.~\ref{sec:prop_model_opt}). 
The PT11 GMF model \citep{2011ApJ...738..192P} is employed for synchrotron radiation losses/production, and we use both the R12 and F98 ISRF models from \citet{2017ApJ...846...67P} for the IC scattering losses and \gray{} production.\footnote{A configuration for each source density and ISRF variation (6 in total) is provided within deeper subdirectories named according to the respective source density models: {\tt SA0/SA50/SA100}.}

\begin{deluxetable}{lccc}[p]
\tabletypesize{\footnotesize} 
\tablecolumns{4}
\tablewidth{0pt}
\tablecaption{Optimised propagation model parameters. \label{tab:ss_params} }
\tablehead{
   & \multicolumn{3}{c}{Source Distribution} \\
   \cline{2-4}
\multicolumn{1}{l}{Parameter} & 
\colhead{SA0} & 
\colhead{SA50} &
\colhead{SA100} 
}
\startdata
\tablenotemark{a}$D_{0,xx}$ [$10^{28}$cm$^2$\,s$^{-1}$] & $4.16$ & $4.43$ & $4.66$ \\
\tablenotemark{a}$\delta_0$ & $0.353$ & $0.346$ & $0.339$ \\
$v_{A}$ [km s$^{-1}$] & $15.3$ & $17.4$ & $19.1$ \smallskip\\
\tablenotemark{b}$\gamma_0$ & $1.398$ & $1.501$ & $1.624$ \\
\tablenotemark{b}$\gamma_1$ & $2.412$ & $2.412$ & $2.418$ \\
\tablenotemark{b}$\rho_1$ [GV] & $3.79$ & $3.97$ & $4.50$ \\
\tablenotemark{b}$\gamma_{0,p}$ & $1.779$ & $1.897$ & $2.00$ \\
\tablenotemark{b}$\gamma_{1,p}$ & $2.444$ & $2.452$ & $2.48$ \\
\tablenotemark{b}$\gamma_{2,p}$ & $2.414$ & $2.415$ & $2.419$ \\
\tablenotemark{b}$\rho_{1,p}$ [GV] & $4.93$ & $7.48$ & $13.5$ \\
\tablenotemark{b}$\rho_{2,p}$ [GV] & $372$ & $282$ & $125$ \\
$\Delta_{\rm He}$ & $0.053$ & $0.045$ & $0.038$ \smallskip\\
\tablenotemark{b} $\gamma_{0,e}$ & $1.435$ & $1.478$ & $1.52$ \\
\tablenotemark{b} $\gamma_{1,e}$ & $2.774$ & $2.784$ & $2.753$ \\
\tablenotemark{b} $\gamma_{2,e}$ & $2.475$ & $2.453$ & $2.422$ \\
\tablenotemark{b} $\rho_{1,e}$ [GV] & $4.15$ & $4.78$ & $5.29$ \\
\tablenotemark{b} $\rho_{2,e}$ [GV] & $79$ & $71$ & $80$ \smallskip\\
\tablenotemark{c}$J_p$  & $4.562$ & $4.545$ & $4.394$ \\
\tablenotemark{c}$J_e$  & $4.545$ & $4.458$ & $4.502$ \smallskip\\
\tablenotemark{d}$q_{0,^{4}\rm He} \hfill [10^{-6}]\qquad\qquad$ & $ 93916$ & $ 92897$  & $ 94248$          \\
\tablenotemark{d}$q_{0,^{12}\rm C} \hfill [10^{-6}]\qquad\qquad$ & $  3121$ & $  3062$ & $  3055$ \\
\tablenotemark{d}$q_{0,^{16}\rm O} \hfill [10^{-6}]\qquad\qquad$ & $  4098$ & $  4041$ & $  4053$ \\
\tablenotemark{d}$q_{0,^{20}\rm Ne} \hfill [10^{-6}]\qquad\qquad$ & $   347$ & $   340$ & $   345$ \\
\tablenotemark{d}$q_{0,^{24}\rm Mg} \hfill [10^{-6}]\qquad\qquad$ & $   761$ & $   752$ & $   764$ \\
\tablenotemark{d}$q_{0,^{28}\rm Si} \hfill [10^{-6}]\qquad\qquad$ & $   848$ & $   842$ & $   856$ \smallskip\\
\tablenotemark{e}$\Phi_{\rm AMS,I}$ [MV] & $ 729$ & $ 741$ & $ 785$ \\
\tablenotemark{e}$\Phi_{\rm AMS,II}$ [MV] & $ 709$ & $ 729$ & $ 778$ \\
\tablenotemark{e}$\Phi_{\rm ACE/CRIS}$ [MV] & $ 359$ & $ 370$ & $ 381$            
\enddata
\tablenotetext{a}{$D(\rho) \propto \beta \rho^{\delta}$, $D_0$ is the normalisation at $\rho_0=4$~GV, $\delta = \delta_0$ for $\rho < \rho_0$ and $\delta = 0.404$ for $\rho > \rho_0$.  
Units are $10^{28}$ cm$^2$ s$^{-1}$.
}
\tablenotetext{b}{The injection spectrum is parameterised as Eq.~(\ref{eq:src_spec}). The spectral shape of the injection spectrum is the same for all species except CR $p$ and He. $\rho_1$, and $\rho_2$ are the same for $p$ and He and $\gamma_{i,\rm He} = \gamma_{i,p}-\Delta_{\rm He}$.}
\tablenotetext{c}{The CR $p$ and e$^-$ fluxes are normalised at the solar system location at a kinetic energy of 100~GeV for the former and 35~GeV for the latter.  $J_p$ is in units of $10^{-9}$ cm$^{-2}$ s$^{-1}$ sr$^{-1}$ MeV$^{-1}$ and $J_e$ is in units of $10^{-10}$ cm$^{-2}$ s$^{-1}$ sr$^{-1}$ MeV$^{-1}$.}
\tablenotetext{d}{The injection spectra for isotopes are normalised relative to the proton injection spectrum at 100~GeV/nuc. The normalisation constants for isotopes not listed here are the same as given in \citet{2016ApJ...824...16J}.}
\tablenotetext{e}{The force-field approximation is used for calculations of the solar modulation and is determined independently for each model and each observing period. $\Phi_{\rm AMS,I}$ and $\Phi_{\rm AMS,II}$ correspond to the 2011-2016 and 2011-2013 observing periods for the AMS-02 instrument, respectively.}
\end{deluxetable}

The calculations use a 3D right-handed spatial grid with the solar system on the positive $X$-axis and $Z=0$~kpc defining the Galactic plane; and, as for Sec.~\ref{sec:prop_model_opt}, we use the IAU-recommended $R_S = 8.5$~kpc for the distance from the Sun to the GC.
As for the previous example, we use the tangent grid function (Sec.~\ref{sec:grids}) for solving the propagation equations.
The parameters for the grid transformation function are chosen so that the $X/Y$ resolution nearby the solar system is $\sim$50~pc, increasing to $\sim$0.5~kpc at the boundary of the Galactic disc, which is at 20~kpc from the GC.
In the $Z$-direction the resolution is 25~pc in the plane, increasing to 0.5~kpc at the boundary of the grid at $|Z_{\rm halo}| = 6$~kpc \citep{2016ApJ...824...16J}.
The kinetic energy grid is logarithmic from 10~MeV to 1~PeV with 32 planes.

The tuning procedure follows that of \citet{2017ApJ...846...67P} and \citet{2018ApJ...856...45J}.
We employ the same set of CR data as \citet[][see their Table~1]{2019ApJ...879...91J}. 
For each of the SA0, SA50, and SA100 models, we make an initial optimisation for the individual propagation model parameters by fitting to the observed spectra of CR nuclei: Be, B, C, O, Mg, Ne, and Si.
These are kept fixed and the injection spectra for electrons, protons, and He nuclei are then fitted together to the data.
The procedure is then iterated until convergence.
(Iteration is required because the proton spectrum affects the normalisation of the heavier species, and hence the propagation parameters.)
Solar modulation is accounted for in this first step by using the force-field approximation with one modulation potential value for each observation period.

\begin{figure*}[tb!]
\centerline{
    \includegraphics[scale=0.87]{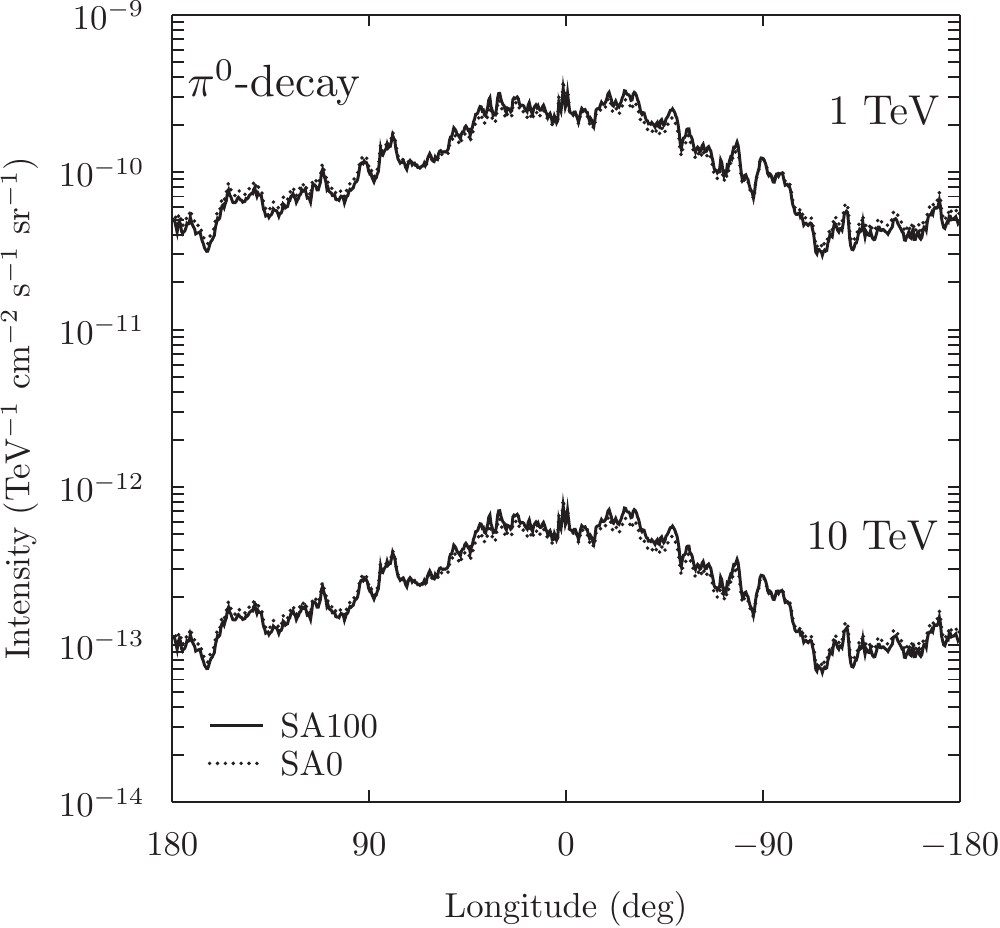}\hfill
    \includegraphics[scale=0.87]{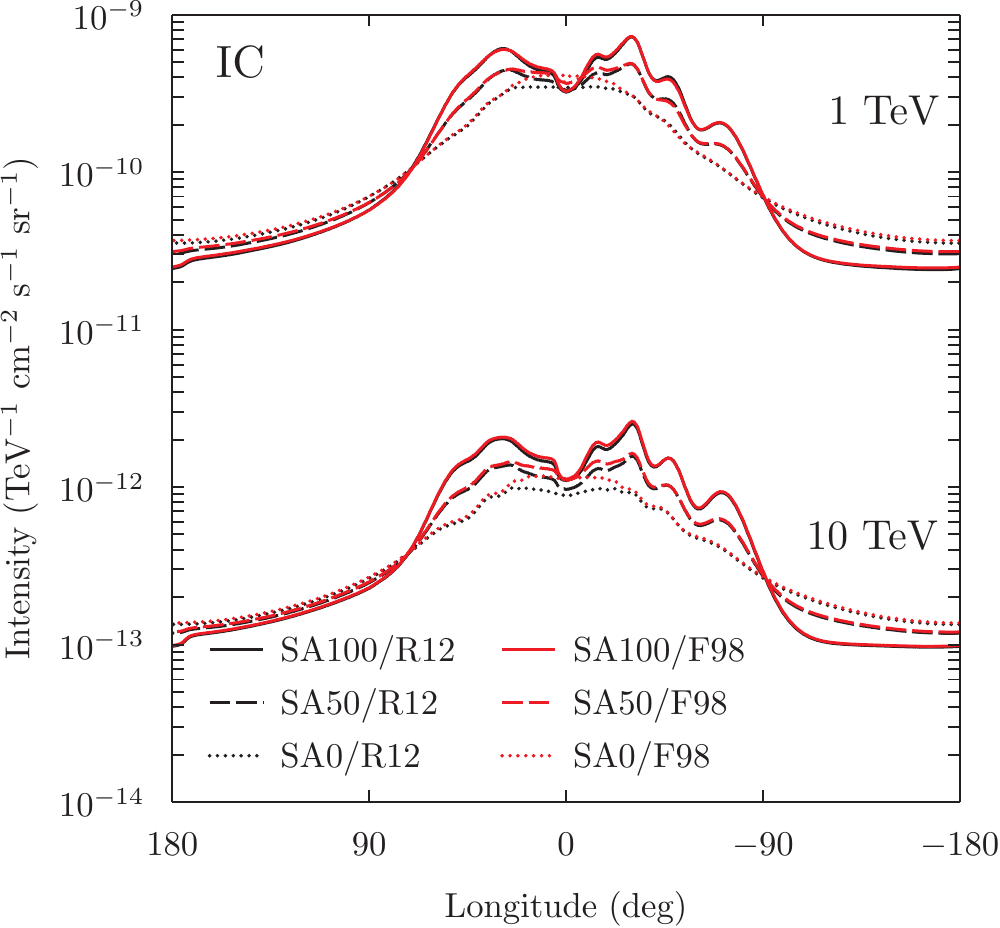}
  }
  \caption{
    VHE \gray{} emission longitude profiles averaged over $-5^\circ<b<5^\circ$ for $\pi^0$ decay (left) and IC scattering (right) for the different modelling configurations.   Line styles: solid, SA100 source density; long dashed, SA50 source density; dotted, SA0 source density.   Line colours: black, R12 ISRF model; red, F98 ISRF model.
    \label{fig:ss_vhe_longprofiles}
  }
\end{figure*}

After the initial optimisation we determine the best-fit model using $\chi^2$ and use it for extrapolation outside of the range covered by the data.
The solution for the SA100 density distribution gives the best-fit model, and its predicted local spectra are used as ``data'' giving coverage for the full CR kinetic energy range (10~MeV to 1~PeV).
We then reoptimise the parameters for the SA0 and SA50 solutions to the SA100 interstellar spectra, as for the first step of the procedure described above (omitting the solar modulation). 
This ensures that the three models give the same local CR spectra, reducing inconsistencies caused by limited data statistics and coverage over the modelled energy range.
Table~\ref{tab:ss_params} lists the parameters that are optimised and their values.

Because of the normalisation for the models, the differences will appear most prominently for the predicted interstellar emissions.
An example is shown in Fig.~\ref{fig:ss_vhe_longprofiles} where VHE \gray{} longitude profiles averaged over $-5^\circ<b<5^\circ$ for $\pi^0$-decay (left) and IC scattering (right) are shown.

For the $\pi^0$-decay profiles, there is minimal variation for the different source density models.
This is due to the slow energy losses and diffusion that smooth out the CR nuclei intensities resulting in low sensitivity to the input source density distribution.
Even for the SA100 model that has no CRs injected $\lesssim$2--3~kpc of the GC, the propagation produces nonzero CR intensities over the inner Galaxy that are somewhat comparable to the SA0 and SA50 models.

For the CR electrons\footnote{Only CR electrons are considered as a primary species for this example.} the different source densities and ISRF distributions produce more marked variations. 
For the $\gtrsim$1~TeV IC emissions the bulk are coming from the IR component of the ISRF, because the UV/optical photons are Klein-Nishina (KN) suppressed.
The higher intensity of UV/optical photons about the spiral arms for the R12 ISRF model does not produce additional structure,\footnote{The ``density squared'' effect that can be due to higher intensity ISRF and CR electrons about arm regions \citep{2017ApJ...846...67P}.} and the major difference is due to the spatial structure of the different CR source densities.
The rapid energy losses mean that the CR electron intensities are more localised about their source regions, and so the spiral structure for the SA50 and SA100 models is evident in the profiles.

While the IR distributions for both R12 and F98 models are fairly axisymmetric about the GC, their intensity is not the same everywhere.
This difference in the IR distributions across the inner Galaxy \citep[the latter is more intense; see, e.g., Fig.~6 of][]{2017ApJ...846...67P} can also be seen in the profiles calculated for the SA0 and SA50 models.
(Because $\gtrsim$1~TeV electrons cool rapidly, their intensities are very low $\lesssim$2--3~kpc from the GC for the SA100 density model and the IC profiles do not differ appreciable for either ISRF model.)

For the smooth axisymmetric SA0 source density there is also structure in the profile toward the inner Galaxy that is more apparent with increasing \gray{} energy.
(It is also there for the SA50 and SA100 models, but superimposed on the spiral structures coming from the source densities, and so less easily discerned.)
This does not come from the MW ISRF, because at the higher energies even the IR component of the ISRF becomes KN suppressed.
Only the spatially invariant CMB is causing the IC energy losses.
Instead the spiral structure for the PBSS magnetic field model affects the CR electron intensity distribution,
because the synchrotron energy losses at higher energies become the more dominant cooling process.
Into the 10s of TeV \gray{} energies and higher, the IC profiles are encoding information for the GMF and CR sources.

\begin{figure}[tb!]
  \includegraphics[scale=0.825]{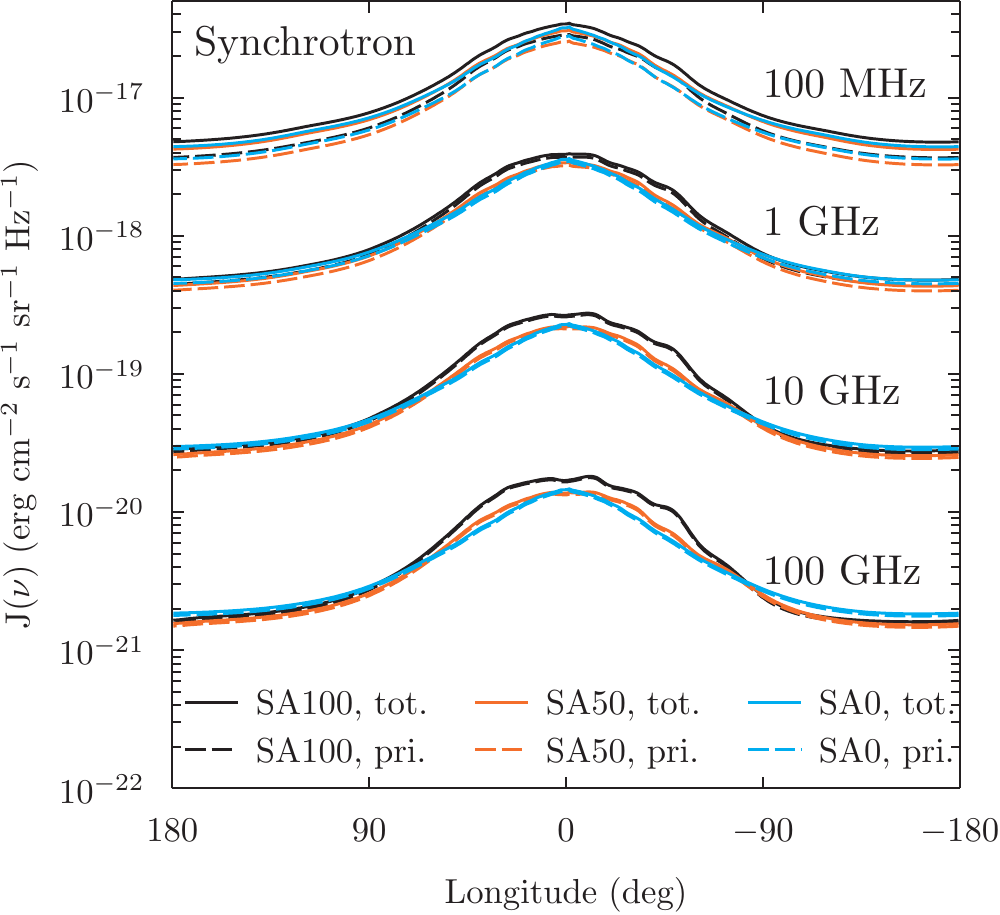}
  \caption{
    Synchrotron radiation longitude profiles averaged over $-5^\circ < b < 5^\circ$ for the SA0, SA50, and SA100 source models used by the example.   Line colours: black, SA100 source density; orange, SA50 source density; cyan, SA0 source density.   Line styles: solid, total electrons, including primary and secondaries produced via inelastic CR nuclei interactions with the gas; long dashed, primary electrons only.
    \label{fig:ss_sync_longprofiles}
  }
\end{figure}

For synchrotron emissions (shown in Fig.~\ref{fig:ss_sync_longprofiles}) the profiles are not as structured.
The CR electrons producing these have energies $\sim$0.1--30~GeV.
  The cooling and diffusion smooths out their distribution much more than for the higher energies.
  Generally, the contribution by the primary CR electrons dominates the profiles.
However, for frequencies $\lesssim$100--200~MHz there is a nonneglible contribution from secondary leptons,\footnote{While not shown in the figure, but included in the example, the profiles for lower frequencies will also reflect the absorption by the free electrons in the ISM. This is modelled using the ionised gas distribution employed for the propagation calculations, and is explicitly accounted for by the LOS integration when generating the intensity sky maps.} which can be seen as a difference between the total and primary only predicted profiles.
The secondary electrons/positrons are produced through inelastic interactions by the CR nuclei with the interstellar gas.
Their contribution is therefore tied also to the intensity of the $\pi^0$-decay emissions.

The rationale for providing this example is that direct comparison between different studies using \GP\ is often complicated, even for experienced users, because of the variation of normalisation conditions and because the modelling configurations employed may have incomplete documentation.
  The models provided here can be used in a variety of ways to develop additional scenarios with the ability to directly evaluate the impact of the changes.
  However, the same methodology described for the normalisation procedure should be followed.
  The extrapolation of spectral models for the CR sources can produce diverging predictions, and so the treatment outside the range covered by the data is important to ensure consistency.
  Optimisation to the best-fit model determined above addresses this problem, and enables a consistent framework for testing predictions for different scenarios across the broad energy range for which the diffuse emissions data are available.

\subsection{Discretised Source Ensemble Interstellar Emission Model}\label{sec:discrete_src}

The steady-state formalism  ignores the reality that the CRs are produced by discrete sources, e.g., SNRs that have finite lifetimes.
  When the source birth rate and active injection lifetime are comparable to timescales for energy losses and diffusion, spatial fluctuations in the CR intensities occur and the assumption that the ISM is prevaded by a temporally invariant CR sea is less clear.
Discretised CR source descriptions have been explored for a long time for modelling the local CR fluxes \citep[e.g.,][]{2003ApJ...582..330H, 2004ApJ...609..173T, 2006AdSpR..37.1909P, 2011JCAP...02..031M, 2012A&A...544A..92B, 2015RAA....15...15L, 2015A&A...573A.134M, 2017A&A...600A..68G, 2018JCAP...11..045M}.  

The fluctuations in the CR intensities also affect the nonthermal emissions \citep[][]{2001AIPC..587..533S,2019ApJ...887..250P}.
Because the other ISM components change over much longer time scales, at the highest energies the diffuse emissions encode the current snapshot of the CR source activity on top of the cumulative emissions from the residual particle clouds produced by sources active in the past.
At lower energies they blend into the large-scale diffuse emissions that are from the pervasive CR sea.

We have included an example that shows how the new release of \GP\ can be employed to model space/time discretised CR source ensembles and investigate the energy-dependent fluctuations in the associated diffuse emissions.  
It employs the SA100 distribution with its best-fit parameters (Sec.~\ref{sec:steady-state}) for the modelling configuration.
The discretised sampler (Sec.~\ref{sec:sampler}) is used to generate a list of discrete regions that have the same source properties (see below) with finite lifetimes distributed over the Galaxy.

The same spatial grid is used as for the steady-state example above.
The energy grid is reduced in range covering 10~GeV to 1~PeV with 32 logarithmically distributed planes.
The nuclei are normalised to data at 100~GeV, while the electrons use data at 35~GeV, which are well contained within this grid.
The upper bound of the energy grid, some 100s of TeV, is set by the validity of the diffusion approximation for the CR propagation, above which the ballistic regime becomes more appropriate \citep[e.g.,][]{2007JCAP...06..027D,2008ICRC....2..195D}.
For the ISM distributions, the R12 ISRF is employed together with the same gas and GMF models as the steady-state example.

\begin{figure*}[tb!]
\centerline{
    \includegraphics[scale=0.87]{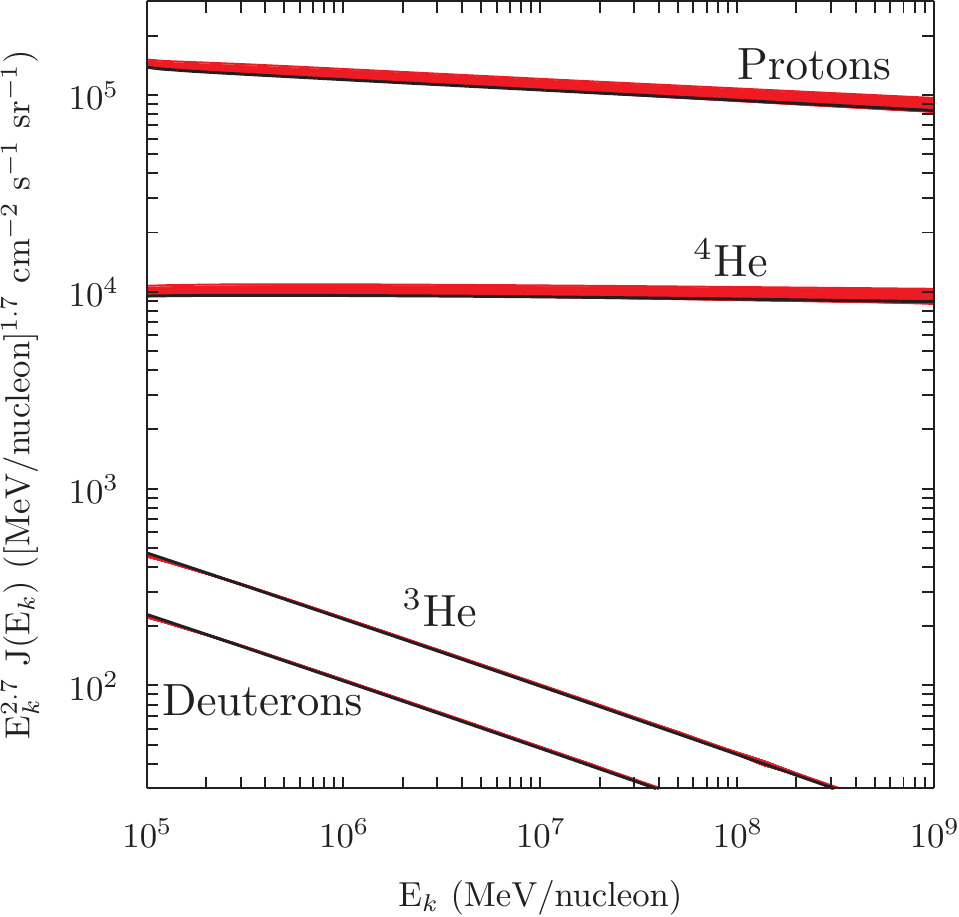}\hfill
    \includegraphics[scale=0.87]{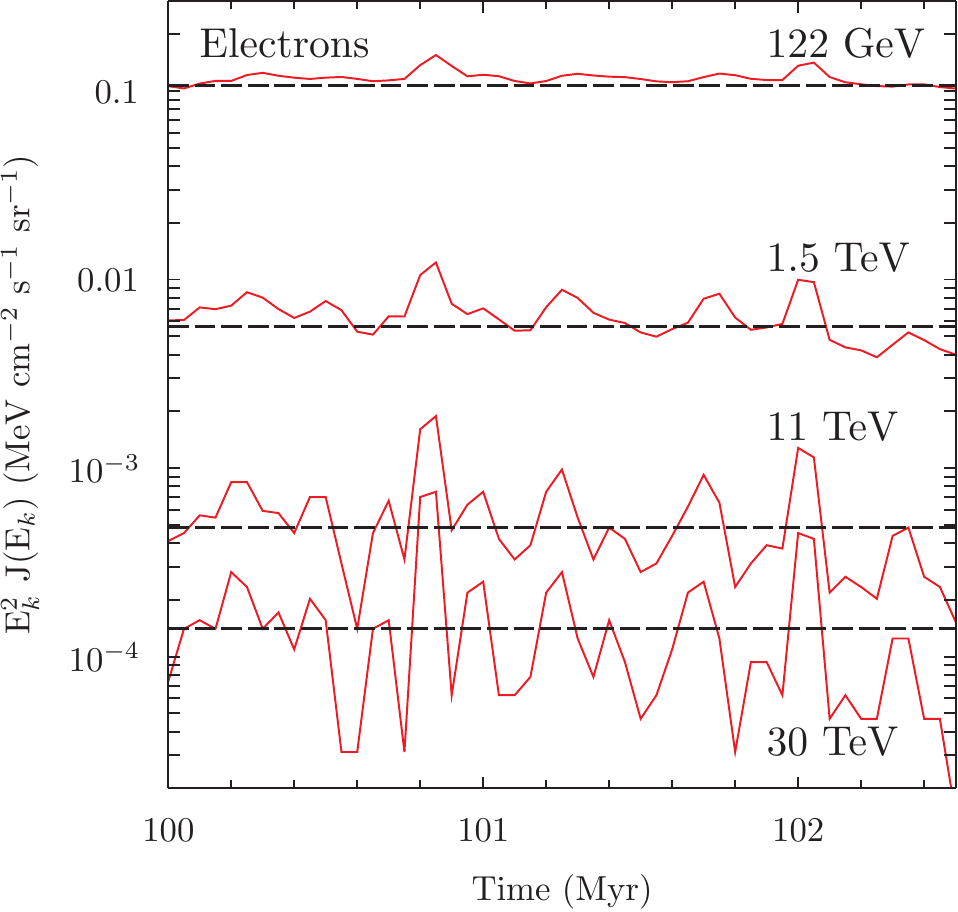}
  }
  \caption{
    CR intensities at the solar system's location for nuclei (left) and primary electrons (right).
    For the nuclei the primary species are protons and helium-4, with the secondary deuterons and helium-3 produced during the propagation through the ISM.
    The individual species' spectral intensities are labelled, but should be sufficiently distinct.
    The steady-state solutions (Sec.~\ref{sec:steady-state}) are shown for all nuclei species as the solid black lines.
    The time-dependent solutions taken at 50~kyr intervals over the 2.5~Myr ``sampling'' interval (see text) are shown as the overlaid red lines.
    For the primary electrons, we use instead a time-series representation to show the fluctuations for the intensities due to the finite source lifetime and rapid cooling.
    In this panel, the steady-state intensity is overlaid as the long dashed line at the respective energies shown.
    The corresponding time-dependent solution intensities at the same energies for the 50~kyr sampling over the 2.5~Myr epoch are shown as the solid red lines.
    \label{fig:sa100_spectra_solar_system}
}
\end{figure*}

Following \citet{2019ApJ...887..250P}, the size of an individual CR injection volume is set to be 50~pc in $X$, $Y$, and $Z$ coordinates, with frequency 0.01~yr$^{-1}$ and active time $10^5$~yr with a constant luminosity over this time and using the same spectral parameters as the smooth density distribution.
The size of the CR injection volumes is chosen to approximate the physical dimension where the CR propagation becomes ``ISM-like''; for smaller sizes the propagation is likely characterised by local effects about the true CR sources rather than in the general ISM~\citep[see, e.g.,][]{2008AdSpR..42..486P,2013ApJ...768...73M,2016MNRAS.461.3552N}.

The simulation epoch for the time-dependent solution is set to be 102.5~Myr.
There is an initial 100~Myr equilibration phase to ensure the particle intensities at their normalisation energies are comparable to the steady-state limit.
Starting with an empty Galaxy, the solution is evolved forward in 100~yr increments for 100~Myr, with the resulting CR distribution at the end of the epoch written out.
This step is time consuming, because it is done for all CR species in the simulation with sufficient time resolution to resolve the diffusion and energy losses for the electrons at the upper end of the energy scale.\footnote{If the solutions for nuclei and electrons are split, time steps suitable for the different species can be used. For simplicity, they are the same for the example as distributed.}
The solution for the CR intensities at the end of the 100~Myr period is then evolved for another 2.5~Myr, sampling the solution for the CR intensities at 50~kyr intervals.
At the end of the 2.5~Myr period, the CR distributions are normalised to the data averaging over a time window of the last 150~kyr of the simulation.
The nonthermal emissions intensity maps for all of the samples taken during the 2.5~Myr are then calculated and stored.

The CR spectral intensities for the time-dependent solution over the 2.5~Myr sampling period after the warm-up phase at the solar system are shown in Fig.~\ref{fig:sa100_spectra_solar_system}.
They are overlaid on the SA100 steady-state solution (depicted as black lines for all species). 
For the nuclei (left panel) it is straightforward to see that the steady-state and time-dependent solutions are fairly coincident for both primary and secondary species.
For the primary protons and helium, only small changes in relative normalisation with the steady-state solution are visible.
This is due to minor perturbations on top of the ``background'' in the time-dependent intensities by the nearby source contributions.
The good agreement for the secondaries shows that for the energies considered for this example the time-dependent solution has effectively ``equilibriated'' even after only $\sim$100~Myr.
The small perturbations in the primary intensities have essentially no effect on the secondaries, as expected \citep[e.g.,][]{2019ApJ...879...91J}.

\begin{figure*}[tb!]
  \centerline{
    \includegraphics[scale=0.75]{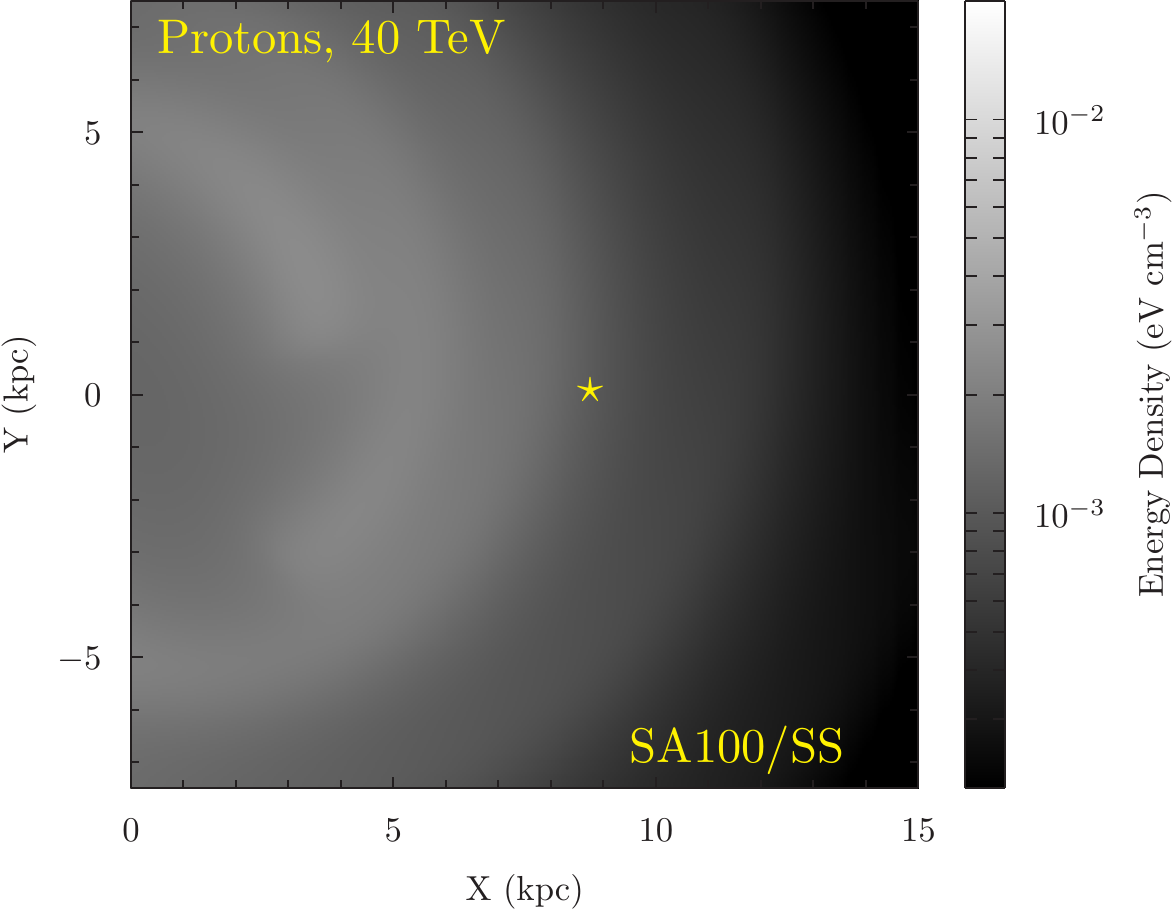}\hfill
    \includegraphics[scale=0.75]{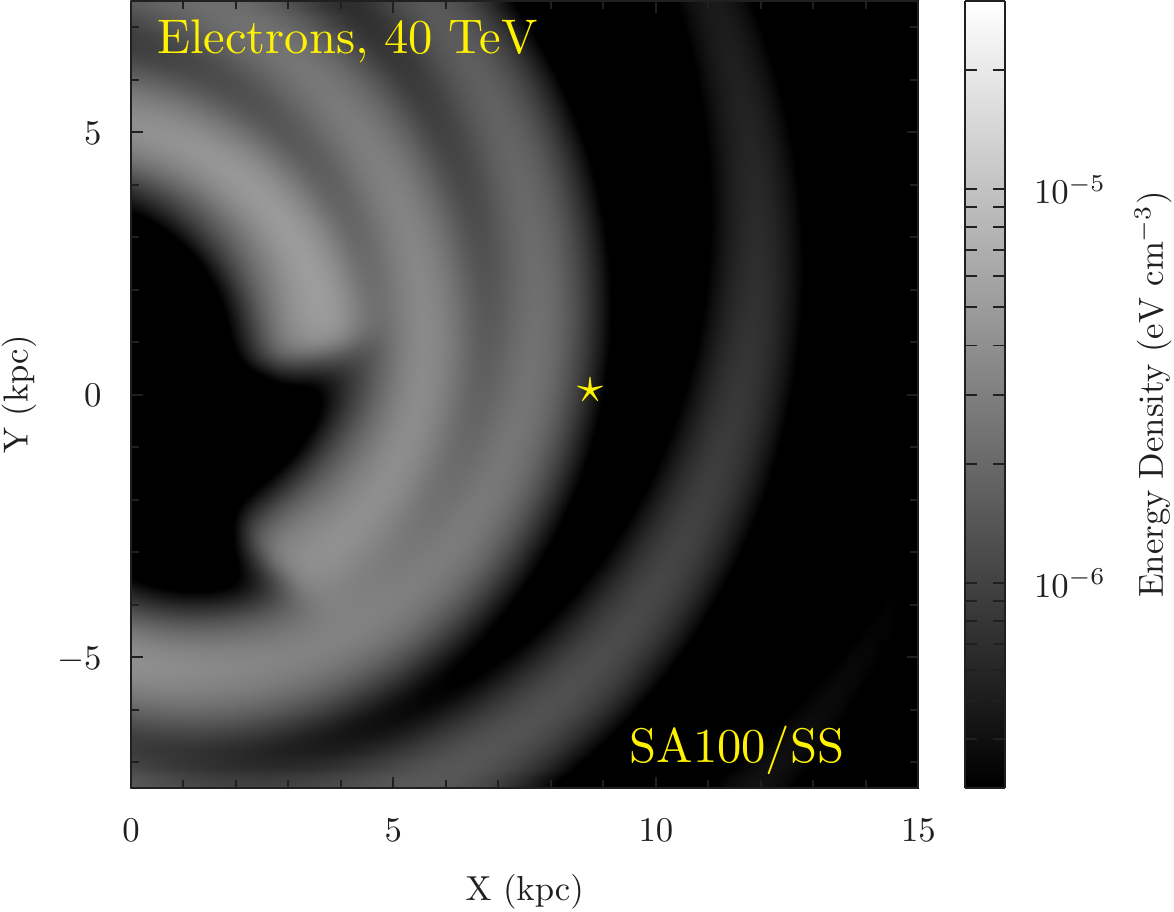}
  }
  \centerline{
    \includegraphics[scale=0.75]{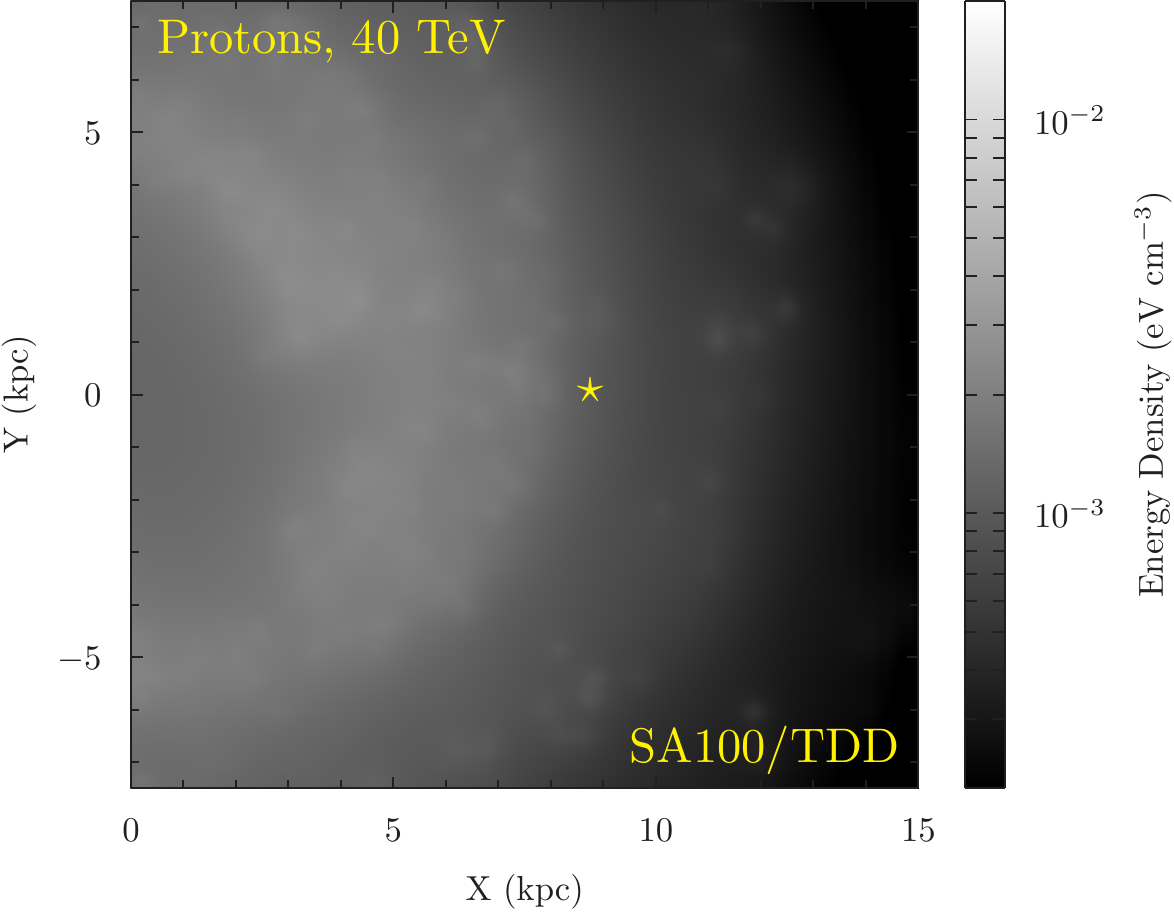}\hfill
    \includegraphics[scale=0.75]{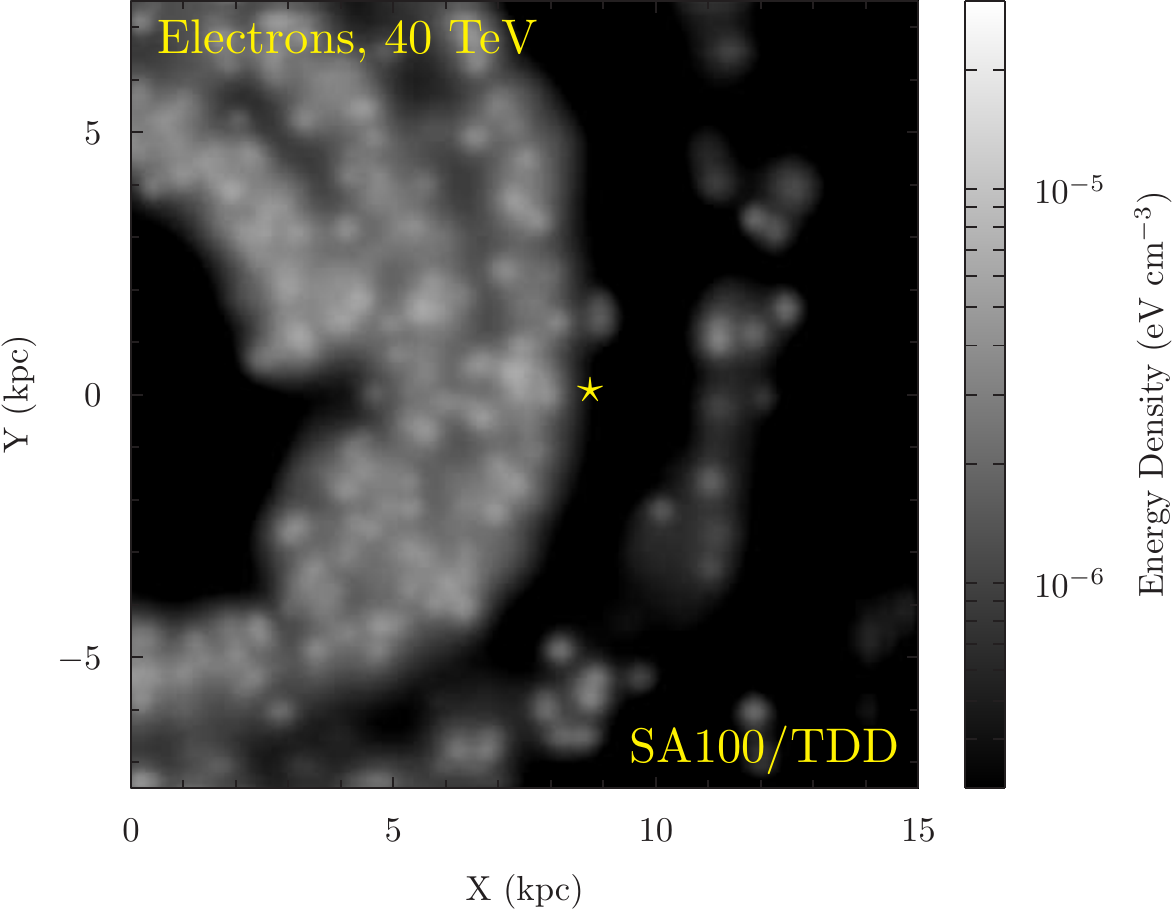}
    }
  \caption{
    The CR proton (left) and electron (right) differential energy densities at the Galactic plane for the SA100 steady-state (SS) solution (top panels) and the time-dependent discrete (TDD) solution (bottom panels).   The solar system's location is marked by the yellow star.   The time-dependent solution sample is taken at 102.5~Myr, at the end of the simulation epoch.
    \label{fig:sa100_energy_density}
  }
\end{figure*}

The CR electrons (right panel) have strongly fluctuating intensities that can be seen via the time series for the 2.5~Myr sampling epoch.
Even down to the normalisation energy (35~GeV, not shown) there are modest fluctuations in the sampled intensities over the steady-state solution.
Strong effects due to the distribution of the discretised sources, their lifetimes, and the propagation/energy losses are seen for energies $\gtrsim$1~TeV.
The finite lifetime of the individual sources, and their proximity to the solar system, coupled with the fast cooling can completely deplete the particles from being measured.
This is why the highest energy shown for the time series is at $\sim$30~TeV.
It is the last energy that has no samples in its time series that go to zero at the solar system.
The effects of a recently active source strongly influencing the particle spectrum at the solar system can also be seen in the time series (e.g., the bumps around 100.85~Myr and 102.05~Myr).

The differences between the VHE steady-state and discretised CR distributions can be gauged in Fig.~\ref{fig:sa100_energy_density}.
  The differential energy densities at 40~TeV for CR protons (left) and electrons (right) are shown.
The top row shows the steady-state solution and the bottom row shows the time-dependent solution at the end of the 2.5~Myr sampling epoch.

For the protons (and other primary nuclei), the effects of individual regions are discernable.
But they do not stand out very much from the accumulated distribution from the past injection and propagation.

For the electrons, the steady-state and time-dependent solutions have a much stronger contrast.
The time-dependent CR electron intensities at VHEs are highly localised.
At the solar system for 40~TeV at the end of the simulation there is a very low (below the scale) energy density caused by a lack of nearby activity within the recent past.
Also, for the region outside the solar circle where there is the outer arm of the source model, the patchy energy density distribution contrasts strongly with the smooth steady-state one that has an easily identifiable arm structure.

\begin{figure*}[tb!]
  \centerline{
    \includegraphics[scale=0.875]{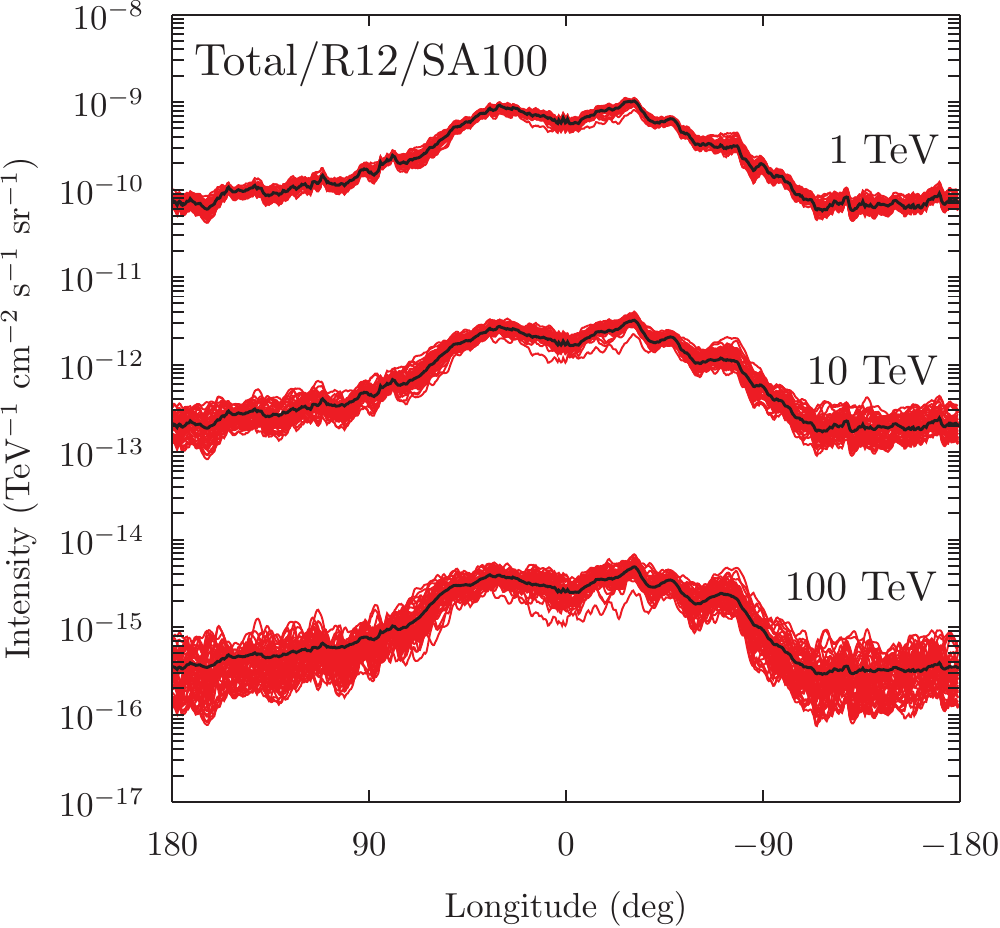}\hfill
    \includegraphics[scale=0.875]{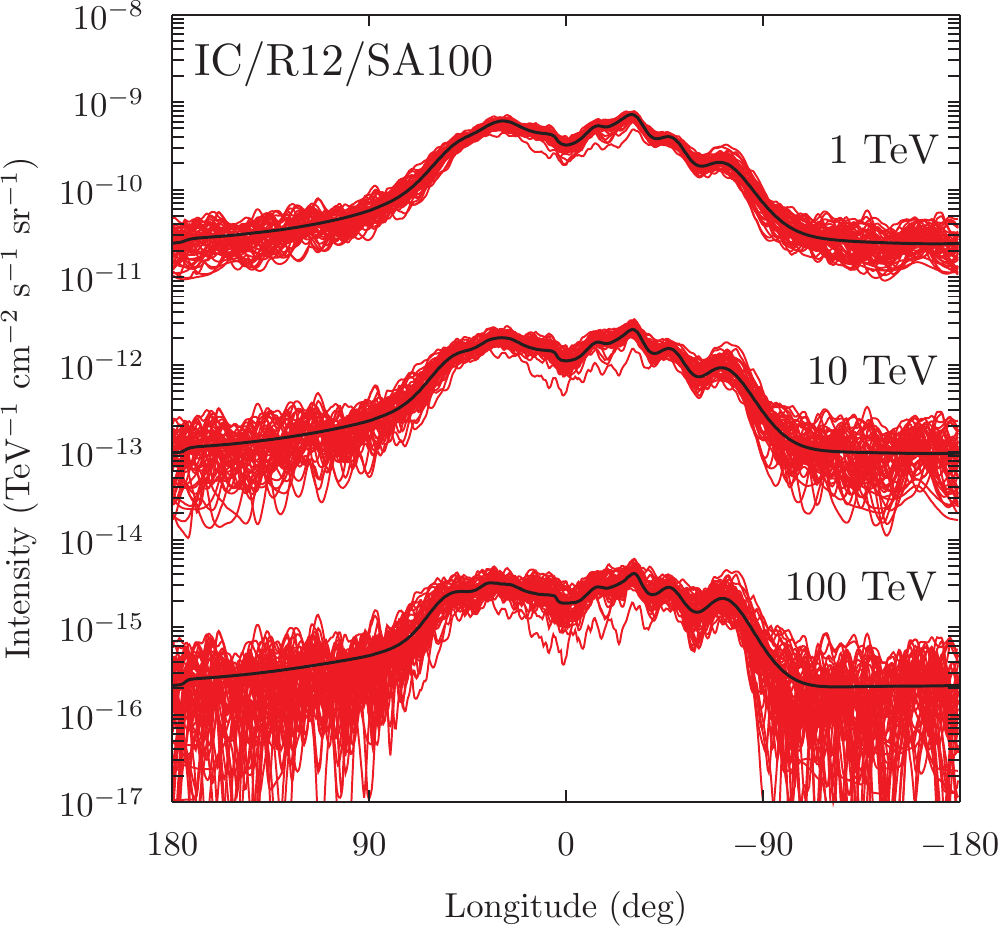}
  }
  \caption{
    VHE \gray{} emission longitude profiles averaged over $-5^\circ<b<5^\circ$ for total (left) (sum of $\pi^0$ decay and IC), and IC scattering (right) for the time-dependent and steady-state SA100 modelling configurations.   Line colours: black, SA100 steady-state solution; red, SA100 time-dependent solution at 50~kyr intervals over the 2.5~Myr sampling duration for the simulation (see text).   Note all profiles include the attendant pair-absorption attenuation calculated using the same R12 ISRF model as for the IC energy losses/\gray{} production.
    \label{fig:sa100_profiles}
}
\end{figure*}

The effect of the inhomogeneous CR intensity distributions on the \gray{} emissions can be seen in Fig.~\ref{fig:sa100_profiles}. The left panel shows the sum of the $\pi^0$ decay and IC component, while the right panel shows the IC component only.
Overlaid also on the individual longitude profiles are the corresponding SA100 steady-state emissions.
Note that the $\gamma\gamma$-absorption effect (pair production) for the R12 ISRF model accounting for the anisotropic radiation field intensity is included for both steady-state and time-dependent profiles.\footnote{\GP\ has since v56 self-consistently calculated the pair-absorption and corresponding secondary electron/positron pair source using the ISRF model employed for the CR propagation and \gray{} production.
  New for this release are the precomputed pair-absorption maps for the Galaxy using the full anisotropic formalism for the R12 and F98 ISRF models, as described by \citet{2018PhRvD..98d1302P}.}
This attenuation generally reduces the \gray{} fluxes $\gtrsim$20~TeV from individual sources as well as emissions from the diffuse ISM for longitudes $\sim$ $\pm$$45^\circ$ of the GC \citep{2006ApJ...640L.155M,2018PhRvD..98d1302P}.

The time-dependent profiles show fluctuations that become stronger with increasing energy.
They are due to the fluctuations in the IC emissions for the samples taken over the 2.5~Myr duration.
Over the inner Galaxy there is a rough balance of a high/low intensity about the steady-state solution, because there is a higher source density and corresponding lower variation in the emissions when averaged along the LOS.
For the outer Galaxy the variations are much stronger, about an order of magnitude, and biased on aggregate downward compared to the steady-state solution.
This is due to the lower probability for individual regions to be active for any one sample of the CR intensity distribution, as described already for Fig.~\ref{fig:sa100_energy_density}.

Such features for the time-dependent models may be crucial interpreting the observations of the VHE \gray{} sky.
For example, the Tibet-AS$\gamma$ data indicate that many of the \gray{s} detected within their $25^\circ<l<100^\circ$, $|b|<5^\circ$ window \citep{2021PhRvL.126n1101A} are not close to the directions of known VHE \gray{} sources, possibly indicating a ``diffuse'' origin.
The data appear inconsistent with steady-state models \citep[e.g.,][]{2018PhRvD..98d3003L,2019JCAP...12..050C}, and a discrete origin has been suggested \citep[e.g.,][]{2021arXiv210402838D,2021ApJ...919...93F}.
A significant contribution by lepton-producing source regions, such as the suggested ``TeV halos'' \citep[][]{2018PhRvL.120l1101L,2019PhRvD.100d3016S,2021arXiv210400014H,2021arXiv210111026S} may be an explanation for the lack of significant correlation between VHE \gray{} and neutrinos \citep[e.g.,][]{2021arXiv210403729Q,2021ApJ...914L...7L}.

\subsection{Inhomogeneous Diffusion}\label{sec:moving_src}

About the individual CR sources the propagation conditions are likely different to the general ISM. 
  Observations of the extended TeV emission around the Geminga and PSR~B0656+14 PWNe by the HAWC experiment \citep{2017Sci...358..911A} indeed show evidence for inhomogeneous diffusion properties nearby the individual sources, extending out to $\sim$50~pc scales.
Similar inhomogeneous CR diffusion has been observed in the Large Magellanic Cloud around the 30 Doradus star-forming region, where an analysis of combined \gray{} (protons) and radio (electrons) observations yielded a diffusion coefficient, averaged over a region with radius 200--300 pc, an order of magnitude smaller than the typical value in the MW \citep{2012ApJ...750..126M}.

To model such scenarios a two-zone approach with so-called ``slow diffusion zones'' (SDZs) about the sources with a transition to ISM conditions has been suggested \citep[e.g.,][]{FangEtAl:2018,2018PhRvD..97l3008P,TangPiran:2018}.
We include with the v57 release an example that shows how treating the inhomogeneous diffusive properties in the space surrounding the ``true'' CR sources\footnote{Compared to the previous example, where the spectral luminosity of the individual source regions combined together the effects of the actual CR source injecting particles into the surrounding space and the evolution of their spectra with the diffusive transport to the $\sim$50~pc boundary where propagation became ``ISM-like.''} can be modelled with \GP.

The example simulates the time-dependent evolution of the CR electron/positron ``cloud'' injected by the Geminga PWN, accounting for the effects of the slower diffusion as well as its proper motion.
It is based on the ``Scenario C'' from the work of \citet{2019ApJ...879...91J}, who considered a collection of scenarios for the diffusive properties about Geminga and its intrinsic source characteristics.
Below we recap the essential elements of the modelling configuration from \citet{2019ApJ...879...91J} and show expected results.

The source model for this example assumes that accelerated electrons and positrons are injected into the ISM in equal numbers with a fraction, $\eta$, of its spin-down power converted to the pairs.
The spectral model for the injected particles is described with a smoothly joined broken power law:
\begin{equation}
  \frac{dn}{dp} \propto E_k^{-\gamma_0}\left[ 1 + \left( \frac{E_k}{E_b}
  \right)^\frac{\gamma_1-\gamma_0}{s} \right]^{-s}.
  \label{eq:injectionSpectrum}
\end{equation}
Here $n$ is the number density of electrons/positrons, $p$ is the particle momentum, $E_k$ is the particle kinetic energy, and $\gamma_1$ is a power-law index at high energies.  
The smoothness parameter $s=0.5$ is assumed constant, as are the low-energy index $\gamma_0 = -1$ and the break energy $E_b=10$~GeV, respectively. 
The low-energy break is used to truncate the source spectrum and ensure physical values for the conversion efficiency, $\eta < 1$.
The injection spectrum is normalised so that the total power injected is given by the expression
\begin{equation}
  L(t) = \eta \dot{E}_0 \left( 1 + \frac{t}{\tau_0} \right)^{-2},
  \label{eq:PulsarPower}
\end{equation}
where $\dot{E}_0$ is the initial spin-down power of the pulsar, and $\tau_0 = 13$~kyr.  
The initial spin-down power is obtained using the current spin-down power of $\dot{E}=3.26\times10^{34}$ erg s$^{-1}$ assuming that the pulsar age is $T_p = 340$~kyr.

For the two-zone diffusion model, the diffusion coefficient in a confined region around the pulsar (the SDZ) is assumed lower than that in the ISM due to the increased turbulence of the magnetic field.
It is further assumed that the stronger turbulence over the region does not change the power spectrum and hence the rigidity dependence of the diffusion coefficient does not vary.
For $r$ the distance from the centre of the SDZ, the spatial dependence of the diffusion coefficient is
\begin{equation}
  D\!=\!\beta \left( \frac{\rho}{\rho_0} \right)^\delta
  \begin{cases}
    D_z, & r < r_z, \\
    \displaystyle D_z \left( \frac{D_0}{D_z} \right)^{\frac{r-r_z}{r_t-r_z}}, & r_z \le r \le r_t, \\
    D_0, & r > r_t 
  \end{cases} 
  \label{eq:diffusion}
\end{equation}
where $\beta=v/c$ is the particle velocity in units of the speed of light, $\rho$ is
the particle rigidity, $\rho_0=4$ GV is the normalisation (reference) rigidity,
$D_0$ is the normalisation of the diffusion coefficient in the general ISM,
and $D_z$ is the normalisation of the diffusion coefficient within the SDZ
with radius $r_z$.
In the transitional layer between $r_z$ and $r_t$, the normalisation of the diffusion coefficient increases exponentially with $r$ from $D_z$ to the interstellar value $D_0$.

As for the other 3D examples, the \GP\ spatial grid is right handed with the GC at the origin, the Sun at $(X_S,Y_S,Z_S)=(8.5, 0, 0)$~kpc, and the $Z$-axis oriented toward the Galactic north pole.
The distance to Geminga has been determined to be 250~pc \citep{FahertyEtAl:2007}, and it is located in this coordinate system at $(8.7407, 0.0651, 0.0186)$~kpc at the current epoch.
Given its estimated age and proper motion \citep[for details, see Sec.~2 of][]{2019ApJ...879...91J}, in our coordinate system Geminga was originally born at $(8.7320, 0.0963, -0.0449)$~kpc.
The ``Scenario C'' has Geminga travelling with constant velocity and the centre of the SDZ following the location of Geminga with its size increasing proportionally to the square root of time, normalised such that the final size of the SDZ is $(r_z,r_t)=(30,50)$~pc.

\begin{figure*}[tb!]
  \centerline{
    \includegraphics[scale=0.525]{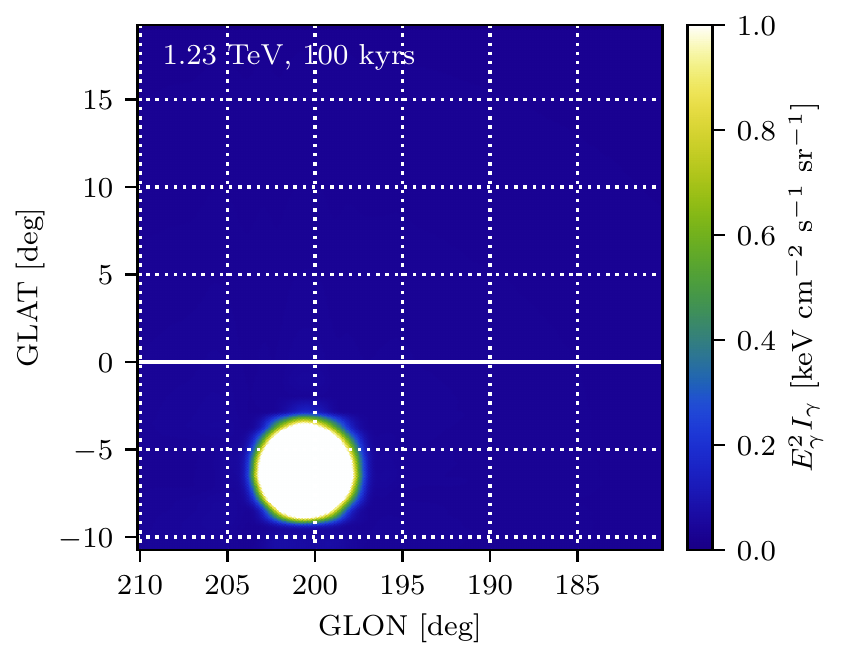}
    \includegraphics[scale=0.525]{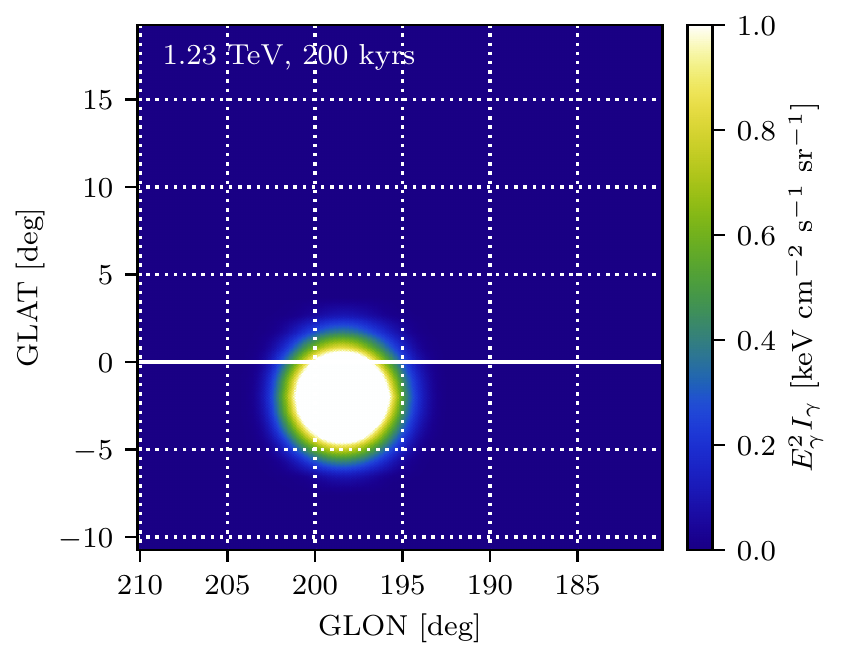}
    \includegraphics[scale=0.525]{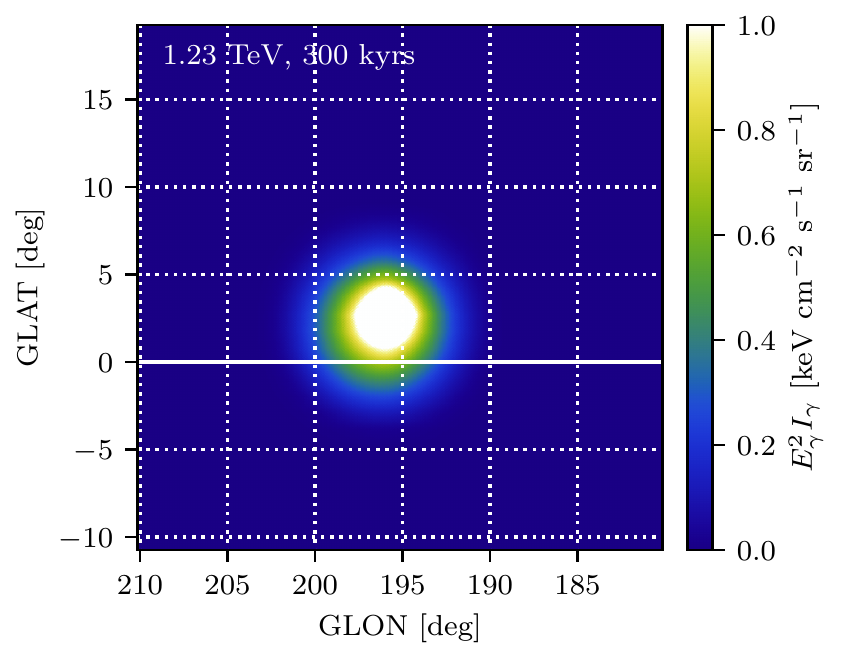}
    \includegraphics[scale=0.525]{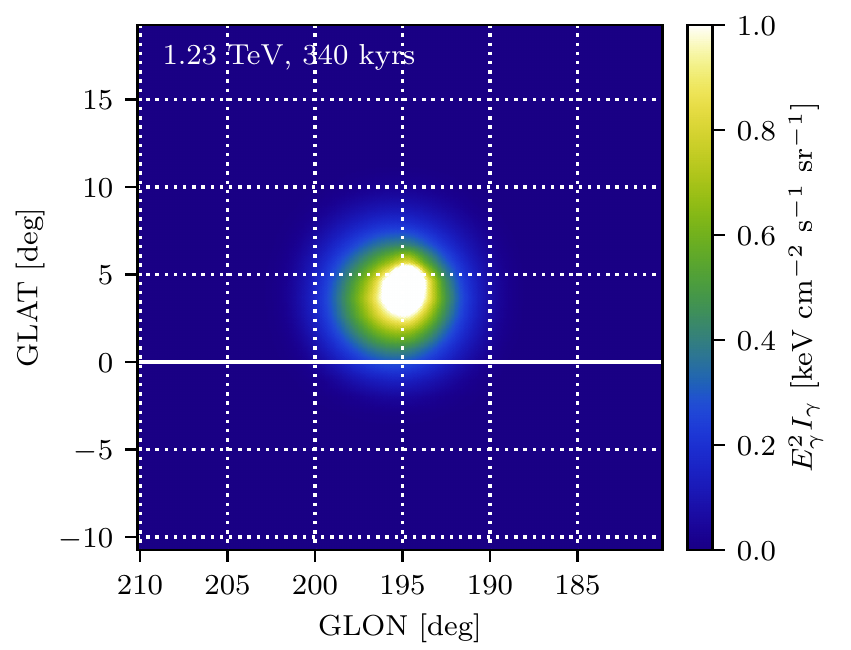}
  }
  \centerline{
    \includegraphics[scale=0.525]{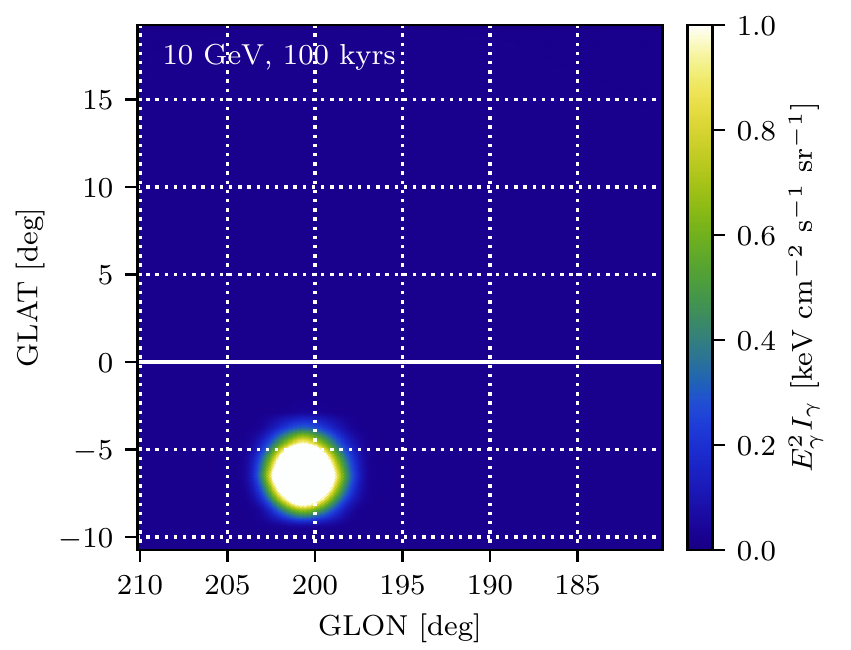}
    \includegraphics[scale=0.525]{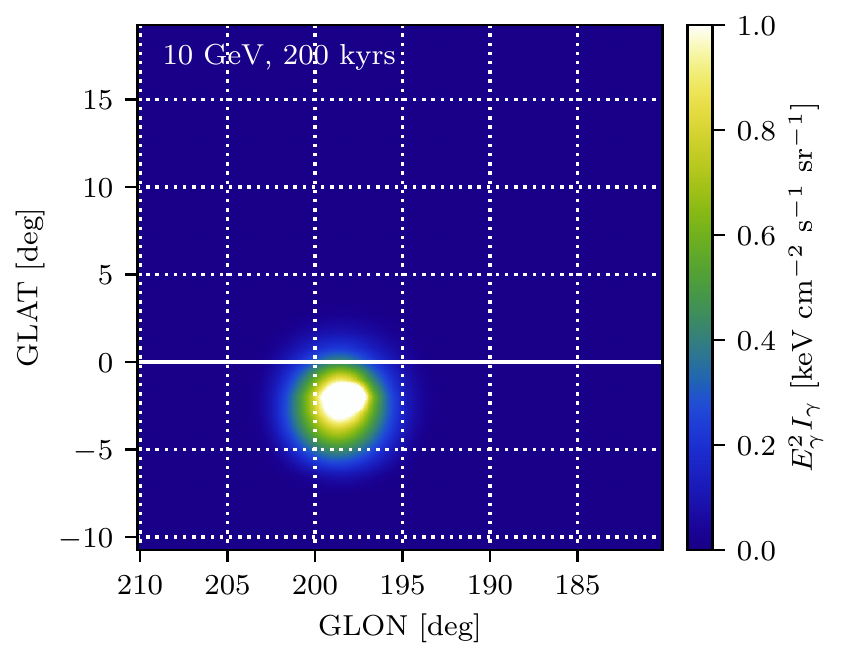}
    \includegraphics[scale=0.525]{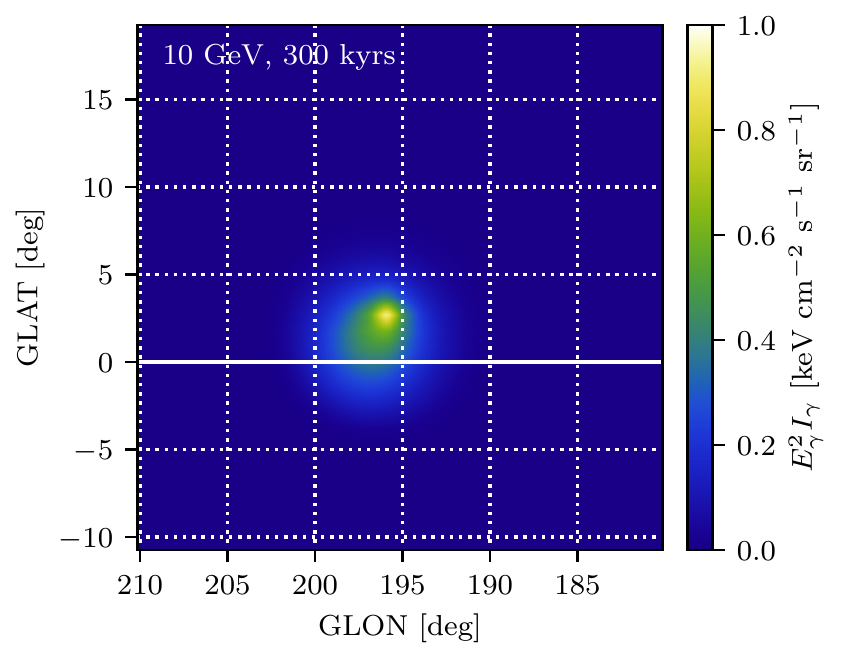}
    \includegraphics[scale=0.525]{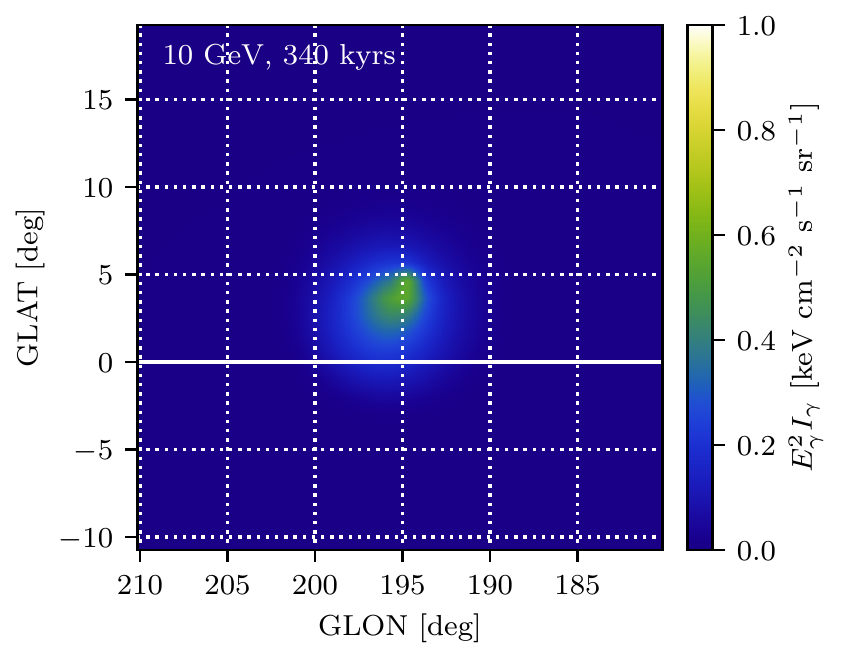}
  }
  \caption{
    IC intensities at (left to right) 100, 200, 300, and 340~kyr, respectively, for the inhomogeneous diffusion example, based on the ``Scenario C'' model described by the work of \citet{2019ApJ...879...91J}.   Upper row is at 1.23~TeV and lower row is at 10~GeV \gray{} energies.   The white solid horizontal line in each figure shows the Galactic plane 
    \label{fig:gemingaC_ic}
}
\end{figure*}

The tangent grid functions are used for the spatial grid, with parameters chosen so that the resolution is 2~pc at the central location but goes up to $0.1$\,kpc at a distance of $700$\,pc from the grid centre.
We take the centre to be at $(8.7358, 0.0962,-0.0124)$ kpc.
This is about halfway between the birthplace and the current location of Geminga for the $X$- and $Y$-axes, but only a quarter of the way for the $Z$-axis. 
This setup provides $\sim$0.2$^\circ$ resolution on the sky for objects located at a distance of 250~pc along the LOS towards the centre of the grid.\footnote{The parameters are selected to enable the configuration to be calculated $\lesssim$12~hr wall clock time, without sampling intermediate times, using a modern well-provisioned laptop, e.g., 8-core, 64GB memory.}
To minimise artificial asymmetry the grid has the current location of Geminga close to the centre of a pixel, and close enough to the centre of the nonequidistant grid so that there is little distortion due to the variable pixel size.

The calculations are performed in a square box with a width of 8~kpc.
This is wide enough so that the boundary conditions do not affect the calculations near the location of the solar system.
The nonequidistant grid allows the boundaries to be extended this far without imposing large computational costs.
A fixed time step of 50~years is used for the calculations.
This is small enough to capture the propagation and energy losses near the upper boundary of the energy grid, which is at 1~PeV.
The energy grid is logarithmic with 16 bins per decade.

For the diffusion constants, values of $D_0 = 4.5 \times 10^{28}$ cm$^2$ s$^{-1}$ and $\delta=0.35$ are assumed, and the value of $D_z = 1.3 \times 10^{26}$ cm$^2$ s$^{-1}$ is selected to be in agreement with the value required to explain the HAWC observations as determined by \citet{2017Sci...358..911A}.
The magnetic field model of \citet{SunEtAl:2008} described by their Eq.~(7) and the R12 ISRF model are used for the synchrotron/IC energy losses and nonthermal emissions production. 
The calculations also include diffusive reacceleration with an Alfv{\'e}n speed $v_A = 17$~km s$^{-1}$, as determined from the fitting to reproduce the secondary-to-primary data.  
To save CPU time the calculations are made using electrons only, because the energy losses and propagation of positrons and electrons are identical.
The IC emission is evaluated from 10~GeV to 40~TeV with 32 logarithmically distributed energy planes. 
The IC intensities are calculated on a HEALPix \citep{2005ApJ...622..759G} order 9 map, giving a resolution of about 0.1$^\circ$.

Figure~\ref{fig:gemingaC_ic} shows the sampled IC intensities (left to right) at snapshots 100, 200, 300, and 340~kyr, respectively.
The upper row shows the intensities at 1.23~TeV, and the lower row is for the 10~GeV \gray{} energies.
It is straightforward to see the progression across the modelled field of view of the \gray{} emission as the source moves along its trajectory.
The \gray{s} at the earlier times come from fairly close about the source location, with a broadening of the emission region as the SDZ expands and the particles escape, diffuse, and cool away from the source trajectory.
The asymmetry in the emissions along the source track (the ``tail'') becomes more pronounced at later times because of the evolution of the escaping particle cloud.
This asymmetrical feature associated with the Geminga proper motion was detected from an analysis of \fermilat\ data \citep{2019PhRvD.100l3015D,2021PhRvD.104h9903D} and may be a feature of the \gray{} emissions about other pulsars \citep[e.g.,][]{2020arXiv201015731Z}.

\section{Future Work}\label{sec:future}

The v57 release is the next version of the \GP\ code, which includes a number of new features to enhance the computational efficiency and utility for modelling the ensemble of CR and nonthermal emissions data.
The features related to high-spatial resolution and time-dependent CR solutions are particularly relevant for the interpretation of the VHE data for CRs and \gray{s} that are now becoming available with high statistics.
This release also includes a set of examples that demonstrate how the code can be applied to model typical use cases.
Development continues and we outline below improvements that are under investigation to be included with the next release.

The \GP\ project has as one its goals enabling reasonable scalability for the systems that the code is deployed on.
So far, this effort has been targeting toward single-unit systems, e.g., from laptops to workstations and individual servers.
To enable this, parrallelisation for \GP\ has been based on OpenMP,\footnote{https://www.openmp.org/} which operates via compiler directives for multithreading and utilisation of vector units.
It is an effective mechanism to adapt codes to take advantage of multiprocessor hardware.
The focus has been on CPUs so far, but with the ubiquity of accelerator cards (e.g., GPUs), even in laptops, it is a natural extension that the parallelisation is made to use these also.
Wide support for different targets via implementation of recent OpenMP specifications is available with the compilers that we support, such as gcc/++\footnote{https://gcc.gnu.org/} and clang/++\footnote{https://clang.llvm.org/}.
We are currently identifying algorithms in \GP\ (e.g., solvers) that may need rearchitecting to take advantage of these heterogeneous computational resources.

The more computationally and memory-resource-intensive 3D time-dependent calculations are leading us to also implement a distributed solution, possibly using the well-established Message Passing Interface (MPI).\footnote{\url{https://www.mpi-forum.org/}}
The combination with the intranode parallelisation (so-called ``hybrid OpenMP/MPI'' parallelisation) will make tracing high spatial and energy resolution for the 3D modelling more computationally tractable.
This will enable \GP\ to treat the very detailed nearby ISM and individual sources, as well as modelling the CRs and nonthermal emissions from the rest of the Galaxy efficiently all within the same framework.

Installation and deployment options are also an area that we are striving to improve. With the v57 release, all code including support libraries (Cfitsio, CCfits, HEALPix, etc.) are built from source using the system tools specified in the installer, with the \GP\ binary and utilities self-contained so that the only dependencies are on the system and tool chain libraries.
This is a straightforward installation method that maintains flexibility to develop user add-ons to the code.
However, to support those users that simply wish to just run a known release of \GP\, it may be more useful to provide a containerised (e.g., Docker\footnote{https://github.com/docker}) distribution. We are examining this possibility, which will likely decrease further the installation overhead, with then straightforward deployment across a wide range of hardware from laptops to cluster/high-performance computing.

Interfacing with other codes has been an essential functionality of \GP\ over many studies.
We have included an example (Sec.~\ref{sec:prop_model_opt}) that illustrates a method to interface with \GP\ as it is exposed as a C++ library.
However, to enhance the code utility we are investigating also offering bindings for other languages, e.g., Python.
This can provide additional flexibility for users to define their own analyses, enabling higher-level work flows with web-based notebook packages, e.g., via IPython/Jupyter.\footnote{https://ipython.org/notebook.html}
We envisage also to continue the process of splitting \GP\ into more self-contained modules.
Exposing inner routines within \GP\ this way will allow users to perform parameter scans via scripting/interpreted languages, or to more easily interface with their favourite fitting package.

Substantial improvements have been made with the v56 and v57 releases for the interstellar gas and radiation field components available for modelling the 3D ISM.
Further optimisations for these models accounting for nearby ISM structure as traced by, e.g., the dust distribution \citep[][]{2019A&A...625A.135L,2019ApJ...887...93G} are in progress.
However, the GMF models included with \GP\ are, so far, taken from the literature.
Optimising for GMF distributions via modelling of synchrotron emission including 3D CR distributions and the other ISM components that can account for features apparent in the all-sky {\it Planck} maps requires a different approach than provided by the hard-coded models.
The galstruct composition scheme introduced with the v56 release provides the functionality to enable this.
Extending the scheme also to vector fields, a much higher degree of flexibility is achieved for defining and optimising GMF model configurations.
  With the fine spatial resolutions attainable now using the grid functions formalism we will, in the future, be able to develop and optimise GMF distributions to better trace the local ISM structure.

  Building a comprehensive model for the diffuse Galactic multiwavelength and multimessenger (CRs, \gray{}, neutrinos) emission for the $\mu$eV--TeV energy range is an ambitious goal.
  Such a model can only be built self-consistently because of interdependencies between the components of the ISM, i.e., gas, GMF, and ISRF, and the diffuse Galactic multiwavelength emission.
  The Galactic distribution of all secondary species, such as $\pi^{0,\pm}$, $e^\pm$, $\bar{p}$, Li, Be, B, F, Sc, and V, is very nonuniform and depends of the distribution of gas (CR target), ionisation and bremsstrahlung energy losses, and the value of the diffusion coefficient.
  Besides, the flux of secondary $e^\pm$, produced in CR interactions with gas, is comparable to the flux of primary $e^-$ below $\sim$1–2~GeV and thus affects the predictions for synchrotron, bremsstrahlung, and IC emission.
  Accounting also for the the effects that a limited source's lifetime has on the major components of the diffuse emission, including the sources in the vicinity of the solar system, is also essential.
  Constructing such a model has been a major objective since the inception of the \GP\ project.
  With the v57 release, together with the future developments outlined above in mind, this is coming closer to realisation.
  

  \vspace{0.45cm}
  Gavin Rowell and Peter Marinos are thanked for constructive comments.
    \galprop\ development is partially funded via NASA grants Nos. NNX17AB48G, 80NSSC22K0477, 80NSSC22K0718.
    Some of the results in this paper have been derived using the 
    HEALPix~\citep{2005ApJ...622..759G} package.


\begin{appendix}

\section{\GP\ v57 Installation}\label{app:A}

An example of a successful \GP\ installation process is given in this section.
  It uses the supplied script, which assumes a Bourne Again SHell (Bash)\footnote{\url{https://www.gnu.org/software/bash/}}.
All packages are built from source using the system tools specified in the installer script \verb+install_galprop.sh+, which is at the top-level directory of the release following decompression of the distribution archive. 
The user needs to edit the variables highlighted at the top of the file \verb+install_galprop.sh+: \verb+MY_CMAKE+, \verb+MY_AUTOCONF+, \verb+MY_CC+, \verb+MY_CXX+, \verb+MY_FC+.
These are to point to \verb+CMake+ and \verb+autoconf+, and the C, C++, and Fortran compilers, respectively. We assume (at minimum) \verb+CMake+~3.17 and \verb+autoconf+~2.69.
Also, the C++ compiler requires C++11 ISO standard implementation. Examples are provided in the \verb+install_galprop.sh+ file for setting these variables on Centos~7 Linux using the \verb+devtoolset-10+ and system \verb+CMake+ and \verb+autoconf+, and on an OSX Macports installation using \verb!clang/++12! and \verb+gfortran11+. The OSX method has been tested with success on 10.15 and 10.16 versions. For these versions of OSX we highly recommend using \verb+clang/+++ because no additional changes should be needed linking with the system C++ library.

We show here the process for the OSX example that is included. For a standard Macports installation, and after the initial extraction of the archive, at the top-level directory the file \verb+install_galprop.sh+ is edited so that these variables are set:
\begin{verbatim}
MY_CMAKE=/opt/local/bin/cmake
MY_AUTOCONF=/opt/local/bin/autoconf
MY_CC=/opt/local/bin/clang-mp-12
MY_CXX=/opt/local/bin/clang++-mp-12
MY_FC=/opt/local/bin/gfortran-mp-11
\end{verbatim}

\noindent
The installer is then run from the same directory. The support libraries for \galtoolslib\ are built first, followed by \galtoolslib, additional packages used only by \GP, and then \GP.
After invocation, the installer reports the progress and, if successful, updates the user shell resource script (assuming adequate write permissions are set) with the full paths for all libraries:
\begin{verbatim}
:$ ./install_galprop.sh
------ Installation of GALPROP cosmic ray propagation code ------
--- Install required libraries ---
Install cfitsio
- Configure cfitsio [done]
- Compile cfitsio [done]
- Local install of cfitsio [done]
Install wcslib
- Configure wcslib [done]
- Compile wcslib [done]
- Local install of wcslib [done]
Install CCfits
- Configure CCfits [done]
- Compile CCfits [done]
- Local install of CCfits [done]
Install Healpix
- Configure Healpix [done]
- Compile Healpix [done]
- Local install of Healpix [done]
Install xerces-c
- Configure xerces-c [done]
- Compile xerces-c [done]
- Local install of xerces-c [done]
Install gsl
- Configure gsl [done]
- Compile gsl [done]
- Local install of gsl [done]
Install boost
- Local install of boost [done] 
Install CLHEP
- Configure CLHEP [done]
- Compile CLHEP [done]
- Local install of CLHEP [done]
Install galtoolslib
- Configure galtoolslib [done]
- Compile galtoolslib [done]
- Local install of galtoolslib [done]
Install eigen3
- Configure eigen3 [done]
- Local install of eigen3 [done]
Install Minuit2 standalone [done]
- Configure Minuit2 standalone [done]
- Compile Minuit2 standalone [done]
- Local install of Minuit2 standalone [done]
- Update .bash_profile [done]
--- Install GALPROP ---
- Configure GALPROP [done]
- Compile GALPROP [done]
- Local install of GALPROP [done]
-> Installation of GALPROP cosmic ray propagation code [done]
 Source ~/.bash_profile to set libraries for current shell to run GALPROP
\end{verbatim}

\noindent
The \GP\ executable is placed in the \verb+GALPROP-57.0.3032/bin+ directory (see Appendix~\ref{app:dir_struct} below).
To run it from the top-level directory for the current shell the resource file should be sourced:
\begin{verbatim}
:$ source ~/.bash_profile
:$ GALPROP-57.0.3032/bin/galprop 
Usage: galprop -r <run> -g <dir> -f <dir> -o <dir> -p <prefix> -c
Where: 
-r <run> (the run number of the galdef file -- required)
-g <dir> (location of the galdef directory -- default ../GALDEF)
-f <dir> (location of the fits directory -- default ../FITS)
-o <dir> (output directory)
-p <prefix> (optional prefix for the output files)
-c (optional selection for new execution path)
\end{verbatim}
\noindent
Other options available from the installer (recompilation, cleaning, etc.) can be queried:
\begin{verbatim}
:$ ./install_galprop.sh -h
Usage: ./install_galprop.sh [-h] [-n]

Install GALPROP on your system.

Optional arguments:
-h        show this help message only.
-n        new installation: compile support libraries and GALPROP.
-r        clean and compile GALPROP with optimisation.
-d        clean and compile GALPROP in debug mode.
-C        clean all builds (support libraries and GALPROP).
-D        delete GALPROP and support libraries from system.
-u        update existing GALPROP build.
\end{verbatim}
\noindent
The procedure also installs a selection of utilities in the \verb+GALPROP-57.0.3032/bin+ directory: \verb+changeGasMaps+, \verb+print_grid+, \verb+optimise_params+.
Assuming consistent galactocentric ring decomposition of the gas column densities from line-intensity information, the executable \verb+changeGasMaps+ can be used to switch the gas column density sky maps used for calculating the $\pi^0$ decay and bremsstrahlung \gray{} intensity maps without rerunning \GP.
The \verb+print_grid+ executable can be used to print out the grid function mapping.
Meanwhile, \verb+optimise_params+ is used by the parameter optimisation example (Sec.~\ref{sec:prop_model_opt}); see its \verb+README+ file in the relevant example subdirectory for usage information.

\section{\GP\ v57 Directory Structure}\label{app:dir_struct}

Following a successful installation the following directory structure is accessible (for the \verb+lib+ subdirectory contents we have suppressed expansion of the tree beyond the top level for the individual support packages):

{\scriptsize
\begin{forest}
  for tree={
        font=\ttfamily,
    grow'=0,
    child anchor=west,
    parent anchor=south,
    anchor=west,
    calign=first,
    edge path={
      \noexpand\path [draw, \forestoption{edge}]
      (!u.south west) +(7.5pt,0) |- node[fill,inner sep=1.25pt] {} (.child anchor)\forestoption{edge label};
    },
    before typesetting nodes={
      if n=1
        {insert before={[,phantom]}}
        {}
    },
    fit=band,
    before computing xy={l=15pt},
  }
  [galprop\_v57
    [DMSPECTRUMDATA]
    [GALDEF]
    [FITS
      [PairAbsorptionMaps
        [F98\_HP5]
        [R12\_HP5]
      ]
    ]
    [GALPROP-57.0.3032
      [bin]
      [include]
      [lib]
      [share]
      ]
    [build]
    [examples
      [parameter\_optimisation]
      [steady\_state
        [SA0]
        [SA50]
        [SA100]
      ]
      [time\_dependent
        [discretised\_ensemble]
        [moving\_source]
      ]
    ]
    [lib
      [build
        [boost\_1\_76\_0]
      [CCfits-2.6]
      [cfitsio-4.0.0]
      [CLHEP-2.4.4.2]
      [eigen-3.4.0]
      [galtoolslib-1.1.1006]
      [gsl-2.7]
      [Healpix\_3.50]
      [Minuit2]
      [wcslib-7.7]
      [xerces-c-3.2.3]
      ]
    ]
    [log]
    [source]
    [utils]
  ]
\end{forest}
}

\noindent
The executables built by the installation procedure are placed in the \verb+GALPROP-57.0.3032/bin+ subdirectory, with headers and libraries placed in the respective \verb+GALPROP-57.0.3032/include+ and \verb+GALPROP-57.0.3032/lib+ subdirectories.
Logging information on the configuration/build/installation process is placed in the \verb+log+ subdirectory.
The \verb+GALDEF+ subdirectory contains a vanilla example configuration with all user-defined parameters.
All examples described above (Sec.~\ref{sec:examples}) are available in the \verb+examples+ subdirectory.

\section{Finite differencing description}\label{app:finite_differences}

The CR propagation equation is
\begin{equation}
  \frac{\partial \psi}{\partial t} = q(\vec{r},p) + \vec{\nabla}\cdot(D_{xx}\vec{\nabla}\psi - \vec{V}\psi)
   + \frac{\partial}{\partial p} p^2 D_{pp} \frac{\partial}{\partial p}\frac{\psi}{p^2} 
   - \frac{\partial}{\partial p} \left[ \dot{p}\psi - \frac{p}{3}(\vec{\nabla}\cdot \vec{V})\psi \right]
   - \frac{\psi}{\tau_f} - \frac{\psi}{\tau_r}
  \label{eq:propagation}
\end{equation}
assuming an isotropic diffusion $D_{xx}$, but it may be nonuniform.
Here, $\psi$ is the CR density per unit momentum $p$, $q(\vec{r},p)$ is the CR source distribution at position $\vec{r}$, $D_{pp}\propto D_{xx}$ is the momentum diffusion constant, $V$ is the velocity of the Galactic wind, and $\tau_f$ and $\tau_r$ are the fragmentation and radioactive decay time scales, respectively.
The equation is solved using finite differences and imposing boundary conditions such that the solution is 0 on the spatial boundary of the grid, except at $R=0$ when running with two spatial dimensions.

 The differencing scheme has been modified slightly from the original scheme described in \cite{1998ApJ...509..212S}.
 In particular, the momentum diffusion is now fully resolved using the chain rule before applying the finite differences and the spatial diffusion at $R=0$ is now handled differently.
 The finite differences now also explicitly handle spatially varying $D_{xx}$ and all the second derivatives are corrected for the grid functions.
 In addition to this, the direct solvers now use finite differences in $\log p$ rather than $p$.
 This is a more natural scheme given that the kinetic energy grid is logarithmic.
 We keep the original linear differences in the other solvers for backwards compatibility.
 Finally, the treatment for the advection velocity, which is defined as 
\begin{equation}
  \vec{V} = 
  \begin{cases}
   \left( V_0 + Z \frac{dV}{dZ}\right) \hat{Z} & Z > 0 \\
   \left( -V_0 + Z \frac{dV}{dZ}\right) \hat{Z} & Z < 0 
  \end{cases}
  \label{eq:advection_velocity}
\end{equation}
has now been improved near the Galactic plane to take into account the rapid change caused by the change in sign in front of $V_0$.

We use only nearest neighbours in the finite differences, so in the end the right-hand side of Eq.~(\ref{eq:propagation}) is approximated as
\begin{equation}
  \frac{\partial \psi_i}{\partial t} \approx \frac{\alpha_1 \psi_{i-1} - \alpha_2 \psi_{i} + \alpha_3 \psi_{i+1}}{\Delta t} + q_i.
  \label{eq:finite_difference}
\end{equation}
The finite differences are described in detail below and the values for $\alpha_i$ derived for each part of Eq.~(\ref{eq:propagation}).  We note that the indexing $i$ will be running over whatever coordinate we are considering at that time.  Also, because of the grid functions, the step size $\Delta_{Q}$ is the gradient of the grid function $Q(\zeta)$.
We will highlight differences compared to previous versions, as well as differences between the direct solvers and the operator splitting solvers.
The values of $\alpha_i$ for different terms in the diffusion equation are summarised in Table~\ref{tab:alphas}.  At boundaries corresponding to the edge of the calculation box (all spatial boundaries except $R=0$), $\psi = 0$.

\subsection{Spatial diffusion}

This is the term $\vec{\nabla}\cdot(D_{xx}\vec{\nabla}\psi)$ and corresponds to the spatial diffusion of the CRs.
As in other places, we fully calculate the derivative analytically before applying the differentiation.
For the 3D $X$, $Y$, and $Z$ coordinates, the derivative expands to
\begin{equation}
  \begin{split}
    \vec{\nabla}\cdot(D_{xx}\vec{\nabla}\psi)_X =& D_{xx}\frac{\partial^2 \psi}{\partial X^2} + \frac{\partial \psi}{\partial X}\frac{\partial D_{xx}}{\partial X}\\
    =& D_{xx}\left( \frac{d\zeta}{d X} \right)^2 \frac{\partial^2 \psi}{\partial \zeta^2} - D_{xx}\left( \frac{d\zeta}{d X} \right)^3\frac{d^2 X}{d \zeta^2}\frac{\partial \psi}{\partial \zeta} + \left( \frac{d\zeta}{d X} \right)^2 \frac{\partial \psi}{\partial \zeta}\frac{\partial D_{xx}}{\partial \zeta},
  \end{split}
  \label{eq:differences_diffusion_X}
\end{equation}
where we have also included the effects of possible nonlinear grid functions $X(\zeta)$.
We use $X$ here as an example; the equations are equivalent for the $Y$ and $Z$ coordinates.
The second derivatives are approximated as
\begin{equation}
  \frac{d^2 f}{d x^2} \approx \frac{f(x+\Delta x) - 2f(x) + f(x-\Delta x)}{(\Delta x)^2},
  \label{eq:differences_second}
\end{equation}
while the first derivatives are approximated with central differencing:
\begin{equation}
    \frac{d f}{d x} \approx \frac{f(x+\Delta x) - f(x-\Delta x)}{\Delta x}.
  \label{eq:central_difference}
\end{equation}
The central differencing is appropriate here because the diffusion process is equally likely in both directions.
At the boundaries, the central difference is still used for $\partial \psi/\partial \zeta$, where it is assumed that $\psi=0$.
For $D_{xx}$, we use forward/backward differencing depending on the boundary.
We use numerical approximation for the derivative of $D_{xx}$, because its value may depend implicitly or explicitly on $\psi$.

  \begin{deluxetable}{F{0.9in}F{0.4in}F{1.7in}F{1.7in}F{1.7in}}[tb!]
  \scriptsize
  \tablecolumns{5}
  \tablecaption{Coefficients for the finite differences in Eq.~(\ref{eq:finite_difference}) \label{tab:alphas}}
  \tablehead{\colhead{Term} & \colhead{Coordinate} & \colhead{$\alpha_1/\Delta_t$}& \colhead{$\alpha_2/\Delta_t$}& \colhead{$\alpha_3/\Delta_t$}}
  \startdata
    $\vec{\nabla}\cdot(D_{xx}\vec{\nabla}\psi)$  & $R>0$ &
    \begin{gather*}
      \frac{D_{xx,i}}{2(\Delta R_i)^2}\left( 2 + \frac{1}{\Delta R_i}\frac{d^2R}{d\zeta^2} -\frac{\Delta R_i}{R_i} \right) \\ 
      - \frac{D_{xx,i+1}-D_{xx,i-1}}{4(\Delta R_i)^2} 
    \end{gather*} 
    &
    \[ 
      \frac{2 D_{xx,i}}{(\Delta R_i)^2} 
    \] 
    &
    \begin{gather*} 
      \frac{D_{xx,i}}{2(\Delta R_i)^2}\left( 2 - \frac{1}{\Delta R_i}\frac{d^2R}{d\zeta^2} + \frac{\Delta R_i}{R_i} \right) \\ 
      + \frac{D_{xx,i+1}-D_{xx,i-1}}{4(\Delta R_i)^2}
    \end{gather*} 
    \\
    & $R=0$ & \nodata &    \[ 
      \frac{2D_{xx,i}}{(\Delta R_i)^2} \left( 1 - \frac{1}{\Delta R_i}\frac{d^2R}{d\zeta^2} \right)
    \]
    & 
    \[
      \frac{2D_{xx,i}}{(\Delta R_i)^2} \left( 1 - \frac{1}{\Delta R_i}\frac{d^2R}{d\zeta^2} \right)
    \]
    \\
    & $X$, $Y$, $Z$ & 
    \begin{gather*}
      \frac{D_{xx,i}}{2(\Delta X_i)^2}\left( 2 + \frac{1}{\Delta X_i}\frac{d^2X}{d\zeta^2} \right) \\
      - \frac{D_{xx,i+1}-D_{xx,i-1}}{4(\Delta X_i)^2}
    \end{gather*} 
    &
    \[
    \frac{2 D_{xx,i}}{(\Delta X_i)^2} 
    \]
    &
    \begin{gather*}
      \frac{D_{xx,i}}{2(\Delta X_i)^2}\left( 2 - \frac{1}{\Delta X_i}\frac{d^2X}{d\zeta^2} \right) \\
      + \frac{D_{xx,i+1}-D_{xx,i-1}}{4(\Delta X_i)^2}
    \end{gather*}
    \\
    $-\vec{\nabla}\cdot(\vec{V}\psi)$ & $Z>0$ & 
    \[ \frac{V_i}{\Delta Z_i} \] & \[ \frac{V_i}{\Delta Z_i} + \frac{dV}{dZ} \] & \nodata \\
    & $Z<0$ & \nodata & \[ \frac{-V_i}{\Delta Z_i} + \frac{dV}{dZ} \] & \[ \frac{-V_i}{\Delta Z_i} \] \\
    & $Z=0$ & \nodata & \[ \frac{dV}{dZ} + \frac{2 V_0}{\Delta Z_i} \]  & \nodata \\
    \[
      \frac{\partial}{\partial p} \left(p^2 D_{pp} \frac{\partial}{\partial p}\frac{\psi}{p^2}\right)
    \]
    & $p$\tablenotemark{a} &  \begin{gather*}
      \frac{2 D_{pp,i}}{p_{-1}^{+1}}\left( \frac{1}{p_{-1}^{+0}} + \frac{1}{p_i} \right) \\
      - \frac{D_{pp,i+1} - D_{pp,i-1}}{(p_{-1}^{+1})^2} 
    \end{gather*}
    &
    \begin{gather*}
      \frac{2D_{pp,i}}{(p_{+0}^{+1})(p_{-1}^{+0})} - \frac{2D_{pp,i}}{p_i^2} \\
      + \frac{2(D_{pp,i+1} - D_{pp,i-1})}{p_i(p_{-1}^{+1})}
    \end{gather*}
    &
    \begin{gather*}
      \frac{2 D_{pp,i}}{p_{-1}^{+1}}\left( \frac{1}{p^{+1}_{+0}} + \frac{1}{p_i} \right) \\
      + \frac{D_{pp,i+1} - D_{pp,i-1}}{(p^{+1}_{-1})^2} 
    \end{gather*}
    \\
    & $\log p$\tablenotemark{b} &
    \begin{gather*}
      \frac{2 D_{pp,i}}{p_i^2 (\logp{+1}{-1}) (\logp{+0}{-1})}\\ - 
      \frac{D_{pp,i+1}-D_{pp,i-1}}{p_i^2 (\logp{+1}{-1})^2} +
      \frac{3 D_{pp,i}}{p_i^2 (\logp{+1}{-1})}
    \end{gather*}
    &
    \begin{gather*}
      \frac{2 D_{pp,i}}{p_i^2 (\logp{+1}{-1})} \left( \frac{1}{\logp{+1}{+0}} + \frac{1}{\logp{+0}{-1}}\right)  \\+ 
      \frac{D_{pp,i+1}-D_{pp,i-1}}{p_i^2 (\logp{+1}{-1})} -
      \frac{2 D_{pp,i}}{p_i^2}
    \end{gather*}
    &
    \begin{gather*}
      \frac{2 D_{pp,i}}{p_i^2 (\logp{+1}{-1}) (\logp{+1}{+0})}\\ +
      \frac{D_{pp,i+1}-D_{pp,i-1}}{p_i^2 (\logp{+1}{-1})^2} -
      \frac{3 D_{pp,i}}{p_i^2 (\logp{+1}{-1})}
    \end{gather*} 
    \\
    \[
      \frac{\partial}{\partial p}(\dot{p}\psi)   \]
    & $p$ & \nodata &
    \[
      \frac{\dot{p}_i}{p^{+1}_{+0}}
    \]
    &
    \[
      \frac{\dot{p}_{i+1}}{p^{+1}_{+0}}
    \]
    \\
    & $\log p$ & \nodata &
    \[
      \frac{\dot{p}_i}{p_i (\logp{+1}{+0})}
    \]
    &
    \[
      \frac{\dot{p}_{i+1}}{p_i (\logp{+1}{+0})}
    \]
  \enddata
  \tablenotetext{a}{$p_{a}^{b} = p_{i+b}-p_{i+a}$}
  \tablenotetext{b}{$\logp{b}{a} = \log(p_{i+b}) - \log(p_{i+b})$}
\end{deluxetable}

For the galactocentric $R$ coordinate, the situation is different, because the differential element is not the same in cylindrical coordinates.
In this case, 
\begin{equation}
  \begin{split}
    \vec{\nabla}\cdot(D_{xx}\vec{\nabla}\psi)_R =& D_{xx}\frac{\partial^2 \psi}{\partial R^2} + \frac{D_{xx}}{R}\frac{\partial \psi}{\partial R} + \frac{\partial \psi}{\partial R}\frac{\partial D_{xx}}{\partial X}\\
    =& D_{xx}\left( \frac{d\zeta}{d R} \right)^2 \frac{\partial^2 \psi}{\partial \zeta^2} - D_{xx}\left( \frac{d\zeta}{d R} \right)^3\frac{d^2 R}{d \zeta^2}\frac{\partial \psi}{\partial \zeta} + \frac{D_{xx}}{R}\frac{d\zeta}{d R}\frac{\partial \psi}{\partial \zeta} + \left( \frac{d\zeta}{d R} \right)^2 \frac{\partial \psi}{\partial \zeta}\frac{\partial D_{xx}}{\partial \zeta}.
  \end{split}
  \label{eq:differences_diffusion_R}
\end{equation}We note that there is a term here with $1/R$, which diverges at $R=0$, unless 
\begin{equation}
  \lim_{R \to 0}\frac{\partial \psi}{\partial \zeta} = 0.
  \label{eq:derivative_limit}
\end{equation}
In that case, we can apply the L'H\^{o}pital's rule to get
\begin{equation}
  \begin{split}
    \vec{\nabla}\cdot(D_{xx}\vec{\nabla}\psi)_{R=0} &= 2D_{xx}\frac{\partial^2 \psi}{\partial R^2} + \frac{\partial \psi}{\partial R}\frac{\partial D_{xx}}{\partial R} \\
    &= 2D_{xx}\left( \frac{d\zeta}{d R} \right)^2 \frac{\partial^2 \psi}{\partial \zeta^2} - 2D_{xx}\left( \frac{d\zeta}{d R} \right)^3\frac{d^2 R}{d \zeta^2}\frac{\partial \psi}{\partial \zeta} + \left( \frac{d\zeta}{d R} \right)^2 \frac{\partial \psi}{\partial \zeta}\frac{\partial D_{xx}}{\partial \zeta}.
  \end{split}
  \label{eq:differences_diffusion_R=0}
\end{equation}
The second derivative needs some attention, because our stencil only allows nearest neighbours.
This we solve by assuming $\psi_{-1} = \psi_0$, resulting in
\begin{equation}
  \frac{\partial^2 \psi}{\partial \zeta^2} \approx_{R=0} \frac{\psi_1 - \psi_0}{(\Delta \zeta)^2}.
  \label{eq:second_derivative_R=0}
\end{equation}
We note that this approximation is somewhat crude and the resulting CR flux near $R=0$ will not be accurate.
For accurate calculations near $R=0$, \GP\ needs to be run with three spatial dimensions.\footnote{For typical runs matching the CR observations, the fractional error using the 2D solution is about 5\%.}
This special treatment at $R=0$ is new for the \galprop\ v57 release.
Earlier versions have assumed that $\psi$ implicitly vanishes for $R<0$.

\subsection{Advection}

Advection of CRs by Galactic winds includes two terms in the diffusion equation, one spatial term and one momentum term.
We will start with the spatial term $-\vec{\nabla}\cdot(\vec{V}\psi)$.  In the \GP\ code, the velocity profile is modelled using Eq.~\eqref{eq:advection_velocity}.
This velocity profile is perpendicular to and away from the Galactic plane.
The term, fully evaluated, is thus
\begin{equation}
  -\vec{\nabla}\cdot(\vec{V}\psi) = -V \frac{\partial \psi}{\partial Z} - \psi \frac{dV}{dZ}
  \label{eq:advection_spatial}
\end{equation}
Due to the change of sign in $V$, the difference is evaluated using forward differencing for $Z<0$ and backward differencing for $Z>0$.
At $V=0$, the formula breaks down, because $V_0$ changes sign.
For the central spatial element, it is assumed that the value of $V_0$ changes sign over its width, and that $V=0$ at the centre.
The term is in that case
\begin{equation}
  -\vec{\nabla}\cdot(\vec{V}\psi)(Z=0) \approx -\psi\left( \frac{dV}{dZ} + \frac{2V_0}{(\Delta Z)(Z=0)} \right),
  \label{eq:advection_spatial_center}
\end{equation}
where $(\Delta Z)(Z=0)$ is the element width at the centre.
This treatment is thus dependent on the width, unless $V_0 = 0$.
The special treatment of the velocity at $Z=0$ is new in \galprop\ v57.

The second term regarding the advection is the momentum term:
\begin{equation}
  \begin{split}
    \frac{\partial}{\partial p}\left( \frac{p}{3}(\vec{\nabla}\cdot \vec{V}) \right) =&
    \frac{1}{3}\frac{dV}{dZ}\left( \psi + p \frac{\partial \psi}{\partial p} \right) \\
    =& \frac{1}{3}\frac{dV}{dZ}\left( \psi +  \frac{\partial \psi}{\partial \log p} \right).
  \end{split}
  \label{eq:advection_momentum}
\end{equation}
The first form uses a linear differential in momentum, while the latter uses a logarithmic differential.  For the finite difference, we use in both cases forward differencing for the partial derivative.

\subsection{Momentum Diffusion (Reacceleration)}

The momentum-space diffusion term,
\[
\frac{\partial}{\partial p} p^2 D_{pp} \frac{\partial}{\partial p}\frac{\psi}{p^2}
\]
is modelling the reacceleration from the second-order Fermi scattering of the CRs as they propagate in the ISM.
We fully evaluate all terms, which results in
\begin{equation}
  \frac{\partial}{\partial p} p^2 D_{pp} \frac{\partial}{\partial p}\frac{\psi}{p^2} = 
  \frac{2\psi}{p}\left( \frac{D_{pp}}{p} - \frac{\partial D_{pp}}{\partial p} \right) + 
  \frac{\partial \psi}{\partial p}\left( \frac{\partial D_{pp}}{\partial p} - \frac{2 D_{pp}}{p} \right) +
  D_{pp}\frac{\partial^2 \psi}{\partial p^2}
  \label{eq:reacceleration_linear}
\end{equation}
for the linear momentum differential and 
\begin{equation}
  \frac{\partial}{\partial p} p^2 D_{pp} \frac{\partial}{\partial p}\frac{\psi}{p^2} =
  \frac{1}{p^2}\left[ 2\psi\left( D_{pp} - \frac{\partial D_{pp}}{\partial \log p} \right) +
    \frac{\partial \psi}{\partial \log p}\left( \frac{\partial D_{pp}}{\partial \log p} - 3 D_{pp} \right) +
    D_{pp} \frac{\partial^2 \psi}{\partial (\log p)^2}
  \right]
  \label{eq:reacceleration_log}
\end{equation}
for the logarithmic momentum differential.
The differentials are all centrally evaluated and it is assumed that $\psi=0$ outside the boundaries.
This approximation is not realistic at the boundaries, and hence the solution will not be accurate close to them.
In practice there should be a buffer of 1--2 elements near the boundaries that are not used for physical interpretation.

\subsection{Energy losses}

This is one of the simplest terms to treat, and we use finite differences for the entire term:
\[
  \frac{\partial}{\partial p}(\dot{p}\psi) = \frac{1}{p}\frac{\partial}{\partial \log p}(\dot{p} \psi).
\]
Here it is appropriate to use forward differencing, and we again assume that $\psi=0$ outside the boundaries.
This will only affect the upper boundary in this case, unlike with the diffusive reacceleration term that is affected at both boundaries.

\subsection{Fragmentation and decay}

This is the simplest term, and for the direct solvers we simply add to $\alpha_2$.
This cannot be done for operator splitting, because the term would be added multiple times.
For operator splitting, a fraction of the term is added instead, a third for two spatial dimensions and a fourth for three spatial dimensions.
In this way, the term is properly accounted for when operator splitting is used.

\section{Solvers and Tests}\label{app:tests}

The \GP\ solvers can be used for obtaining either steady-state or time-dependent solutions for the CR distributions.
Selecting the solver is done with the configuration parameter {\tt solution\_method}.
Table~\ref{tab:solvers} gives the list of solvers, which are currently implemented that we recommend for use.
Numbers smaller than given in the table are for nonvectorised versions of the operator splitting methods, which are considerably slower than their vectorised counterparts.
These are considered deprecated and the nonvectorised solvers will be removed in some future \GP\ release.

\begin{deluxetable}{ccc}[h!]
\tablecolumns{3}
\tablewidth{0pc}
\tablecaption{Solvers and their Respective Values for the {\tt solution\_method} Parameter \label{tab:solvers} }
\tablehead{\colhead{Solver} & \colhead{Scheme} & \colhead{{\tt solution\_method}}}
\startdata
Operator splitting & Crank-Nicholson & 4 \\
Operator splitting & Explicit & 5 \\
Operator splitting & Implicit & 7 \\
Eigen BiCGStab & Diagonal & 10 \\
Eigen BiCGStab & IncompleteLUT & 11 
\enddata
\end{deluxetable}

\GP\ has integrated tests to evaluate numerical and analytic correspondence for steady-state solutions since early in its development.
  The analytic formulation from \citet{1974Ap&SS..29..305B} is used, where the solution is provided for CR electrons diffusing through a halo with an energy-dependent diffusion coefficient and continuous synchrotron/IC losses for a uniform disc of sources.
  For earlier releases, the test was made only up to $\sim$1--10~TeV energies.
  The general agreement over all solvers used for steady-state solutions was found to be better than $\sim$1\%.
  With a suitable choice of parameters the analytic solution can also be evaluated with sufficient precision up to the $\sim$1~PeV energies.
  Testing for the v57 release using the solver methods 4 (CN) and 10 (BiC), which are the most practical for obtaining steady-state solutions, show a similar level of agreement: better than $\sim$1\% up to 1~PeV, with the numerical solutions essentially identical.
  Sample configurations used to evaluate steady-state correspondence for numerical solutions with the solver methods 4 (CN) and 10 (BiC) are provided in the {\tt GALDEF} directory of the distribution.
    
With this release we include also an analytic time-dependent case that can be used as a reference for evaluating the effects on numerical solutions of parameter choices for the solvers and spatial/energy grid resolution for these modelling scenarios.
   Below the analytic solution and its implementation are described, and illustrative results using the \GP\ solvers for sample configurations with the grid function formalism are given.

The time-dependent analytic formulation uses the results of \citet{1995PhRvD..52.3265A}, who derived the kernel for the density of electrons/positrons at radius $r$ and time $t$ for injection spectrum $Q(\gamma = E_e/m_e c^2)$:

\begin{equation}
  N(r,t,\gamma) = \frac{Q(\gamma_t) B(\gamma_t)}{\pi^{3/2} B(\gamma) r_{\rm diff} ^3} \exp \left( -\frac{r^2}{r^2_{\rm diff}} \right)
\label{AppTst:gfn}
\end{equation}

\noindent
where $B(\gamma)$ is the energy-loss term, $\gamma_t$ is the initial (dimensionless) energy of particles that are cooled down to energy $\gamma$ over time $t$, and $r_{\rm diff}(\gamma, t) = 2 \sqrt{\Delta u}$ is an effective diffusion horizon that the particles propagate over by time $t$ after injection by the source.
The function $\Delta u$ has the form

\begin{equation}
  \Delta u(\gamma,\gamma_t) = \int _\gamma ^{\gamma_t} \frac{D(x) dx}{B(x)},
  \label{AppTst:ufn}
\end{equation}

\noindent
where $D(\gamma)$ is the diffusion coefficient.
This solution is implemented in \GP\ so that it can output the particle density for a user-specified combination of diffusion coefficient and energy losses, spatial/energy grid, and source model.

To compare the analytic and numeric solutions, a simple modelling configuration is used.
  A spatially stationary point source injecting at a constant rate is assumed.
  For the injection spectrum, the spectral model used for the inhomogeneous diffusion example (Eq.~\ref{eq:injectionSpectrum} emitting with constant luminosity) is used.
  For the energy losses, we include synchrotron radiation only.
  Because the analytic solution is for an homogeneous ISM, we set throughout the simulation volume the strength of the GMF to be 2~$\mu$G, and use an isotropic diffusion coefficient with normalisation $4.5\times10^{28}$~cm$^2$~s$^{-1}$ at reference rigidity $\rho_0 = 4$~GV and index for its rigidity dependence $\delta = 0.35$.

The solutions are generated using the above model configuration for a simulation volume of $\pm1$~kpc with the source at the centre of the grid.
We use the tangent grid function with three central pixel sizes (4, 8, and 16~pc) and show the convergence properties for the solutions obtained with the CN and method 5 (Exp) solvers (the BiC method produces very close results to the CN one, so we do not show it in the comparisons below).
As for the steady-state tests, all configuration files to reproduce these runs are made available in the {\tt GALDEF} directory of the distribution.
  
\begin{figure*}[tb!]
  \centerline{
    \includegraphics[scale=0.85]{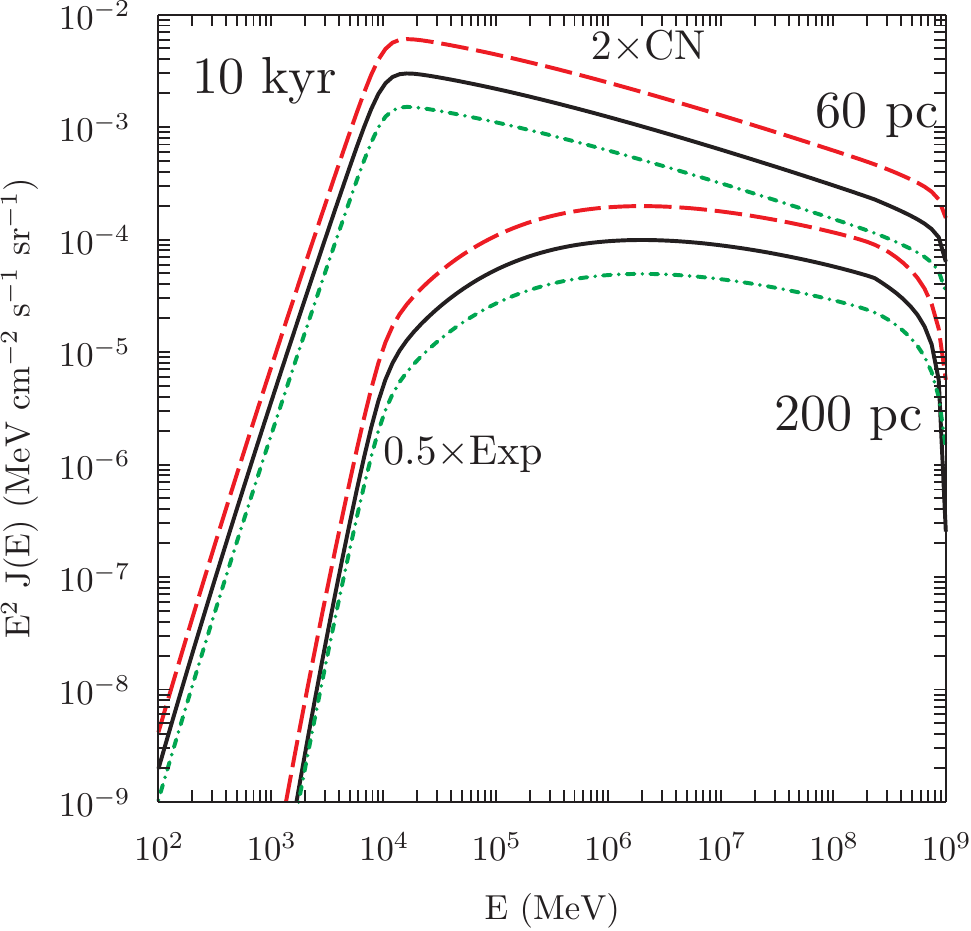}
    \includegraphics[scale=0.85]{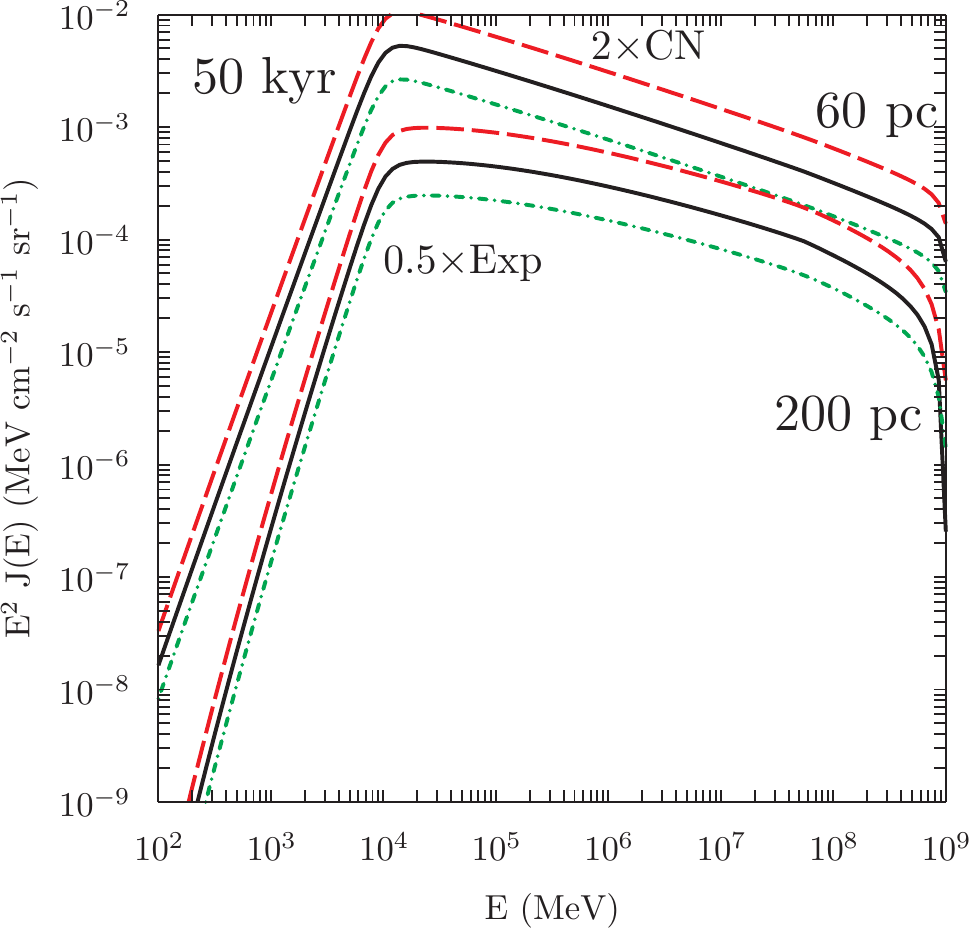}
  }
  \caption{
    Comparison between analytic and numerical solutions for 10~kyr (left) and 50~kyr (right) for 60 and 200~pc locations from the grid centre. Line coding: black solid, analytic; red long dashed, CN; green dash-dotted, Exp. The numeric solutions have been multiplied by factors 2 (CN) and 0.5 (Exp) for clarity.
    \label{AppTst:f1}
  }
\end{figure*}

Figure~\ref{AppTst:f1} shows the CN and Exp solutions together with the analytic at 10~kyr (left) and 50~kyr (right) for 60~pc and 200~pc distances from the grid centre for the tangent grid with an 8~pc central pixel.
Because the numerical solutions are very close to the analytic one, they are offset by factors of 2 (CN) and 0.5 (Exp) for clarity.
The correspondence is already quite good after 10~kyr over most of the energy range, and becomes very good after 50~kyr with only slight deviations for energies $\gtrsim$10~TeV.

\begin{figure*}[tb!]
  \centerline{
    \includegraphics[scale=0.85]{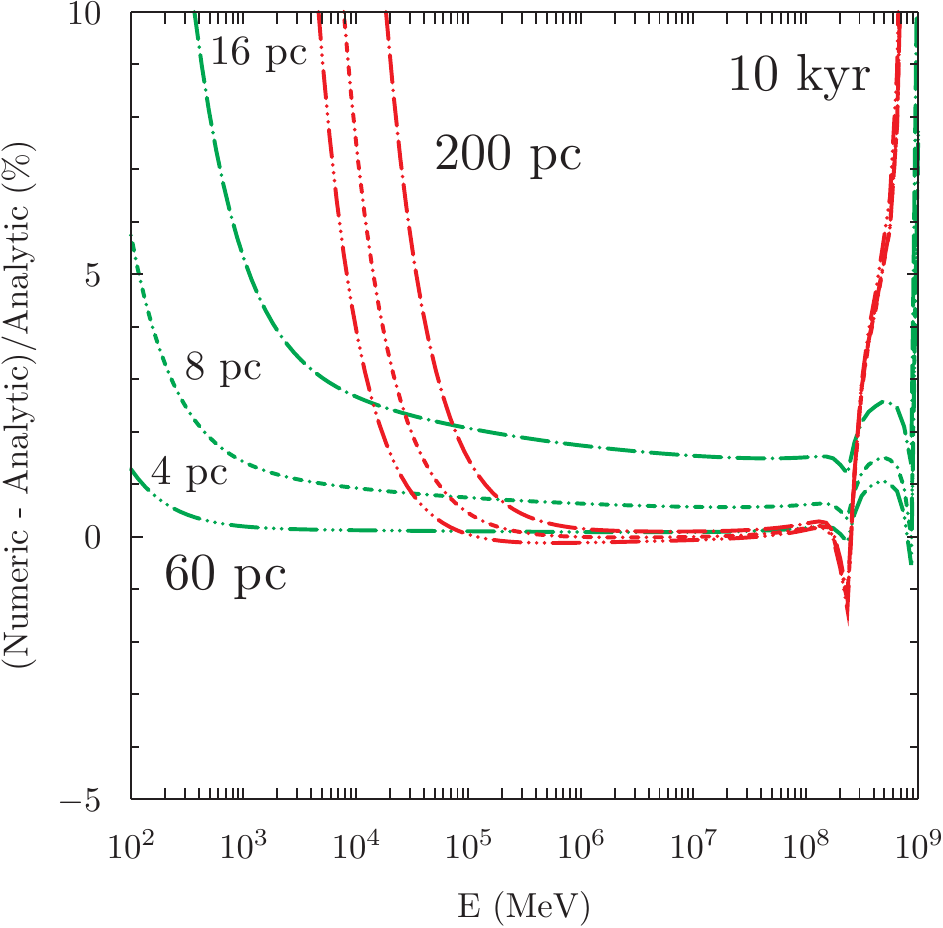}
    \includegraphics[scale=0.85]{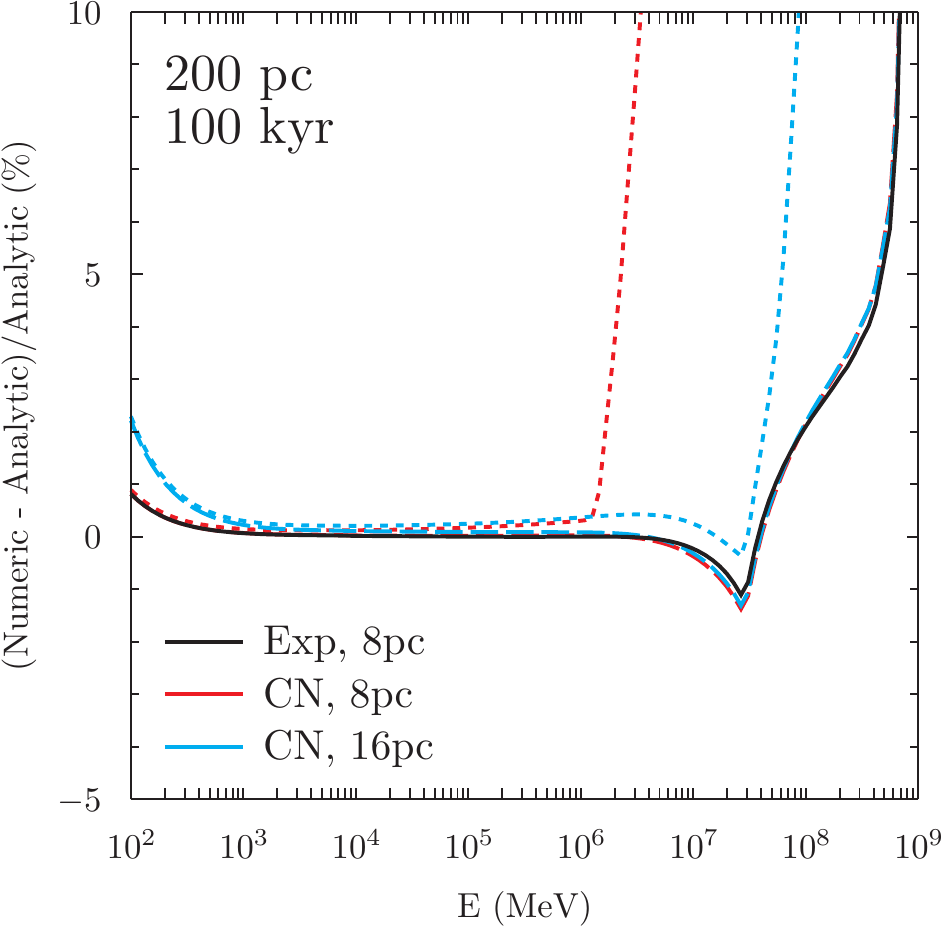}
  }
  \caption{
    The relative difference for solutions generated using the solvers for different grid discretisations and time/spatial sampling.
      Left panel: Exp solver solutions at 60~pc (green lines) and 200~pc (red lines) after 10~kyr using tangent grid discretisation with the central pixels 4 (triple dotted-dashed), 8 (dot short dashed), and 16~pc (dot long dashed), respectively.
Right panel: CN (red, cyan lines) and Exp (black line) solver solutions at 200~pc after 100~kyr. Short dashed lines are for the CN solver using time stepping of 100~yr, while long dashed lines use time stepping of 10~yr. The different colour lines for the CN solutions are for tangent grid discretistations with central pixel 8~pc (red) and 16~pc (cyan), respectively. The Exp solution is shown only for the 8~pc discretisation because the 16~pc case is essentially the same.
    \label{AppTst:f2}
  }
\end{figure*}

The convergence properties can be seen in Fig.~\ref{AppTst:f2}, which shows the relative differences between the numerical solutions and analytic for varying central pixel sizes and solver parameters.
For the left panel, the Exp solution after 10~kyr is shown for varying tangent grid central pixel spacing at the sampling locations used for Fig.~\ref{AppTst:f1}.
The Exp solver is used for numerical solution in this panel, because it automatically adjusts the time stepping to resolve the finest scales in the model configuration and is the most accurate, albeit at the cost of taking the most time.

There are features in the numerical/analytic correspondence at early times that depend on the spatial resolution employed.
Because the numerical solution assumes linear interpolation between grid points, an artificial gradient exists in the pixels nearby the source right at the start.
This results in the diffusion initially being slightly faster for the numerical solution than the analytic one.
It is a numerical artifact that is quickly reduced after the particles diffuse into the space, so is only visible for $r > r_{\rm diff}$.
Increased spatial resolution mitigates the effect, and the cooling of the particles from high to low energies also decreases its impact.
Consequently, the relative difference between numeric and analytic solutions becomes significantly reduced at later times.

For the early times, up to the $\sim$100~TeV the numerical solutions are within $\sim$1\%--2\% of the analytic one when using the finer grid resolutions.
For the higher energies, the solvers do not completely reproduce the break that can be seen in the respective spectra shown in Fig.~\ref{AppTst:f1} around $\sim$200~TeV.
However, close to the source the agreement is still very good, and only reaches a maximum difference $\lesssim$10\% for the last energy bin.
Meanwhile, further away the analytic solution is steeply falling and the numerical solutions differ from the analytic by up to $\sim$10\%, except for the highest energy bins, where it is larger.

The effect of different time stepping and grid spacing for the CN solver is shown in the right panel of Fig.~\ref{AppTst:f2}.
The relative differences with the analytic solution at 200~pc after 100~kyr for the CN solver using time steps of 100~yr (long dashed) and 10~yr (short dashed) for 8~pc (red) and 16~pc (cyan) central tangent grid pixels.
Also shown as a reference for the most accurate numerical solution is that for the Exp solver for the 8~pc central spacing tangent grid (black line). (The corresponding solution for the 16~pc grid is not shown because it is essentially the same as the 8~pc case.)
At low energies the numerical solutions are approaching the analytic one as the particles diffuse throughout the simulation volume.
Finer grid spacings effectively reduce the numerical artifact described above, while also the particles cooling from higher to lower energies also decrease the impact of the effect as the solution is evolved.
Longer run-time simulations to several hundred thousand years (not shown) have a $\lesssim$0.5\% relative difference at these energies across all central pixel sizes for the grid shown, indicating that the solutions converge reasonably quickly.

At higher energies for the CN solver, the coarser grid and time stepping provide a fairly reasonable accuracy up to $\sim$few 10s of TeV energies.
This contrasts with the accuracy using a finer grid spacing, which is only comparable to that using the 16~pc central pixel up to $\sim$1~TeV.
The reason for this is that for smaller spatial grid spacing, finer time stepping is required to adequately resolve the solution.
However, the coarser spatial grid and time stepping generally require longer run times to achieve their minimum relative difference from the analytic case.
  
Meanwhile, using a finer time stepping for either central grid pixel size with the CN solver produces a solution very similar to that obtained using the Exp solver.
The run time is, however, considerably shorter.
The relative difference for either solver remains fairly good ($\lesssim$5\%) even into the $\gtrsim$50~TeV energy range for this time sample, where the spectrum is steeply falling.

The solvers implemented with this release of \GP\ perform well up to $\sim$1~PeV for both steady-state and time-dependent modelling scenarios.
For steady-state models, we recommend either the CN or BiC solver depending on the hardware that a user has available.
The CN solver parallelises across the available resources, while the BiC solver does not utilise the resources as well.
However, the latter is possibly more efficient on certain classes of hardware (e.g., laptops) because it uses less resources and avoids the iteration of the CN solver for obtaining steady-state solutions.

For time-dependent models, we recommend the CN or Exp solver, depending on the particular scenario.
If best accuracy is required at the expense of longer run times, then the Exp solver will provide the most accurate solution and use the least amount of memory.
However, if a lower level of accuracy can be tolerated, then the CN solver with appropriate time stepping is likely a more useful option and will yield a solution quicker, but use more memory.
The BiC solver can be used, also, but its solution properties are similar to the CN solver, while being slower due to its lower overall utilisation of the computational resources.

For either steady-state or time-dependent models we recommend for any of the solvers the use of the grid functions over the uniform/linear grids.
We have found the tangent grid function to have the widest utility.
It allows much finer resolution close to `important' regions, while maintaining reasonable resolution across a simulation volume.
For a tangent grid function with given central pixel spacing, the run time and memory usage compared to the same calculation made using a linear grid with spacing as the central pixel is reduced by about an order of magnitude.
    
\section{Nonuniform Grid Description}\label{app:grid_fns}

The subsections below describe the grid functions currently implemented with the v57 release of the \GP\ framework (options {\tt linear, tan, step}).
There are three input parameters that are always given for the grid functions implemented in \GP: $Q_{\rm min}$, $Q_{\rm max}$, and $\Delta_Q$.
Here $Q$ is a generalised spatial coordinate that can be $X,Y,Z$ or galactocentric $R$, depending on the spatial dimensionality.
Because the solution of the propagation equations depends more strongly on the actual values of the boundaries, $Q_{\rm min}$ and $Q_{\rm max}$ are fixed to their input while the value of $\Delta_Q$ is adjusted internally to be as close to that supplied by the user so that pixels always fall on both $Q_{\rm min}$ and $Q_{\rm max}$.
By default, the number of points in the grid is adjusted to the nearest odd number (to have a pixel in the centre of a uniform grid) or, in the case of the vectorised solver, to the closest multiple of 8.

\subsection{Linear}

This is the trivial function
\begin{equation}
  Q(\zeta) = \Delta_Q \zeta + Q_{\rm min}
  \label{eq:linear}
\end{equation}
that simply scales the uniform grid to a step size different from 1.  $\Delta_Q$ is adjusted so that $Q(N-1) = Q_{\rm max}$, where $N$ is the number of points in the grid:
\begin{equation}
  \Delta_Q = \frac{Q_{\rm max}-Q_{\rm min}}{N-1}.
  \label{eq:linear_delta_adjustment}
\end{equation}

\subsection{Tangent}

This function is designed to have the maximum resolution $\Delta_Q$ at a certain point $Q_0$ that is then increased to $\lambda \Delta_Q$ at another point $Q_{\rm ref}$.
We assume that $\lambda > 1$.
The grid function is
\begin{equation}
  Q(\zeta) = \frac{\Delta_Q}{a} \tan\left( a[\zeta - \zeta_0] \right) + Q_0,
  \label{eq:tan}
\end{equation}
where $a$ is a scale parameter that is determined from the condition that $dQ/d\zeta(Q_{\rm ref}) = \lambda \Delta_Q$.  The first derivative is
\begin{equation}
  \frac{dQ}{d\zeta} = \frac{\Delta_Q}{\cos^2\left( a[\zeta - \zeta_0] \right)}
  \label{eq:tan_dxdz}
\end{equation}
and the second derivative is
\begin{equation}
  \frac{d^2Q}{d\zeta^2} = \frac{2a \Delta_Q  \tan\left( a[\zeta - \zeta_0] \right)}{\cos^2\left( a[\zeta - \zeta_0] \right)}.
  \label{eq:tan_d2xdz2}
\end{equation}
The value of $a$ is determined from the equation
\begin{equation}
  a = \frac{\Delta_Q}{Q_{ref}-Q_0} \tan\left( \arccos\left[ 1/\sqrt{\lambda} \right] \right)
  \label{eq:tan_a}
\end{equation}
Requiring that $Q(0) = Q_{\rm min}$ gives
\begin{equation}
  \zeta_0 = \frac{-1}{a} \arctan\left( \frac{a[Q_{\rm min} - Q_0]}{\Delta_Q}  \right),
  \label{eq:tan_z0}
\end{equation}
and that $Q(N-1) = Q_{\rm max}$ gives
\begin{equation}
  N-1 = \frac{1}{a} \left\{ \arctan\left( \frac{a[Q_{\rm max}-Q_0]}{\Delta_Q} \right) - \arctan\left( \frac{a[Q_{\rm min} - Q_0]}{\Delta_Q}  \right) \right\}.
  \label{eq:tan_n1}
\end{equation}
Eqs. (\ref{eq:tan_a}) and (\ref{eq:tan_n1}) can be merged to find
\begin{equation}
  a = \frac{1}{N-1}\left\{ \arctan\left( \frac{Q_{\rm max}-Q_0}{Q_{ref}-Q_0} \tan\left[ \arccos\left( 1/\sqrt{\lambda} \right) \right] \right)
  - \arctan\left( \frac{Q_{\rm min}-Q_0}{Q_{ref}-Q_0} \tan\left[ \arccos\left( 1/\sqrt{\lambda} \right) \right] \right) \right\},
  \label{eq:tan_a2}
\end{equation}
which can be used with Eqs.~(\ref{eq:tan_a}) and (\ref{eq:tan_z0}) to set $a$, $\Delta_Q$, and $\zeta_0$ given the parameters $Q_{\rm min}$, $Q_{\rm max}$, $Q_0$, $Q_{ref}$, $\lambda$, and $N$.  The first five are user provided, while $N$ is determined from the user-provided $\Delta_Q$ as described above.

\subsection{Step}

This grid is designed to have the step size $\Delta_Q$ everywhere, except for a small region with $Q_0-Q_B < Q < Q_0+Q_B$ where the step size is $\epsilon \Delta_Q$.  We assume $0 < \epsilon < 1$.  The grid function is
\begin{equation}
    Q(\zeta) = \Delta_Q \left\{ \frac{1-\epsilon}{2a} \Big( \log\big[ \cosh\left\{ a(\zeta-\zeta_0-b) \right\} \big] - \log\big[ \cosh\left\{ a(\zeta-\zeta_0-b) \right\} \big]  \Big)
  + \zeta - \zeta_0 \right\} + Q_0,
  \label{eq:step}
\end{equation}
where $\epsilon$, $a$, and $Q_0$ are user-defined parameters.  The value of $b$ is determined from the condition that $Q(b-\zeta_0) = Q_0 - Q_B$.  The first derivative of the function is
\begin{equation}
  \frac{dQ}{d\zeta} = \Delta_Q\left\{ \frac{1-\epsilon}{2} \Big( \tanh\left[ a(\zeta-\zeta_0-b) \right] - \tanh\left[ a(\zeta-\zeta_0+b) \right]  \Big) + 1 \right\},
  \label{eq:step_dxdz}
\end{equation}
and the second derivative is
\begin{equation}
  \frac{d^2Q}{d\zeta^2}= \Delta_Q \frac{a(1-\epsilon)}{2} \left\{ \sech^2\left( a[\zeta-\zeta_0-b] \right) - \sech^2\left( a[\zeta-\zeta_0+b] \right)  \right\}.
  \label{eq:step_d2xdz2}
\end{equation}
To determine the value of $b$, we set $Q(b-\zeta_0) = Q_0 - Q_B$, resulting in
\begin{equation}
  b \approx \frac{1}{\epsilon}\left[ \frac{Q_B}{\Delta_Q} - \frac{1-\epsilon}{2a} \log(2) \right],
  \label{eq:step_b}
\end{equation}
where we have used that $\log(\cosh[x]) \approx |x| - \log(2)$ for $x \gg 1$.  In practice $x>3$ is enough.  Here we have assumed $2ab \gg 1$.
As always, we must have $Q(0) = Q_{\rm min}$ and $Q(N-1) = Q_{\rm max}$. These two conditions result in 
\begin{equation}
  \zeta_0 \approx \frac{Q_0 - Q_{min}}{\Delta_Q} + (1-\epsilon)b,
  \label{eq:step_z0}
\end{equation}
and
\begin{equation}
  N-1 \approx \frac{Q_{\rm max}-Q_{\rm min}}{\Delta_Q} + \frac{2(1-\epsilon)}{\epsilon}\left[ \frac{Q_B}{\Delta_Q} - \frac{1-\epsilon}{2a}\log(2) \right]
  \label{eq:step_n1}
\end{equation}
where we have assumed $a(\zeta_0-b) \gg 1$ and $a(N-1-\zeta_0 -b) \gg 1$, respectively.  The last equation can be solved for $\Delta_Q$ to give
\begin{equation}
  \Delta_Q \approx \frac{Q_{max} - Q_{\rm min} + \left(2 Q_B[1-\epsilon]\right)/\epsilon}{N-1 + \log(2) (1-\epsilon)^2/(a\epsilon)}.
  \label{eq:step_dx}
\end{equation}
Along with Eqs. (\ref{eq:step_b}) and (\ref{eq:step_z0}), this sets the value of $\Delta_Q$, $b$, and $\zeta_0$ given the values of the parameters  $Q_{\rm min}$, $Q_{\rm max}$, $Q_0$, $Q_B$, $a$, $\epsilon$, and $N$.  The first six are user defined while the last one is determined from the user-provided $\Delta_Q$.

Note that the value of $a$ determines how rapidly the step size goes from $\epsilon \Delta_Q$ to $\Delta_Q$.
Smaller values mean a slower change.
For more stable solutions, we advise for $a < 1$, e.g., $a=0.5$ is appropriate for $\epsilon = 0.1$.
The approximations above also require that $Q_{\rm max}-Q_B \gg \Delta_Q$, $Q_B - Q_{\rm min} \gg \Delta_Q$, and $Q_B \gg \epsilon\Delta_Q$.
That is, there should be at least 10 pixels on either side of the region with finer resolution, and at least 10 pixels in the finer-resolution region.

\section{Updated formalism for calculating inelastic cross sections for light nuclei projectiles}\label{app:xs}

There are several parameterisations of the reaction cross sections for $p,d$, $^3$He, and $^4$He projectiles \citep{1996NIMPB.117..347T,1996PhRvC..54.1329W,1997lrc..reptQ....T, 1999NIMPB.155..349T, 1999STIN...0004259T}.
The many modifications to the original formalism by \citeauthor{1996NIMPB.117..347T} are scattered across the literature \citep[see][and references therein]{2021NJPh...23j1201L}.
Here we provide a description, including corrections, to the options available with this release of \GP.

\subsection{$p+A$, Wellisch \& Axen 1996}\label{Wellisch}

The original formalism was published by \citet{1996PhRvC..54.1329W}, but typos prevented it from working properly.
We contacted the authors, who provided us with their original subroutine (Wellisch, private comm.).
This parameterisation is valid for proton energies from 6.8~MeV to $>$10 GeV for target charges $Z_t>5$. 

Here we provide the corrected formulas; for motivation and justification, we refer the reader to the original publication.
The reaction (inelastic) cross section is
\begin{equation}
\sigma_R (T_p) = 10 \sigma_R^0 f_1 f_2 \,\, {\rm mb},
\end{equation}
where
\begin{eqnarray}
\sigma_R^0(T_p) &=& \pi r_0^2 f_{\rm corr}(T_p) \ln(A_t-Z_t) \left[1+A_t^{1/3}-b_0(1-A_t^{-1/3})\right],\nonumber\\
f_1(T_p) &=& \displaystyle \frac{1}{1+e^{-P_1(\log_{10}T_p+P_2)}},\nonumber \\
f_2(T_p) &=& 1+P_3 \left( 1 -\frac{1}{1+e^{-P_4(\log_{10}T_p+P_5)} }\right),\nonumber\\
b_0 &=& 2.247-0.915(1-A_t^{-1/3}),\nonumber\\
f_{\rm corr}(T_p) &=& \frac{1-0.15e^{-T_p}}{1-0.0007A_t},\nonumber\\
\end{eqnarray}
\begin{eqnarray}
P_1 &=& 8-\frac{8}{A_t}-0.008A_t,\nonumber\\
P_2 &=& 2 \left(1.17-\frac{2.7}{A_t}-0.0014A_t\right),\nonumber\\
P_3 &=& 0.8+\frac{18}{A_t}-0.002A_t,\nonumber\\
P_4 &=& 5.6-0.016A_t,\nonumber\\
P_5 &=& 1.37\left(1+\frac{1}{A_t}\right),\nonumber
\end{eqnarray}
here $T_p$ is the proton kinetic energy in GeV, $Z_t$ and $A_t$ are the charge and the mass number of the target nucleus, correspondingly, and $r_0=1.36$~fm.

\begin{deluxetable}{cccclc}[tb!]
\tablecaption{Parameters of the inelastic cross sections\label{tab:TGD}}
\tablewidth{0pt}
\tablehead{
\colhead{Projectile} & \colhead{Target} & \colhead{$T_1$} & \colhead{$G$}& \colhead{$D$} & \colhead{Reference} 
}
\startdata
$p$ & $d$                       & 23    &\nodata& $1.85+{0.16}/\left\{1+\exp\left[(500-T_p)/200\right]\right\}$ & [1]\\
$p$ & $^{3}$He  & 58 &\nodata& 1.70 & [2]\\ 
$p$ & $^{4}$He  & 40 &\nodata& 2.05 & [2]\\ 
$p$ & $^{6}$Li          & 40 &\nodata& 2.05 & [2]\\ 
$p$ & $^{7}$Li          & 37 &\nodata& 2.15 & [2]\\ 
$p$ & $A_t>7$           & 40 &\nodata& 2.05 & [2]\\ 
$d$ & $^{4}$He  & 23 &\nodata& $1.65+{0.22}/\left\{1+\exp\left[(500-T_p)/200\right]\right\}$ & [2]\\ 
$d$ & $A_t$             & 23 &\nodata& $1.65+{0.10}/\left\{1+\exp\left[(500-T_p)/200\right]\right\}$ & [1]\\ 
$^3$He & $A_t$  & 40 &\nodata& 1.55 & [1]\\ 
$\alpha$ & $A_t$        & 40 &75 
& $2.77-8.0\times10^{-3}A_t+1.8\times10^{-5}A_t^2-0.8/\left\{1+\exp\left[(250-T_p)/G\right]\right\}$ & [1]\\
$\alpha$ & $^4$He       & 40 &300& Same as for $\alpha+A_t$ system & [1]\\
$\alpha$ & Be           & 25 &300& Same as for $\alpha+A_t$ system & [1]\\
$\alpha$ & N            & 40 &500& Same as for $\alpha+A_t$ system & [1]\\
$\alpha$ & Fe           & 40 &300& Same as for $\alpha+A_t$ system & [1]\\
$\alpha$ & $6\le Z_t\le14$ & 40 &50 
& $2.20-8.0\times10^{-3}A_t+1.8\times10^{-5}A_t^2-0.3/\left\{1+\exp\left[(120-T_p)/G\right]\right\}$ & [2,3]\\
\enddata
\tablerefs{[1] \citet{1999STIN...0004259T}, [2] \citet{2021NJPh...23j1201L}, [3] \citet{2019PhRvC..99a4603H}.}
\end{deluxetable}

\subsection{$p+A$, $A+A$, updated Tripathi et al. formalism}\label{Tripathi}

The original formalism was published by \citet{1996NIMPB.117..347T, 1997lrc..reptQ....T, 1999NIMPB.155..349T, 1999STIN...0004259T}, but the description contains some confusing statements.
Since then the formalism has been tested and updated to provide better agreement with available data \citep[][and references therein]{2021NJPh...23j1201L}, but its pieces are scattered in the literature and are difficult to reconcile. Here we provide a short and coherent description.

The inelastic cross section can be parameterised as
\begin{equation}
\sigma_R = 10 \pi r_0^2 (A_p^{1/3}+A_t^{1/3}+\delta_E)^2 \left(1-R_c\frac{B}{T_{cm}}\right)X_m \,\, {\rm mb},
\end{equation}
where $r_0=1.1$~fm, $Z_p$ and $A_p$ are the charge and the mass number of the projectile nucleus, correspondingly, $T_{cms}$ is the kinetic energy in the centre-of-mass (CMS) system:
\begin{eqnarray}
T_{cms} &=& \sqrt{s}-M_p-M_t, \nonumber\\
s &=& (M_p+M_t)^2+2M_t A_p T_p,\nonumber
\end{eqnarray}
where $T_p$ is the kinetic energy per nucleon in the laboratory system (LS), $M_p$ and $M_t$ are the masses of the projectile and target nuclei, $X_m=1$ is the optical model multiplier, and the terms $\delta_E$ and $B$ are given by the following expressions:
\begin{eqnarray}
\delta_E &=& 1.85S+\frac{0.16S}{T_{cms}^{1/3}}-C_E+0.91\frac{Z_p(A_t-2Z_t)}{A_pA_t},\nonumber\\
C_E &=& D\left(1-e^{-T_p/T_1}\right)-0.292e^{-T_p/792} \cos\left(0.229T_p^{0.453}\right),\nonumber\\
S &=& \frac{A_p^{1/3} A_t^{1/3}}{A_p^{1/3}+A_t^{1/3}},\nonumber\\
B &=& 1.44\frac{Z_p Z_t}{R}, \nonumber\\
R &=& \mathcal{R}(A_p)+\mathcal{R}(A_t)+1.2 \frac{A_p^{1/3}+A_t^{1/3}}{T_{cms}^{1/3}}, \nonumber\\
\mathcal{R}(A_i) &=& \sqrt{\frac{5}{3}}\ r_{\rm rms,i}. \nonumber
\end{eqnarray}
Values or formulas for $D, T_1,$ and $R_c$ are listed in Tables \ref{tab:TGD} and \ref{tab:RC}. $r_{\rm rms,i}$ is the charge density distribution parameter, taken as an average of several estimates from \citet{1987ADNDT..36..495D}, and is listed in Table \ref{tab:ri}.
For other isotopes it can be calculated as
\begin{equation}
r_{\rm rms,i}=0.84 A_i^{1/3}+ 0.55, \quad A_i>26.
\end{equation}
\begin{deluxetable}{cccc}[tb!]
\tablecaption{Coulomb multiplier $R_c$\label{tab:RC}}
\tablewidth{0pt}
\tablehead{
\colhead{Projectile} & \colhead{Target} & \colhead{$R_c$} & \colhead{Reference} 
}
\startdata
$p$ & $d$               &13.5   & [1]\\ 
$p$ & $^{3}$He  & 21            & [1]\\ 
$p$ & $^{4}$He  & 27            & [1]\\ 
$p$ & Li                        & 2.2   & [1]\\ 
$p$ & C                 & 3.5   & [2]\\ 
$d$ & $d$               &13.5   & [1]\\ 
$d$ & $^{4}$He  &13.5   & [1]\\ 
$d$ & C                         & 6.0           & [1]\\ 
$\alpha$ & Ta           & 0.6           & [1]\\ 
$\alpha$ & Au           & 0.6           & [1]\\ 
\multicolumn{2}{r}{All other cases} & 1 & [1]\\
\enddata
\tablerefs{[1] \citet{1999STIN...0004259T}, [2] \citet{2021NJPh...23j1201L}.}
\end{deluxetable}

\begin{deluxetable}{ccccccccccccccc}[tb!]
\tablecaption{Nuclear charge density distribution parameters\label{tab:ri}}
\tablewidth{0pt}
\tablehead{
\colhead{} & \multicolumn{14}{c}{Averaged values from \citet{1987ADNDT..36..495D}} 
}
\startdata
Isotope & $^1$H& $^2$H& $^3$He& $^4$He& $^6$Li& $^7$Li& $^9$Be& $^{10}$B& $^{11}$B& $^{12}$C& $^{13}$C& $^{14}$N& $^{15}$N& $^{16}$O\\
$r_{rms,i}$ & 0.850& 2.106& 1.899& 1.681& 2.557& 2.40& 2.51& 2.45& 2.395& 2.469& 2.440& 2.548& 2.654& 2.728\\
\hline
Isotope & $^{17}$O& $^{18}$O& $^{19}$F& $^{20}$Ne& $^{22}$Ne& $^{23}$Na& $^{24}$Mg& $^{25}$Mg& $^{26}$Mg& $^{27}$Al& $^{28}$Si& $^{29}$Si& $^{30}$Si& $^{31}$P\\
$r_{rms,i}$ & 2.662& 2.727& 2.900& 3.012& 2.969& 2.94& 3.047& 3.057& 3.06& 3.048& 3.114& 3.105& 3.176& 3.188\\
\hline
Isotope & $^{32}$S& $^{34}$S& $^{36}$S& $^{35}$Cl& $^{37}$Cl& $^{36}$Ar& $^{40}$Ar& $^{39}$K& $^{40}$Ca& $^{48}$Ca& $^{48}$Ti& $^{50}$Ti& $^{51}$V& $^{50}$Cr\\
$r_{rms,i}$ & 3.242& 3.281& 3.278& 3.388& 3.384& 3.327& 3.432& 3.404& 3.470& 3.461& 3.655& 3.573& 3.598& 3.669\\
\hline
Isotope & $^{52}$Cr & $^{53}$Cr& $^{54}$Cr& $^{55}$Mn& $^{54}$Fe& $^{56}$Fe& $^{58}$Fe& $^{59}$Co& $^{58}$Ni& $^{60}$Ni& $^{61}$Ni& $^{62}$Ni& $^{64}$Ni& \nodata\\
$r_{rms,i}$ & 3.647& 3.726& 3.713& 3.68& 3.696& 3.750& 3.775& 3.813& 3.768& 3.795& 3.806& 3.826& 3.867& \nodata\\
\enddata
\end{deluxetable}

\subsection{Barashenkov and Polansky}\label{BarPol}

The \GP\ code includes the two Barashenkov and Polansky (BarPol) routines for calculation of the total (elastic+inelastic) and inelastic cross sections for reactions $A_p+A_t$: SIGHAD for $A_p=p,n,\pi$ and SIGION for $A_p>1$.
For the $p+A$ cross section, the SIGHAD input parameters are $Z_p=A_p=1$, $4\le A_t\le239$, and 1 MeV $\le T_p\le10^6$ MeV.
Meanwhile, SIGION works for heavier projectiles $A_p>1$, and targets $4\le A_t<240$.
The {\it total} kinetic energy must be in the range 1 MeV $\le T_p \le 10^6$ MeV.

\begin{figure*}[tb!]
\includegraphics[width=0.33\linewidth]{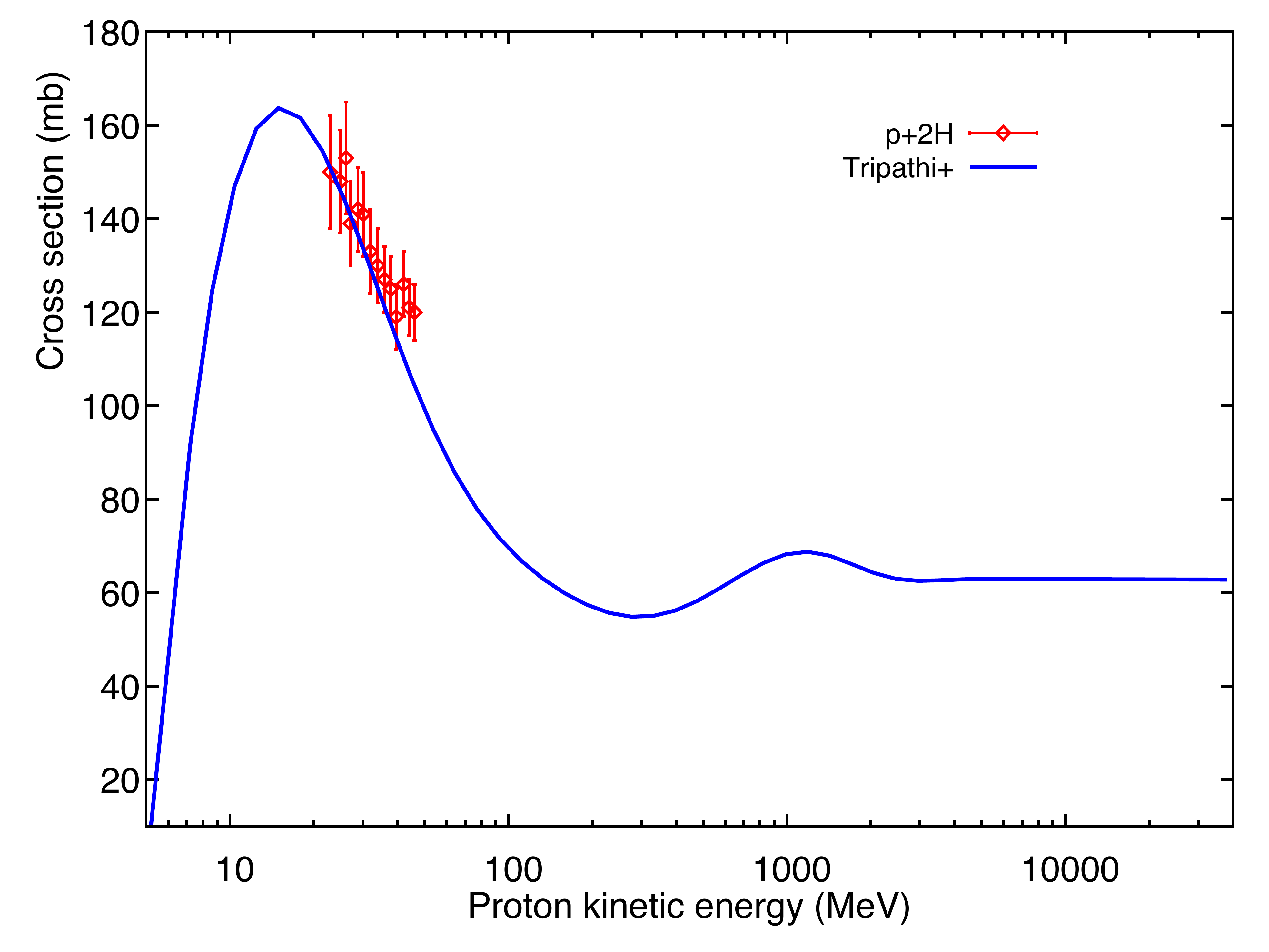}
\includegraphics[width=0.33\linewidth]{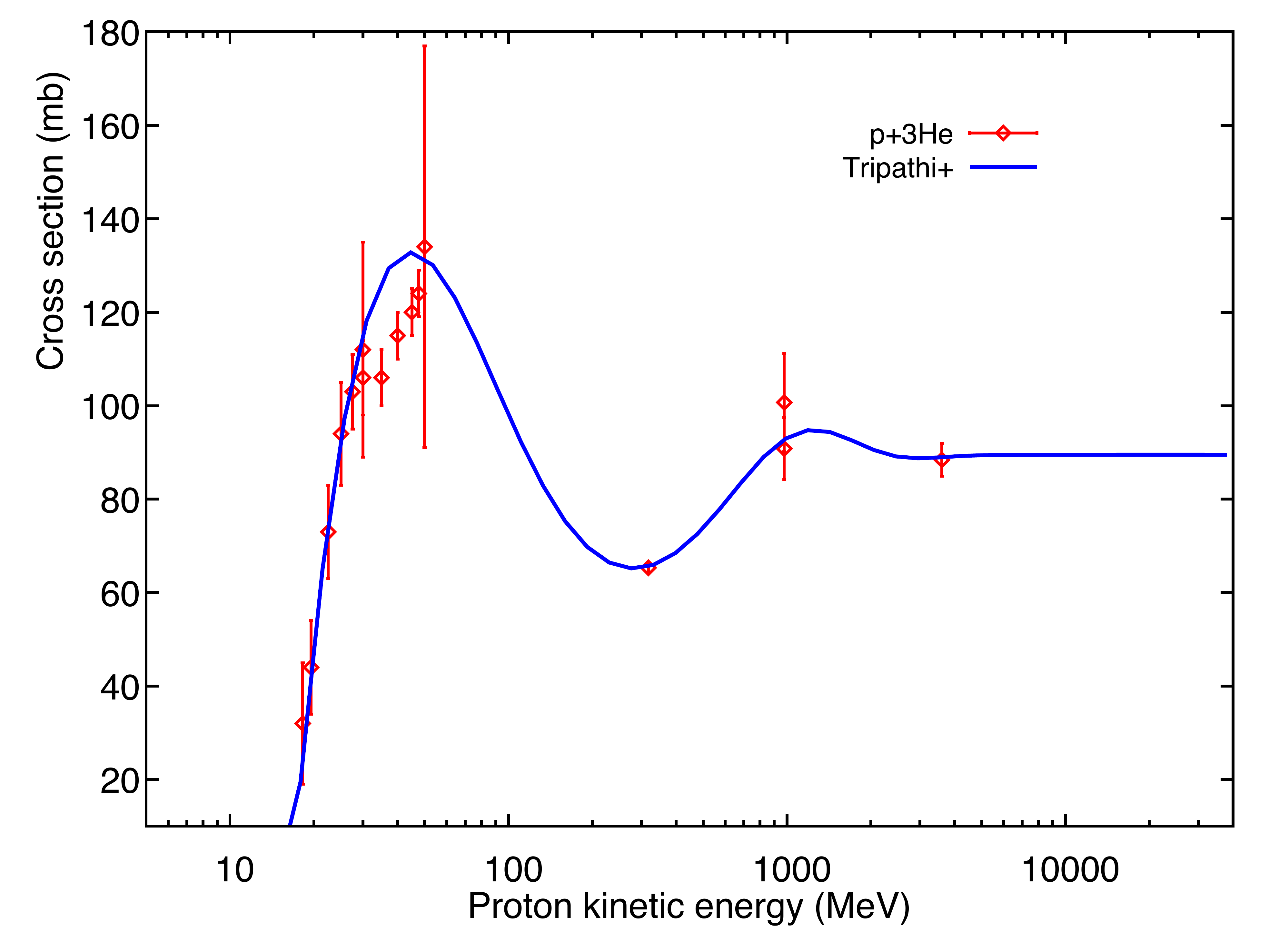}
\includegraphics[width=0.33\linewidth]{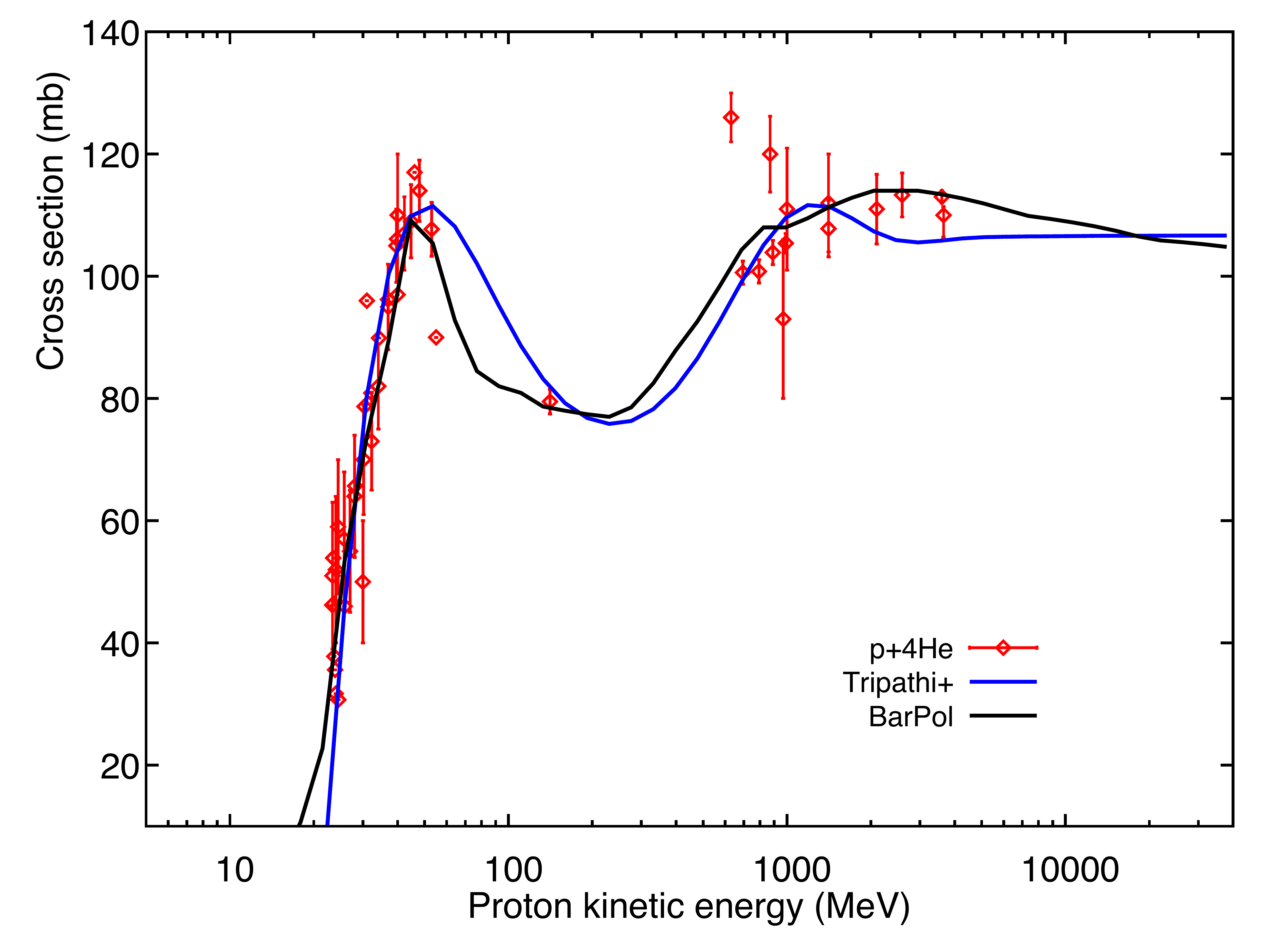}\\
\includegraphics[width=0.33\linewidth]{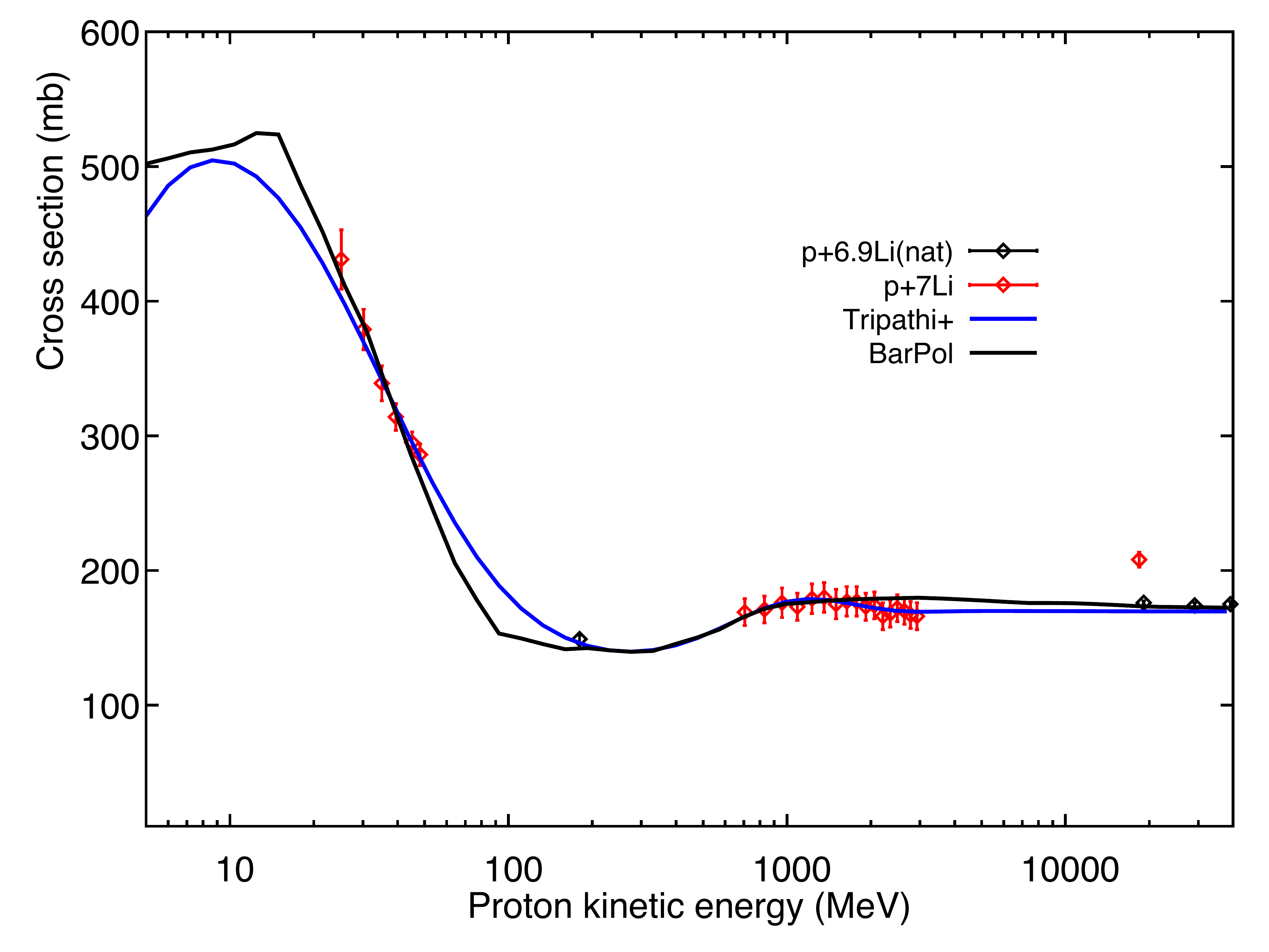}
\includegraphics[width=0.33\linewidth]{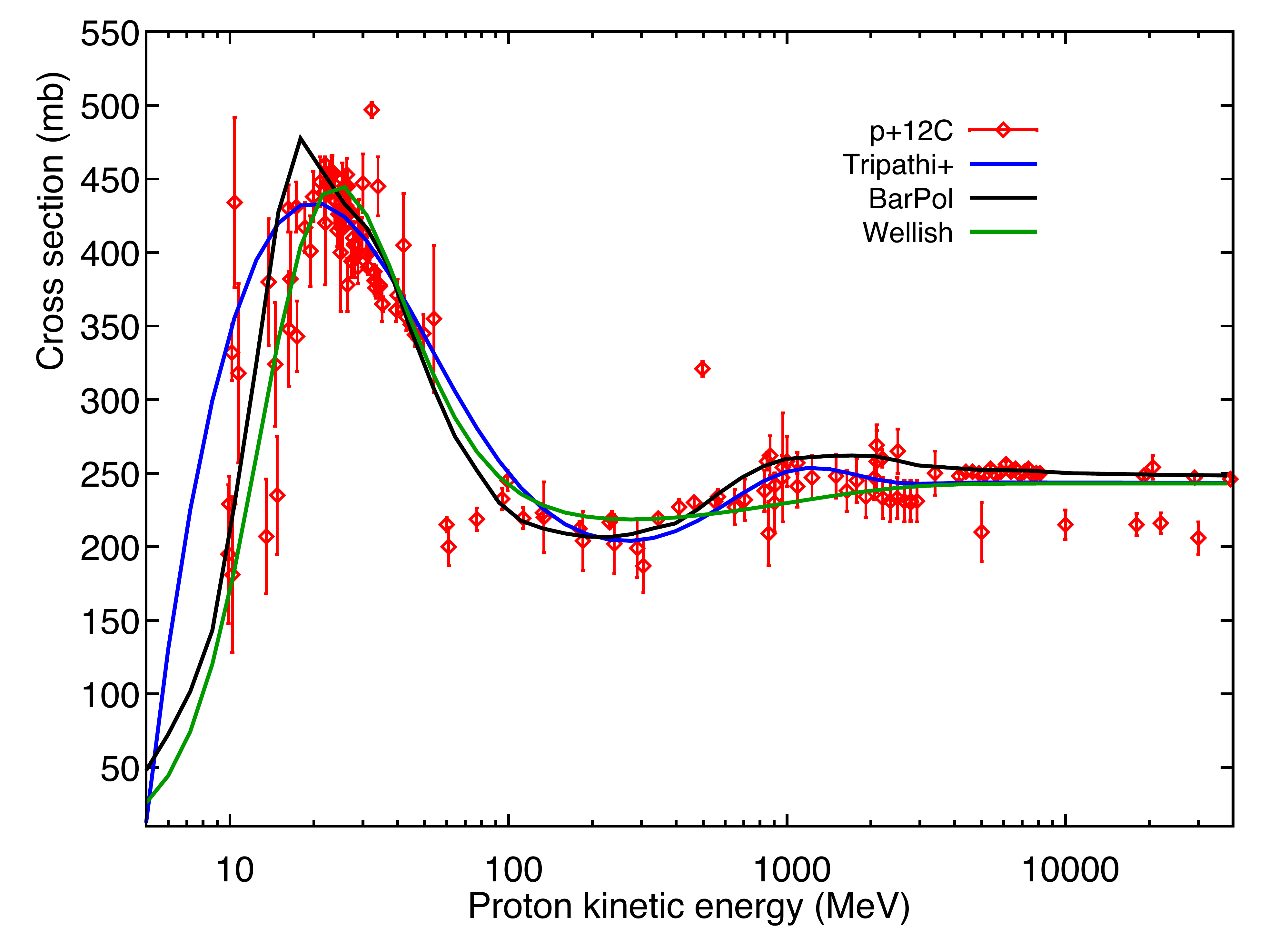}
\includegraphics[width=0.33\linewidth]{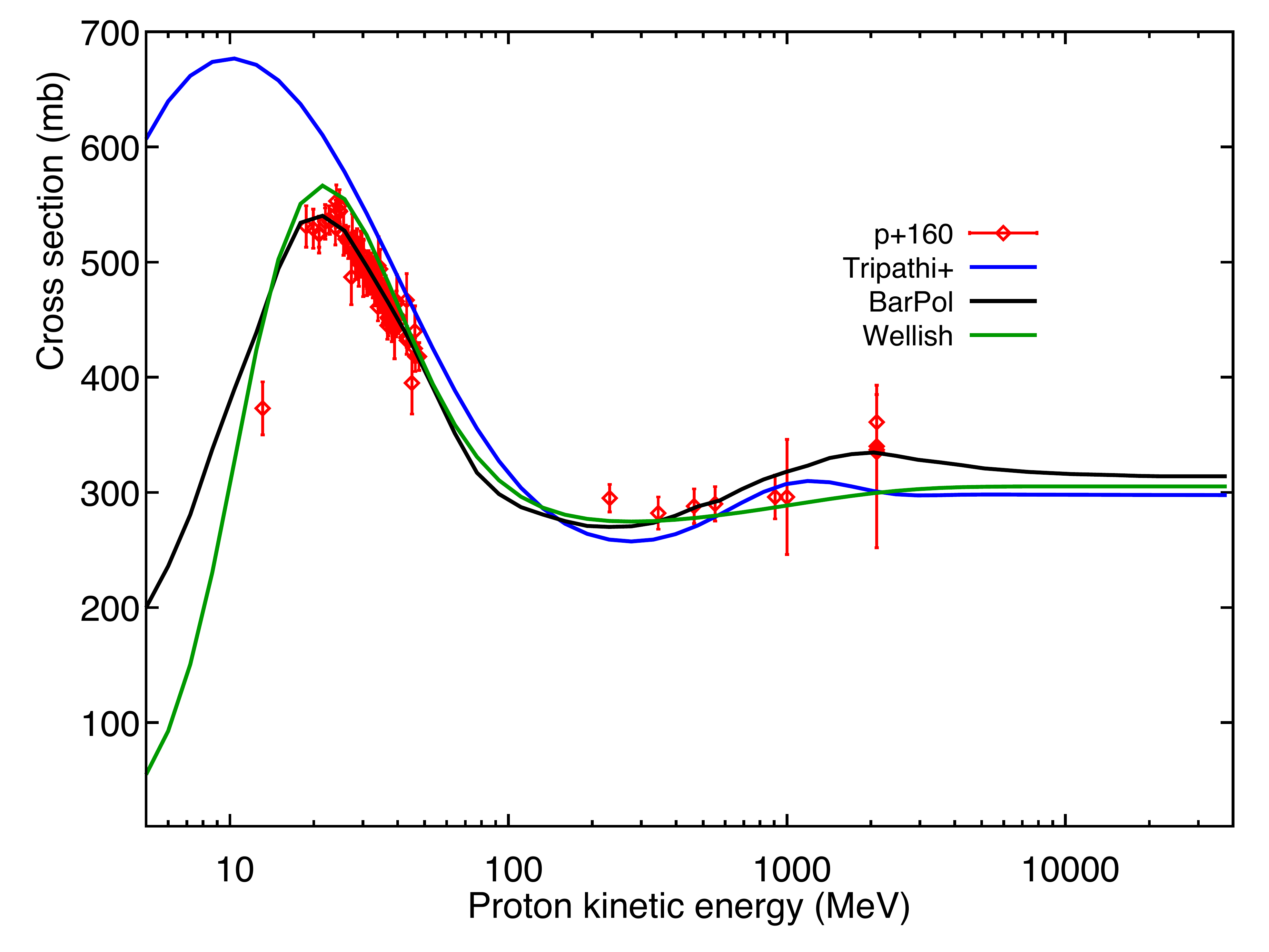}\\
\includegraphics[width=0.33\linewidth]{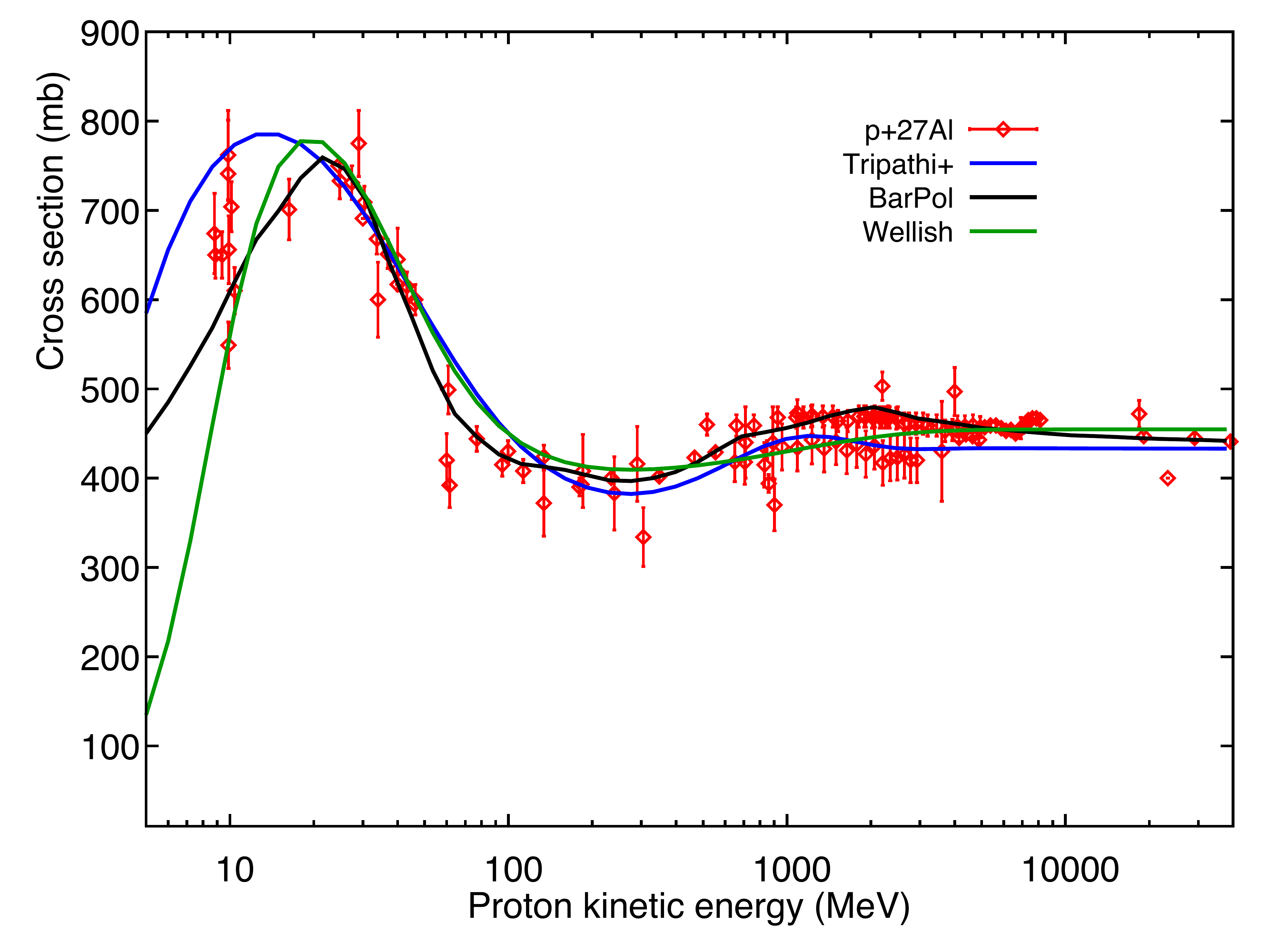}
\includegraphics[width=0.33\linewidth]{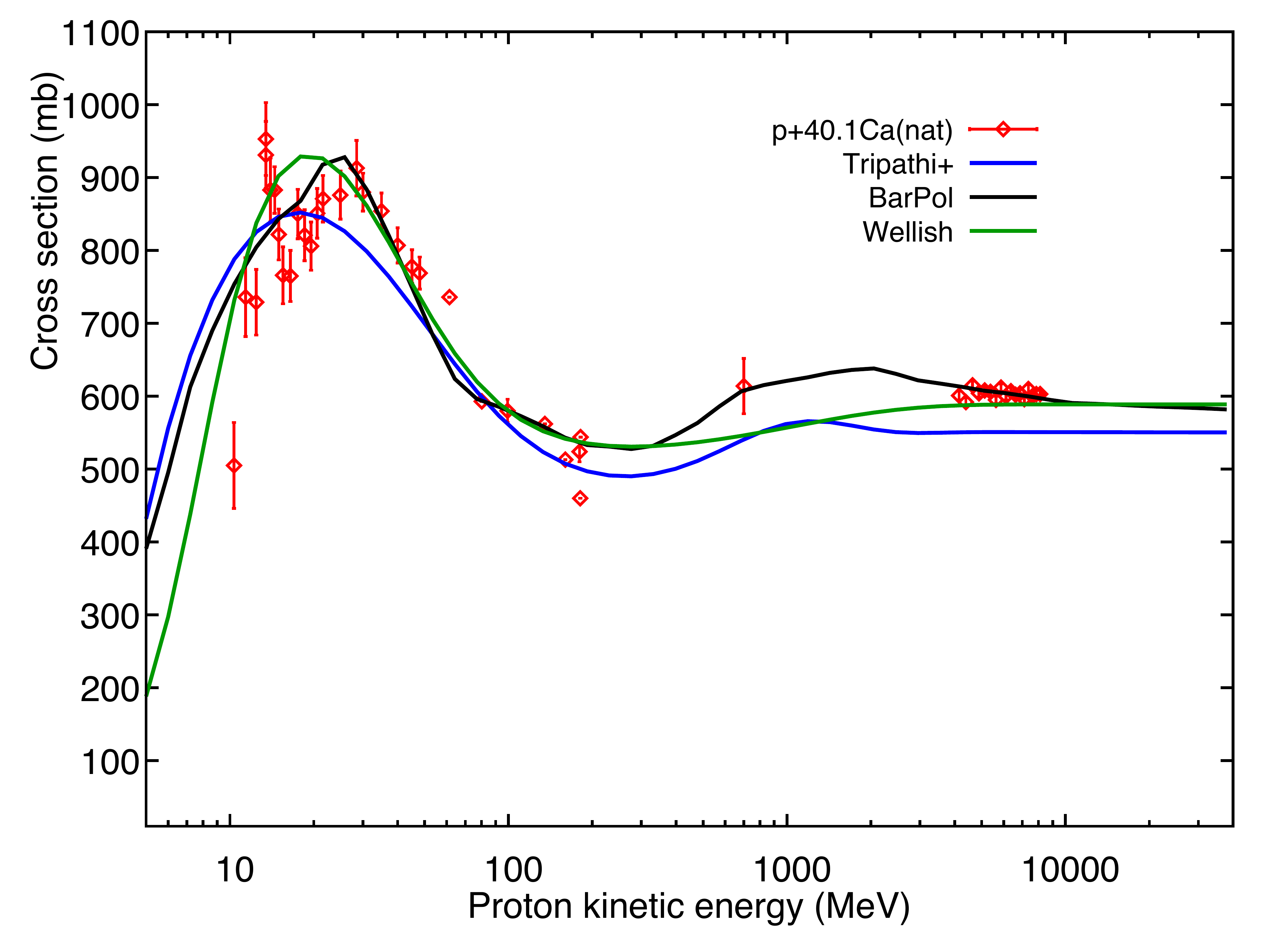}
\includegraphics[width=0.33\linewidth]{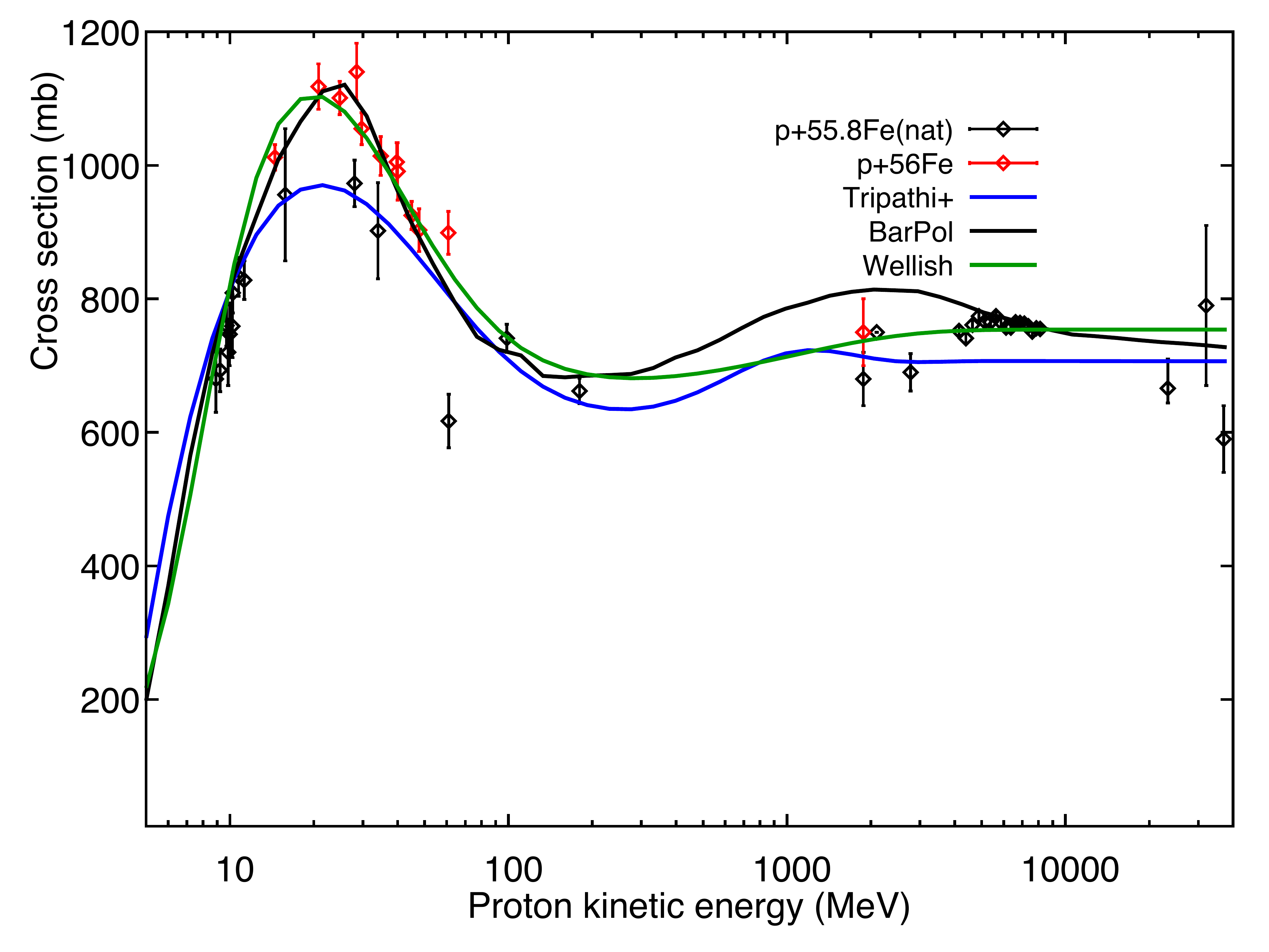}
\caption{Inelastic cross sections for the reactions $p+d$, $p+^3$He, $p+^4$He, $p+^7$Li, $p+^{12}$C, $p+^{16}$O, $p+^{27}$Al, $p+^{40}$Ca, and $p+^{56}$F. The data are taken from the database posted at {https://www.oecd-nea.org/dbdata/bara.html}. Data for $p+d$ reaction are taken from \citet{1973LNC_8_319C}. Lines: black, Barashenkov \& Polansky routines (Appendix~\ref{BarPol}); blue, modified Tripathi et al. formalism (Appendix~\ref{Tripathi}); green, corrected Wellisch \& Axen formalism (Appendix~\ref{Wellisch}). 
\label{sigma_pA}}
\end{figure*}

\begin{figure*}[tb!]
\includegraphics[width=0.33\linewidth]{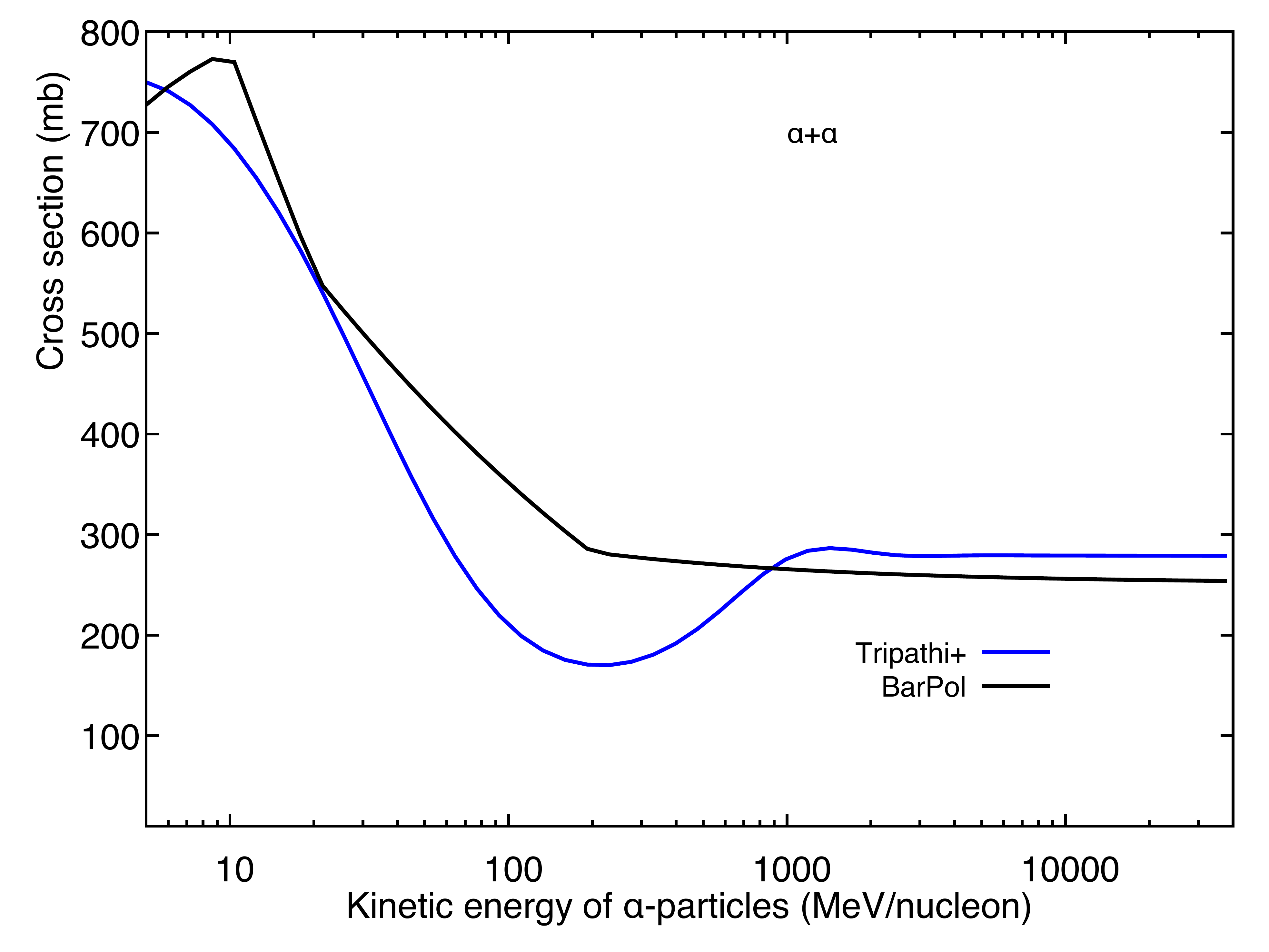}
\includegraphics[width=0.33\linewidth]{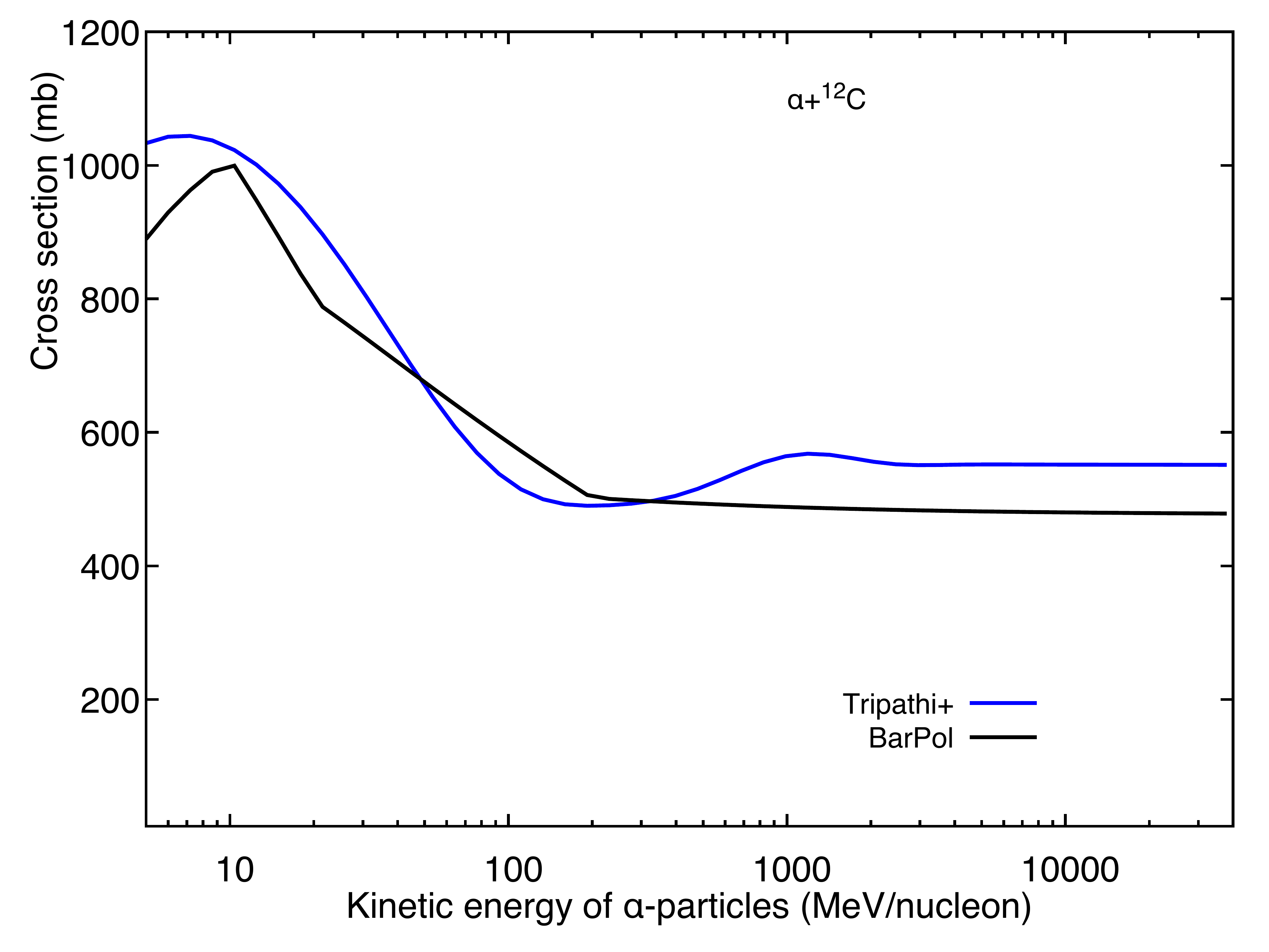}
\includegraphics[width=0.33\linewidth]{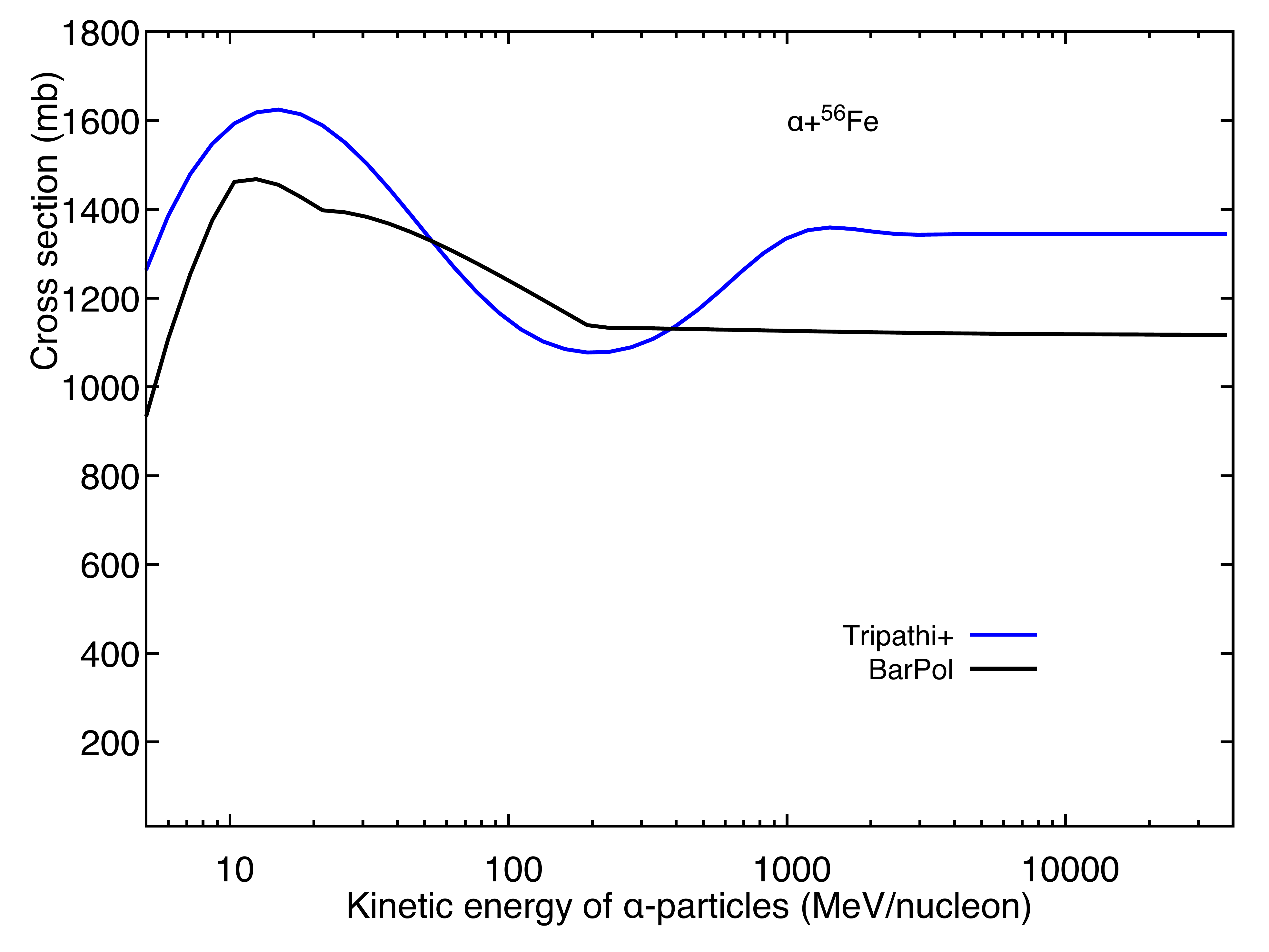}
\caption{Inelastic cross sections for the reactions $\alpha+\alpha$, $\alpha+^{12}$C, and $\alpha+^{56}$F. Lines: black, Barashenkov \& Polansky routines (Appendix~\ref{BarPol}); blue, modified Tripathi et al. formalism (Appendix~\ref{Tripathi}). 
\label{sigma_aA}}
\end{figure*}

\subsection{Summary}
Example plots are shown in Figs.~\ref{sigma_pA} and \ref{sigma_aA}.
The modified \citet{1996NIMPB.117..347T} formalism (T+) works better for light nuclei, while the BarPol routines give better agreement with data for intermediate and heavy nuclei.
The corrected \citet{1996PhRvC..54.1329W} formalism (WA) works adequately for $p+A$ interactions for its validity range ($A_t>5$).
Therefore, we suggest to use the following options:
\begin{itemize}
\item ``T+'': T+ can be used for the full range of projectile and target nuclei;
\item ``WA/T+'': T+, $A_p>1\, ||\, A_t\le 4\, $ and WA, $A_p=1\, {\rm and}\, A_t> 5$;
\item ``BarPol/T+'': T+, $A_t\le 4$ and BarPol, $A_t> 4$;
\item ``WA/T+/BarPol'': T+, $A_t\le 5$ and WA, $A_t>5$ and BarPol, $A_p>1\, {\rm and}\, A_t> 5$.
\end{itemize}

\section{Fragmentation of $^{3,4}$He to $^3$He, $^3$H, $^2$H, and production of 2H in $pp$-reaction}\label{app:frag}

This appendix describes the parameterisations of the cross sections for $^4$He fragmentation into $^3$He, $t$, and $d$.
The following channels are distinguished in the literature: $^4$He$(p,pnX)^3$He, $^4$He$(p,dX)^3$He, $^4$He$(p,ppX)^3$H, $^4$He$(p,ppnX)d$, $^4$He$(p,dX)d$.
They are considered separately because the physics involved is different.
We also describe deuteron $^2$H production in the $pp$ reaction.

As a basis for our parameterisation we use formulations provided by \citet{1993STIN...9417666C}, but with the parameters fitted to the data.
The $Q$-values and threshold energies of the reactions are calculated using
\begin{eqnarray}
Q&=&M_i-M_f,\\
E_{\rm th}&=&\frac{1}{2M_t}(Q^2-2QM_i),\nonumber
\end{eqnarray}
where $M_i=M_p+M_t$ are the rest masses of the initial particles, projectile and target, and $M_f$ is the mass of the final products.
The numerical values of $Q$ and $E_{\rm th}$ are listed in Table \ref{tab:threshold}.

\begin{deluxetable}{ccc}[tb!]
\tablecaption{Threshold energy\label{tab:threshold}}
\tablewidth{0pt}
\tablehead{
\colhead{Reaction} & \colhead{$Q$-value} & \colhead{$E_{\rm th}$}
}
\startdata
$^4$He$(p,dX)^3$He      & $-18.3530$  & 23.0181\\
$^4$He$(p,2pX)^3$H      & $-19.8140$  & 24.8543\\
$^4$He$(p,pnX)^3$He     & $-20.5776$  & 25.8143\\
$^4$He$(p,dX)d$         & $-23.8467$  & 29.9258\\
$^4$He$(p,2pnX)d$               & $-26.0713$  & 32.7252\\
$^3$He$(p,2pX)d$                & $-5.49365$  & 7.3344\\
\enddata
\end{deluxetable}
\begin{figure*}[tb!]
\includegraphics[width=0.33\linewidth]{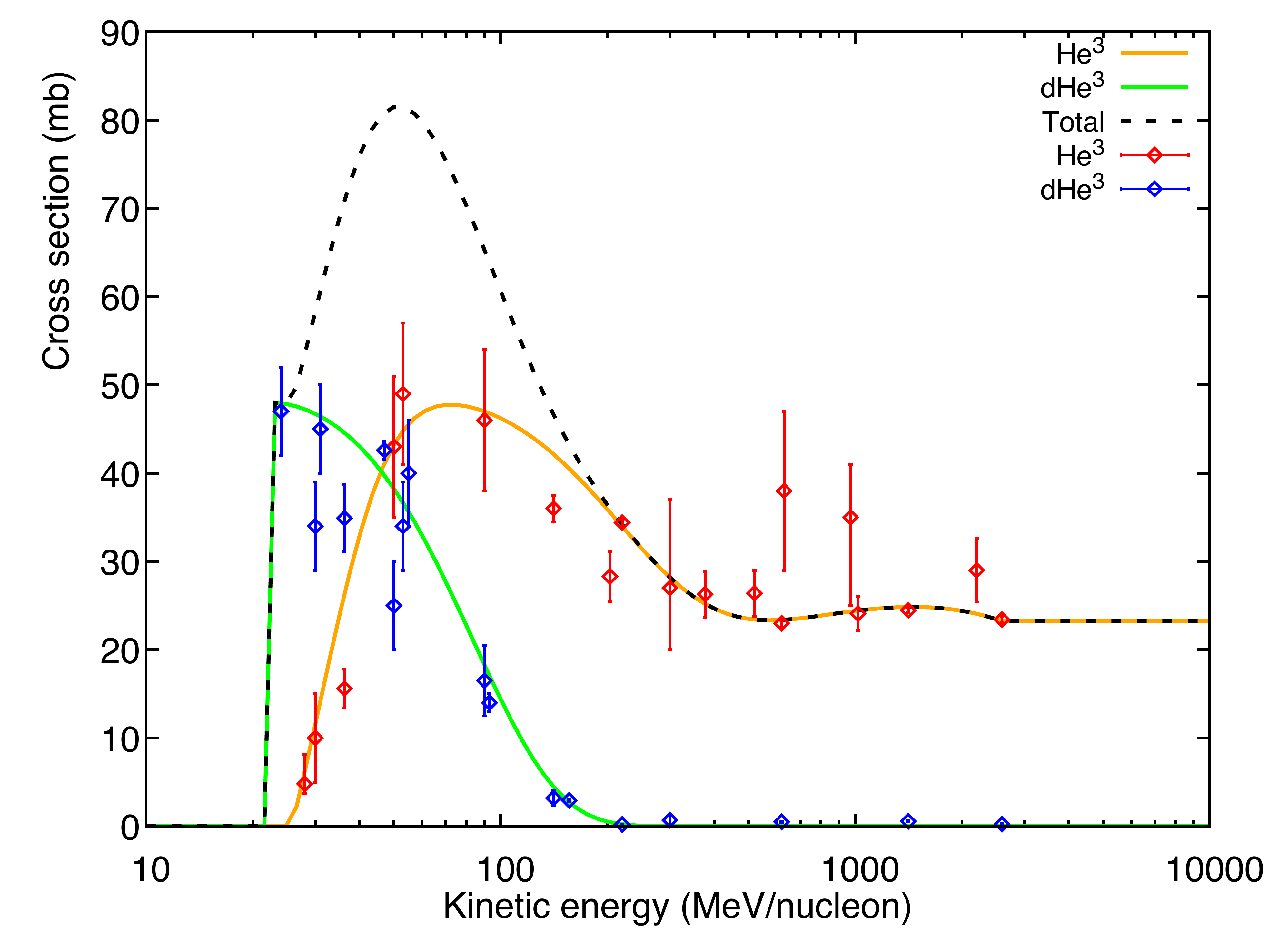}
\includegraphics[width=0.33\linewidth]{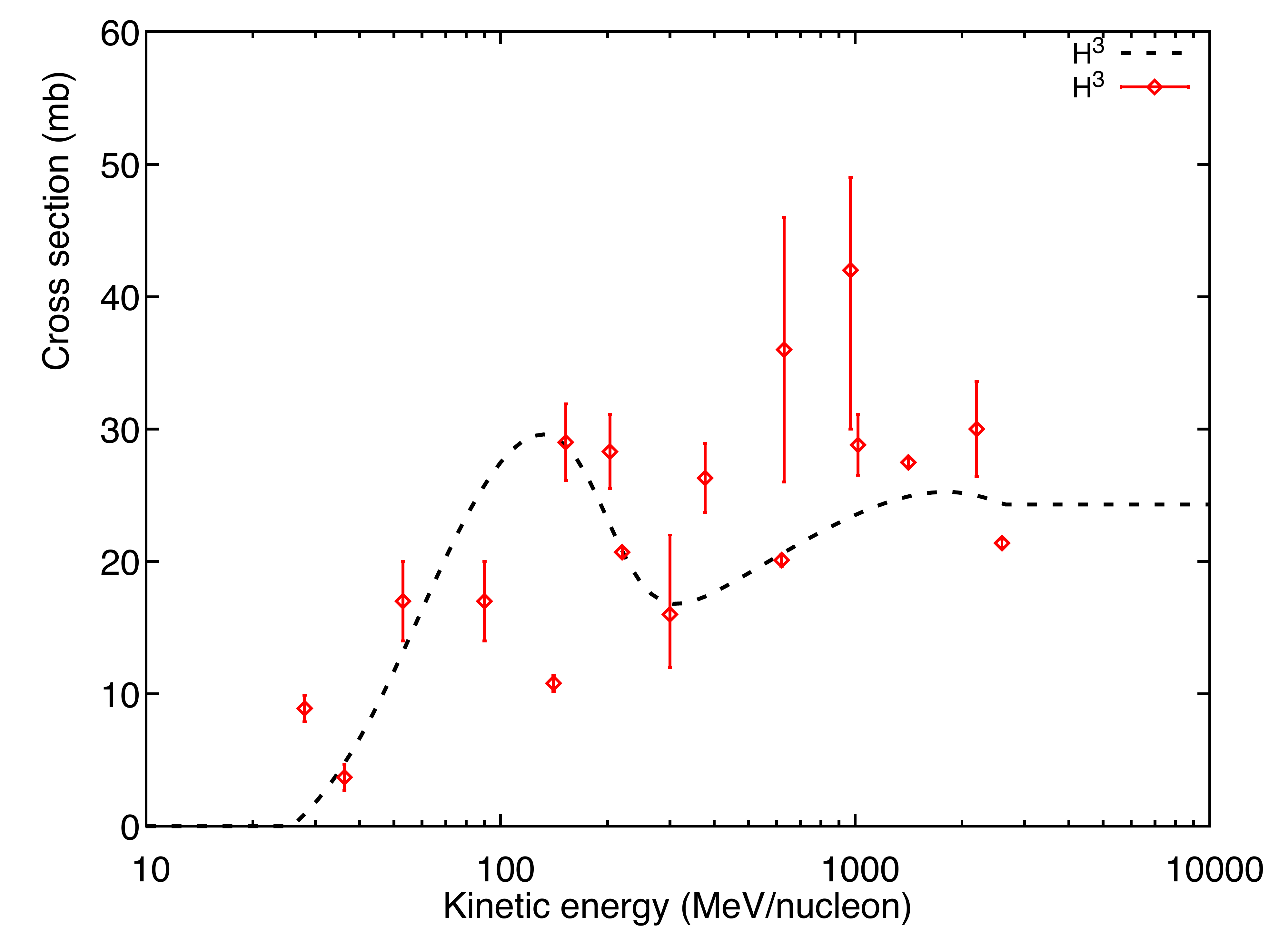}
\includegraphics[width=0.33\linewidth]{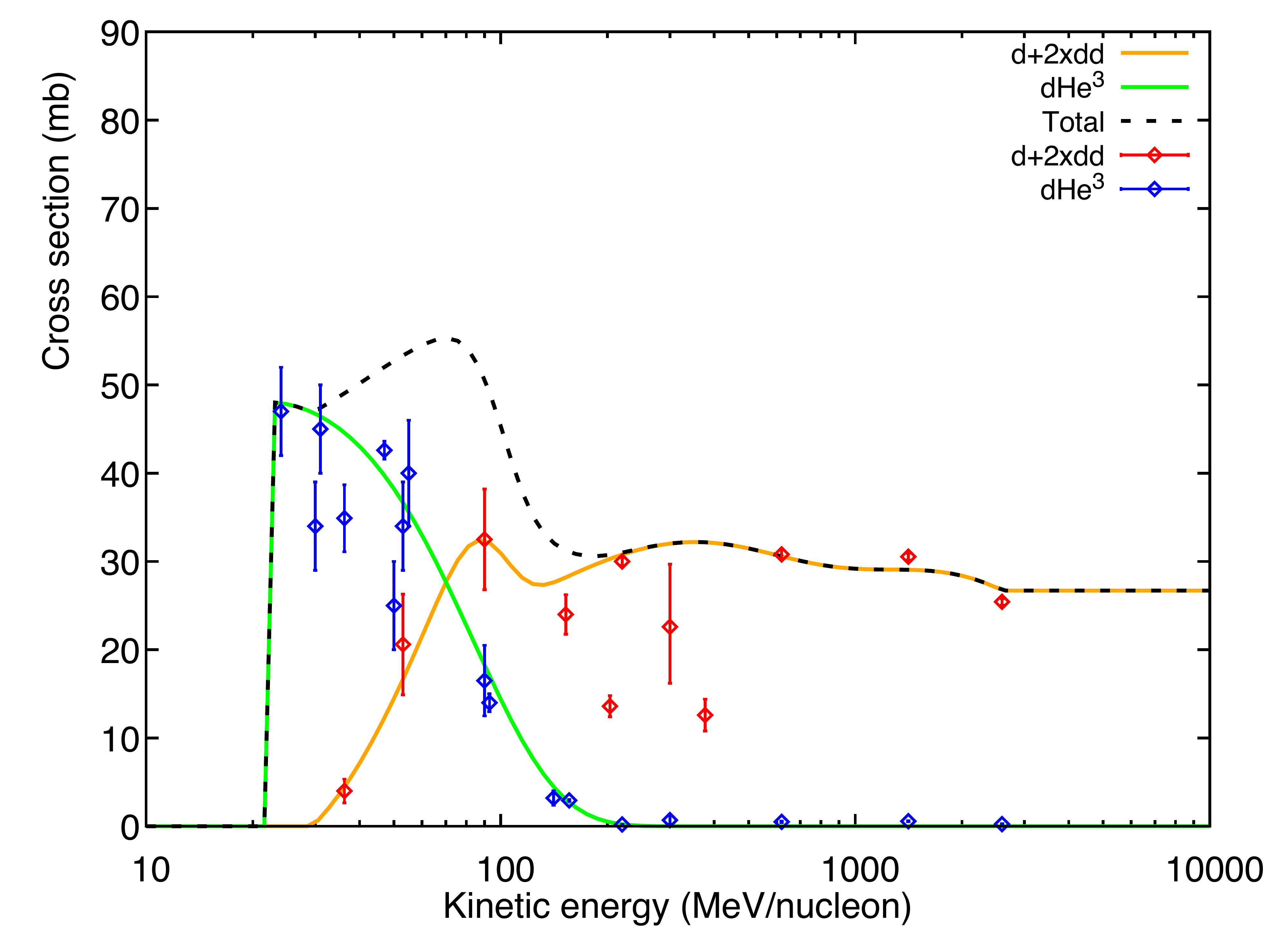}
\caption{{\it Left:} Fragmentation $\sigma_{^3\rm He}$ and pickup $\sigma_{d^3\rm He}$ cross sections. Black dashed line is the adapted total cross section. Fragmentation measurements are taken from: \citet{1972A&AS....7..417M}--a review, \citet{1969ApJ...158..711G}, \citet{1972NuPhA.181..329N}, \citet{1973PhRvC...7.2209J}, \citet{1977NuPhA.285..461B}, \citet{1990AIPC..203..294W}, \citet{1993ZPhyC..60..421G}, \citet{1993PAN....56..536A,1994NuPhA.569..753A}, \citet{2001PAN....64..907B}.
Pickup measurements are taken from: \citet{1972A&AS....7..417M}--a review, \citet{1969ApJ...158..711G}, \citet{1972NuPhA.181..329N}, \citet{1973PhRvC...7.2209J}, \citet{1993ZPhyC..60..421G}, \citet{1993PAN....56..536A,1994NuPhA.569..753A}, \citet{2001PAN....64..907B}.
{\it Middle:} Tritium production cross section. The data are taken from: \citet{1972A&AS....7..417M}--a review, \citet{1973PhRvC...7.2209J}, \citet{1972NuPhA.181..329N}, \citet{1977NuPhA.285..461B}, \citet{1990AIPC..203..294W}, \citet{1993ZPhyC..60..421G}, \citet{1993PAN....56..536A, 1994NuPhA.569..753A}, \citet{2001PAN....64..907B}.
{\it Right:} Deuterium production cross section. The data are taken from: \citet{1972A&AS....7..417M}--a review, \citet{1973PhRvC...7.2209J}, \citet{1990AIPC..203..294W}, \citet{1993ZPhyC..60..421G}, \citet{1993PAN....56..536A, 1994NuPhA.569..753A}, \citet{2001PAN....64..907B}.
\label{pHe_fig1}}
\end{figure*}

\subsection{$^3$He}\label{sec:dHe3}

The cross section for production of $^3$He consists of two components that are parametrised independently.
The fragmentation cross section $^4$He$(p,pnX)^3$He:
\begin{eqnarray}\label{4He-p_pn-3He}
&&\sigma_{^3\rm He} = 42.5 e^{-\frac{E_k-832}{4005}}\left[\frac{2}{1+\exp\left[(E_{\rm th}-E_k)/8.67\right]}-1\right]\\
&&\times\left[1-\frac{0.428}{1+3.27\exp(-E_k/106)}\right]^3 \left[1-1.66\left(\frac{E_k}{571}\right)^{1/2}\right] \,\, {\rm mb},\nonumber
\end{eqnarray}
where $\sigma_{^3\rm He}$ is assumed a constant above 2600~MeV,
and the pickup cross section $^4$He$(p,dX)^3$He:
\begin{equation} \label{dHe3}
\sigma_{d^3\rm He} = 48 \exp\left[-(E_k-E_{\rm th})^{1.59}/830\right] \,\, {\rm mb},
\end{equation}
where $E_k$ is the LS kinetic energy per nucleon, and $E_{\rm th}$ is the threshold energy of the corresponding reaction (Table \ref{tab:threshold}).
The cross sections along with available experimental data are plotted in Fig.~\ref{pHe_fig1} (left).

\subsection{Tritium}

A similar parameterisation was used also for the reaction $^4$He$(p,ppX)^3$H:
\begin{eqnarray}
&&\sigma_{^3\rm H} = 0.049 e^{-\frac{E_k-606}{3600}}\left[\frac{2}{1+\exp\left[(E_{\rm th}-E_k)/20.7\right]}-1\right]\\
&&\times\left[1-\frac{0.33}{1+142\exp(-E_k/38)}\right]^3 \left[1-5.90\left(\frac{E_k}{0.011}\right)^{1/2}\right] \,\, {\rm mb},\nonumber
\end{eqnarray}
where $\sigma_{^3\rm H}$ is assumed a constant above 2600 MeV, $E_{\rm th}$ again is the threshold energy of the corresponding reaction (Table \ref{tab:threshold}), and the parameters are adjusted to the data (see Fig.~\ref{pHe_fig1} middle).
\subsection{Deuterium}

Deuterium production through $^4$He fragmentation is calculated as a sum of three channels: $^4$He$(p,dX)^3$He + $^4$He$(p,ppnX)d$ + $2\times ^4$He$(p,pdX)d$.
Note the factor of 2 in the last term.
The first one, the pickup cross section, is exactly the same as Eq.~(\ref{dHe3}) in Sec.~\ref{sec:dHe3}:
\begin{equation}
\sigma_{d^3\rm He} = 48 \exp\left[-(E_k-E_{\rm th})^{1.59}/830\right] \,\, {\rm mb}.
\end{equation}
A sum of two other cross sections can be approximated with the following expression:
\begin{eqnarray}\label{d}
&&\sigma_{d} = 0.049 e^{-\frac{E_k-881}{3136}}\left[\frac{2}{1+\exp\left[(E_{\rm th}-E_k)/13\right]}-1\right]\\
&&\times\left[1-\frac{0.31}{1+2.52\exp(-E_k/247)}\right]^3 \left[1-5.90\left(\frac{E_k}{0.011}\right)^{1/2}\right]\nonumber\\
&&+10e^{-\frac{(82.5-E_k)^2}{756}} \,\, {\rm mb},\nonumber
\end{eqnarray}
where $\sigma_{d}$ is assumed a constant above 2600~MeV, and the threshold energy $E_{\rm th}$ corresponds to the reaction $^4$He$(p,pdX)d$.
The parameters of Eq.~(\ref{d}) are adjusted to the data (see Fig.~\ref{pHe_fig1} right).
Note that two red points $13.6\pm1.2$~mb at 203.3~MeV~nucleon$^{-1}$ and $12.6\pm1.8$~mb 377.1~MeV~nucleon$^{-1}$ \citep{1990AIPC..203..294W} are disregarded because they are contradicting to more recent data $30.0\pm0.45$~mb at 220~MeV~nucleon$^{-1}$ \citep{1994NuPhA.569..753A}, and $30.80\pm0.61$~mb at 620~MeV~nucleon$^{-1}$ \citep{2001PAN....64..907B}.

\begin{figure}[tb!]
\includegraphics[width=0.99\linewidth]{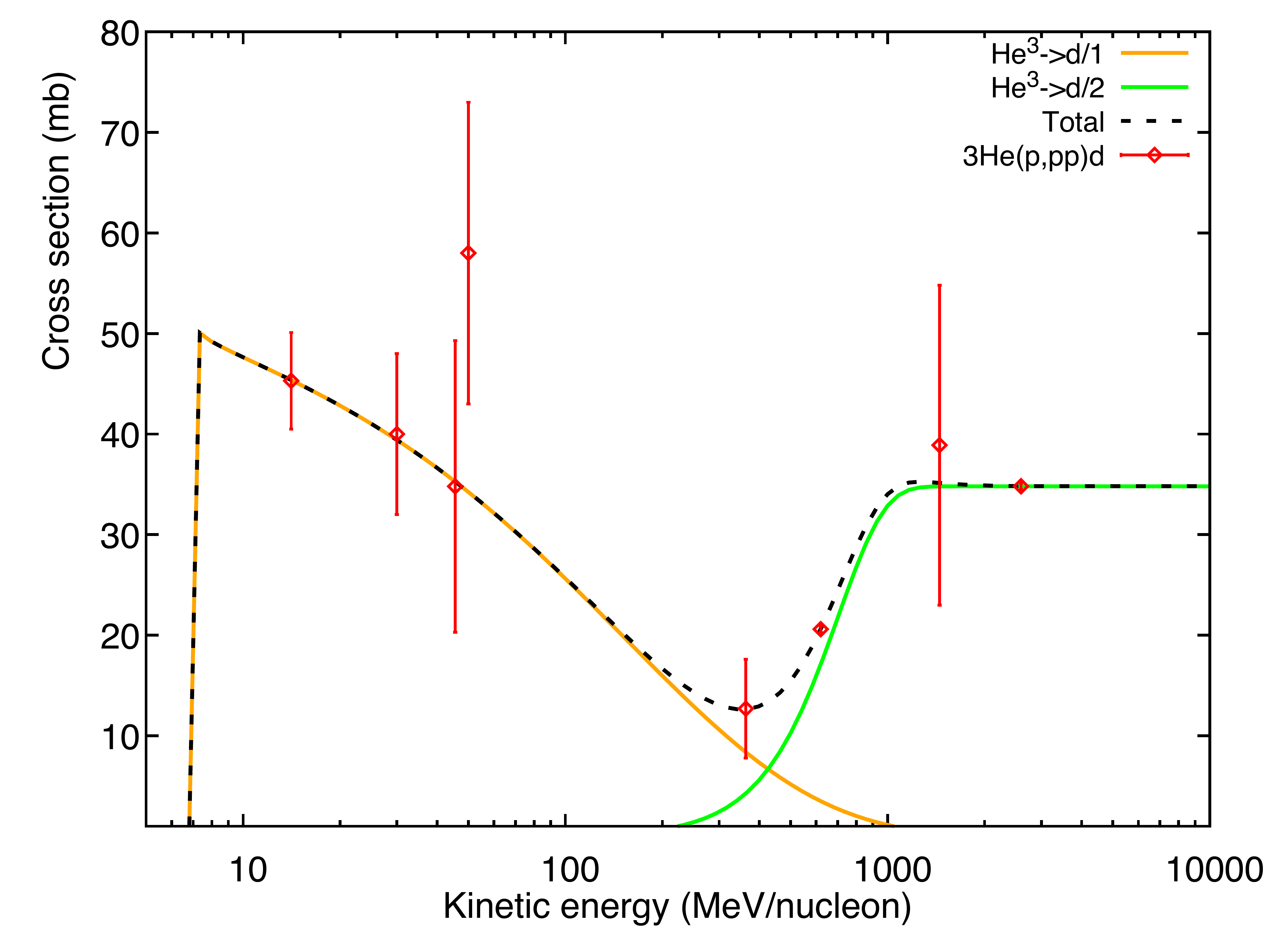}
\caption{A fit to the cross section $\sigma_{^3{\rm He}\to d}$ together with data. Coloured lines show separate components, and the black dashed line is the adapted total cross section. Data points are taken from \citet[][Figure 4]{1972A&AS....7..417M}, \citet{1969ApJ...158..711G}, \citet{1986NuPhA.451..701B}, and \citet{1993ZPhyC..60..421G}.
\label{pHe_fig2}}
\end{figure}

Fragmentation of $^3$He also produces deuterium in the reaction $^3$He$(p,ppX)d$, which we approximate using two terms:
\begin{equation}
\sigma_{^3{\rm He}\to d}= 50.1e^{-\frac{(E_k-E_{\rm th})^{0.73}}{40.7}} +34.8\left[1-e^{-\frac{(E_k-E_{\rm th})^{3.02}}{3.9\times10^8}}\right] \,\, {\rm mb},
\end{equation}
where the corresponding value of $E_{\rm th}$ is listed in Table \ref{tab:threshold}.
The plot is shown in Fig.~\ref{pHe_fig2}.

\subsection{Deuteron production cross section via $pp\to\pi^++d$} 

The dominant CR proton component means that deteuron production in $pp$ collisions can be an important channel. 
Its cross section has a maximum at $T_p$$\approx$560 MeV, and sharply drops for lower and higher energies.
It is therefore important only in a limited energy range, but the process yield can be nonnegligible.
The cross sections of reaction $pp\to\pi^++d$ and inverse reaction $\pi^++d\to pp$ were a matter of particular interest as their ratio presented a way to determine the spin of pions \citep{LockMeasday1970}.

The cross sections of these reactions are connected through a simple formula derived using the principle of detailed balance: 
\begin{equation}\label{balance}
\frac{\sigma_{pp}}{\sigma_{\pi d}}= 2 \frac{(2S_{\pi}+1)(2S_{d}+1)}{(2S_{p}+1)^2} \frac{{p'}_{\pi}^2}{{p'}_{p}^2}=\frac{3}{2}\frac{{p'}_{\pi}^2}{{p'}_{p}^2},
\end{equation}
where $\sigma_{pp}$ and $\sigma_{\pi d}$ are the cross sections of corresponding reactions, $S_{p,\pi, d}$ is the spin of a corresponding particle, and ${p'}_{p, \pi}$ are the CMS system momenta of projectiles ($p, \pi^+$).
An additional factor of 2 is added due to the identical particles (protons) in the final state of the inverse reaction.
The cross sections have to be evaluated at the same $s$, the invariant CMS energy squared.

Abundant data exist for both reactions.
Converting the cross section $\sigma_{\pi d}$ to $\sigma_{pp}$ using Eq.~(\ref{balance}) expands the energy range, where the data are available.
Calculations of ${p'}_{p, \pi}$ are done as follows:
\begin{eqnarray}\label{momenta}
s&=&(m_{\pi}+m_{d})^2+2 m_{d} T_{\pi}, \nonumber \\
{E'}_{\pi}&=&\frac{1}{2}\frac{(s+m_{\pi}^2-m_{d}^2)}{\sqrt{s}}, \nonumber \\
{E'}_{p}&=&\frac{\sqrt{s}}{2},  \\      
{p'}_{\pi}^2&=&{E'}_{\pi}^2 - m_{\pi}^2, \nonumber \\
{p'}_{p}^2&=&{E'}_{p}^2 - m_{p}^2, \nonumber 
\end{eqnarray}
where the prime sign ($'$) marks the CMS variables, ${E'}_{p, \pi}$ are the energies of corresponding particles, $T_{\pi}$ is the kinetic energy of pions in the LS, and $m_{p, \pi, d}$ are the masses. The corresponding LS kinetic energy of a proton can be calculated as
\begin{equation}
T_{p}=\frac{s-4m_p^2}{2m_{p}}, 
\end{equation}
where $s$ is determined by $T_{\pi}$ in Eq.~(\ref{momenta}).

\begin{figure}[tb!]
\includegraphics[width=1\linewidth]{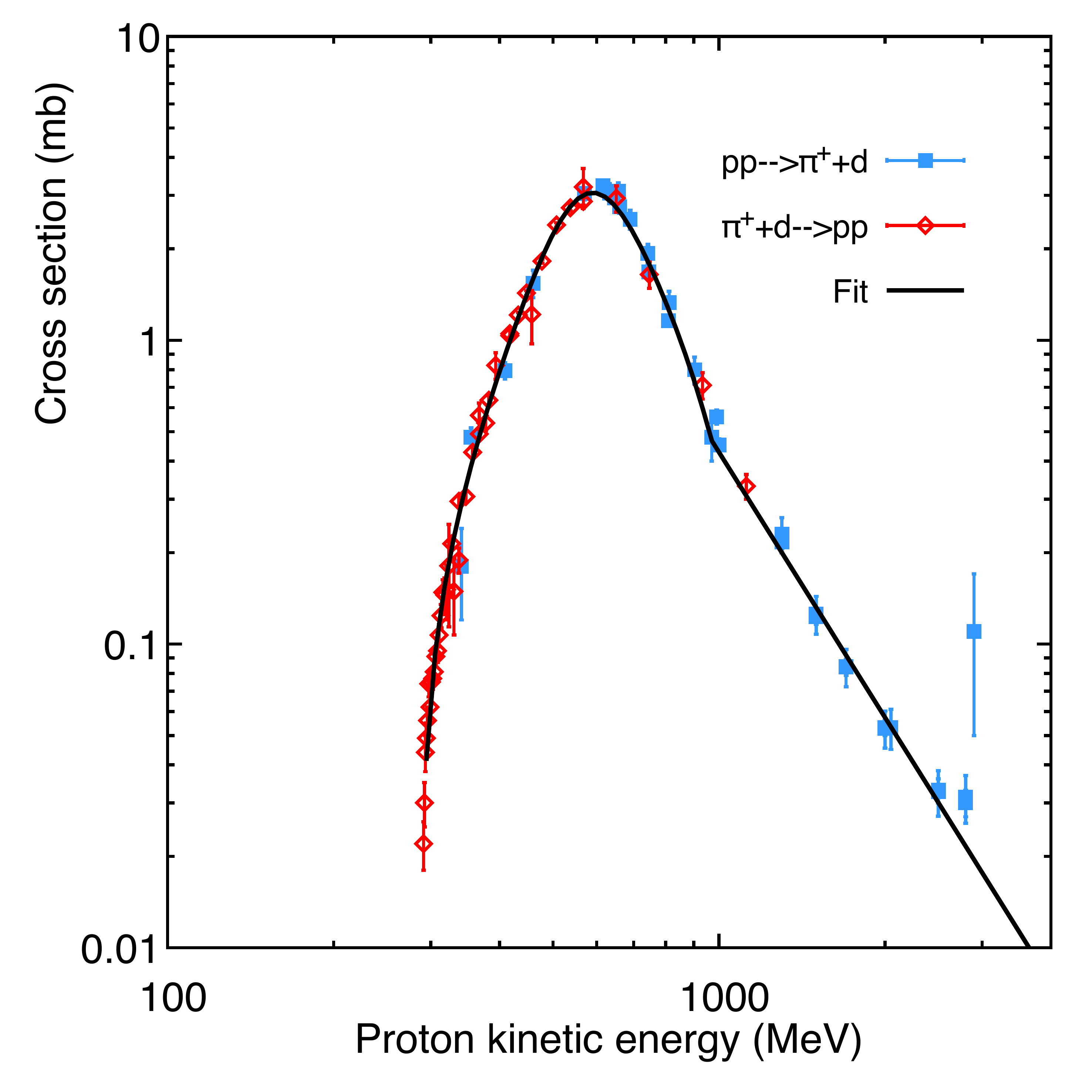}
\caption{Cross section data for the reaction $pp\to\pi^++d$ \citep[blue squares;][]{1953PhRv...91..677C, 1955Doklady_100_673M, 1955Doklady_100_677M, 1958PhRv..109.1733S, 1958JETP...34..767N, 1963PhRvL..11..474T, 1964PhRvL..13...59O, 1964PhRv..133.1017B, 1964PhL....11..253C, 1968PhRv..167.1232H, 1970NuPhB..20..413R}, and for the inverse reaction $\pi^++d\to pp$ \citep[red open diamonds;][]{1951PhRv...84..581D, 1957PhRv..105..247R, 1967PhRv..154.1305R,1976NuPhA.256..387A,  1981PhRvC..24..552R, 1982PhRvC..25.2540B, 1983PhRvC..27.1685R}, which were converted using Eq.~(\ref{balance}). The solid black curve is the fit for Eq.~(\ref{parameterization1}).
\label{sigma_pp}}
\end{figure}

Some measurements of $\sigma_{\pi d}$ are presented versus the (${p'}_{\pi}/m_{\pi}$) variable, the CMS pion momentum in units of $m_{\pi}$.
In this case, a corresponding CMS proton momentum ${p'}_{p}$ can be derived using Eq.~(\ref{momenta}), with $s$ calculated from
\begin{equation}\label{momenta1}
         s=\left[({p'}_{\pi}^2+m_{\pi}^2)^{1/2}+({p'}_{\pi}^2+m_{d}^2)^{1/2}\right]^2.
\end{equation}
To parameterise to the data we employ an expression for $\sigma_{\pi d}$ given by \citet{1983PhRvC..28..926R}:
\begin{equation}\label{parameterization}
\sigma_{\pi d}=a+\frac{b}{\sqrt{T_{\pi}}} +\frac{c}{(\sqrt{s}-E_r)^2+d},
\end{equation}
that we converted to $\sigma_{pp}$ using Eq.~(\ref{balance}), refitted to all available data for both reactions, and added a power-law tail to reproduce the high-energy data:
\begin{equation}\label{parameterization1}
\sigma_{pp}=
\begin{cases}
\displaystyle \frac{3}{2}\frac{{p'}_{\pi}^2}{{p'}_{p}^2} \sigma_{\pi d}\ \ {\rm mb},  &288\ {\rm MeV} <T_{p}\le 970\ {\rm MeV},\\
\displaystyle q\left(\frac{T_{p}}{E_{\rm norm}}\right)^{-\delta}\ \ {\rm mb}, &970\ {\rm MeV}<T_{p}\le 4000\ {\rm MeV},\\
0, & T_{p}> 4000\ {\rm MeV},\\
\end{cases}
\end{equation}
where the true threshold energy of the reaction is $T_{\rm th}=287.52$ MeV, and constant values derived from the fit are $a=-1.294$, $b=3.271$, $c=7.808\times10^4$, $d=6035$, $E_r=2136$, $q=0.4675$, $E_{\rm norm}=970$ MeV, and $\delta=2.9$.
Fig.~\ref{sigma_pp} shows the data collected from the literature together with the fit.

\begin{figure*}[tb!]
\includegraphics[width=0.49\linewidth]{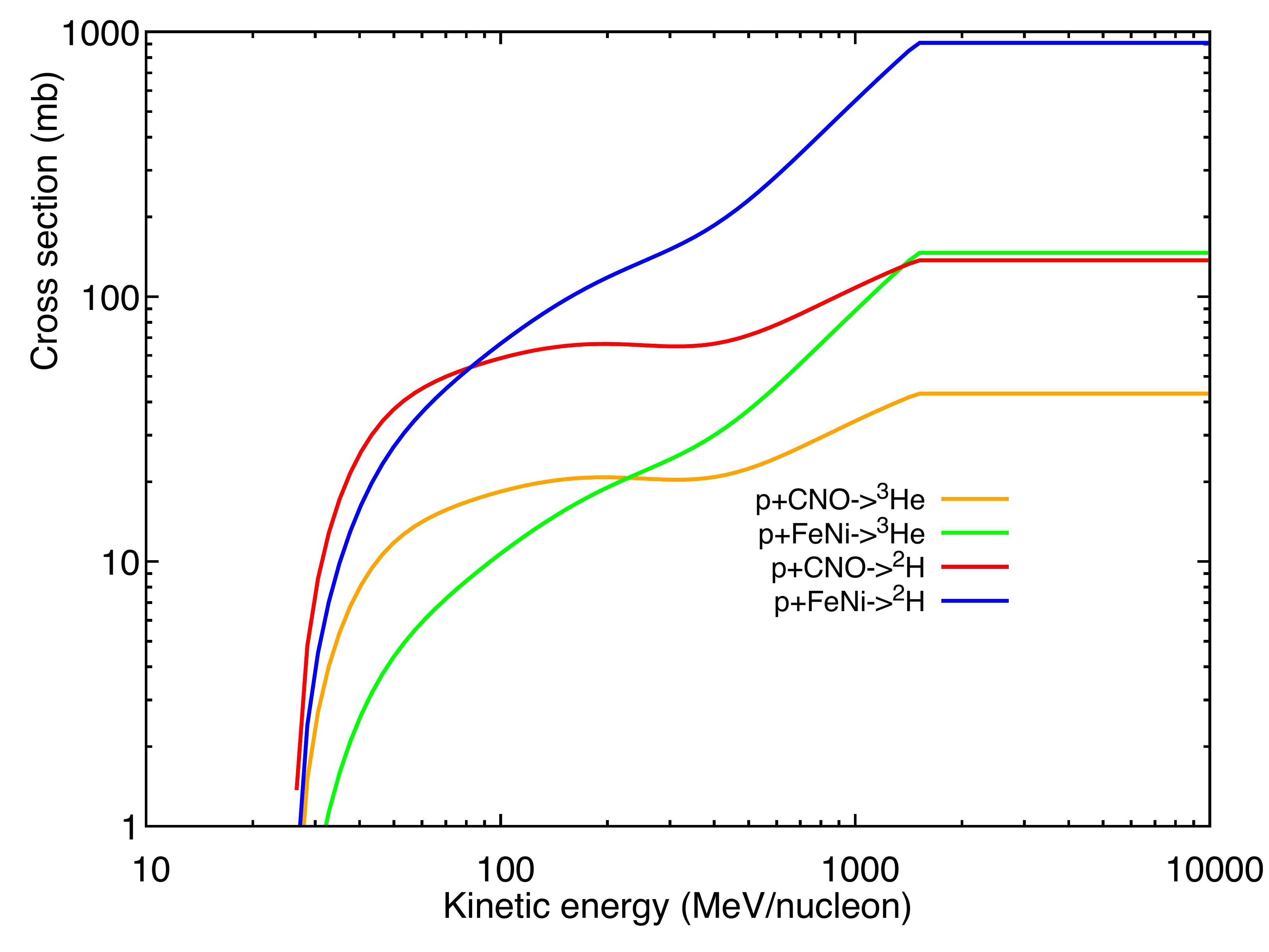}
\includegraphics[width=0.49\linewidth]{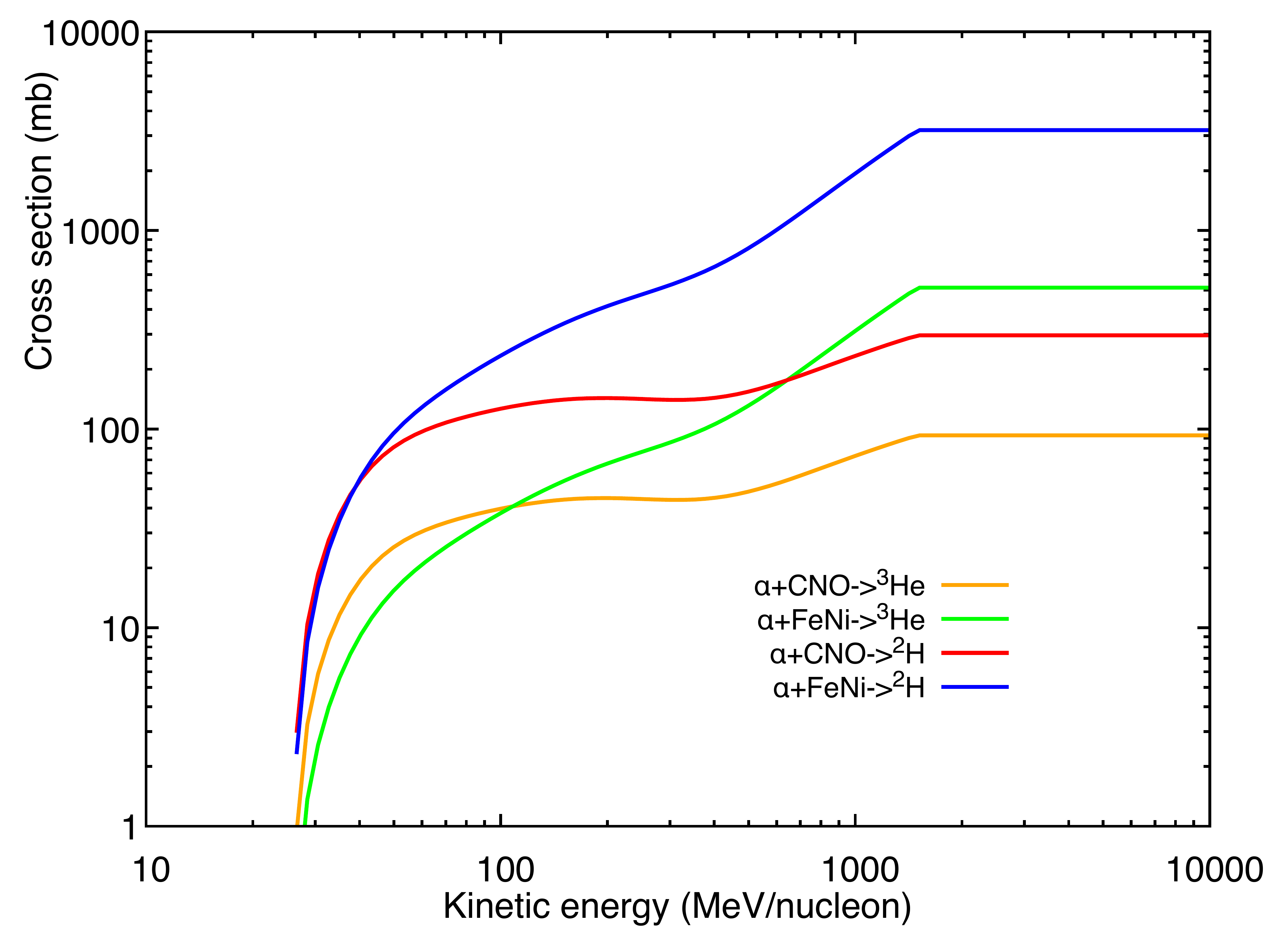}
\caption{Left: calculations of production of $^3$He, $^3$H, and $^2$H in $p+A$ $(A=12, 58)$ reactions.
Right: calculations of production of $^3$He, $^3$H, and $^2$H in $\alpha+A$ $(A=12, 58)$ reactions.
Calculations are done using the formalism described by \citet{2012A&A...539A..88C} and reproduced in Appendix~\ref{cs:heavy}.
\label{sigma_AA}}
\end{figure*}

\subsubsection{Differential cross section of deuteron production}

The range of LS kinetic energies of the produced deuteron in the $pp$ reaction, $(T_{d\,min},T_{d\,max})$, can be calculated as 
\begin{eqnarray}
T_{d\,min}&=&\gamma_{cms}E_d'-\beta_{cms}\gamma_{cms}p_d'-m_d,\\ 
T_{d\,max}&=&\gamma_{cms}E_d'+\beta_{cms}\gamma_{cms}p_d'-m_d,\nonumber 
\end{eqnarray}
where $T_{d\,min},T_{d\,max}$ are the minimum and maximum kinetic energy, and
\begin{eqnarray}
E_d'&=&\frac{s+m_d^2-m_\pi^2}{2 \sqrt{s}},\nonumber\\ 
p_d'&=&(E_d'^2-m_d^2)^{1/2},\nonumber\\                  
s&=&4m_{p}^2+2 m_{p} T_{p},\\
\gamma_{cms}&=&\frac{\sqrt{s}}{2m_p},\nonumber\\     
\beta_{cms}\gamma_{cms} &=& (\gamma_{cms}^2-1)^{1/2}.\nonumber
\end{eqnarray}
Here $T_p$ is the kinetic energy of the projectile proton, and $\gamma_{\rm cms},\beta_{\rm cms}$ are the Lorentz factor and speed of the CMS system.

As the first approximation, we assume the deuteron LS energy distribution is uniform in kinetic or total energy:
\begin{eqnarray}
\frac{d\sigma_{pp}(T_p,T_d)}{dT_{d}}&=&\frac{\sigma_{pp}(T_p,T_d)}{T_{d\,max}-T_{d\,min}},\ \  T_{d\,min}\le T_{d}\le T_{d\,max},\nonumber\\
\\
\frac{d\sigma_{pp}(T_p,E_d)}{dE_{d}}&=&\frac{\sigma_{pp}(T_p,E_d)}{E_{d\,max}-E_{d\,min}},\ \  E_{d\,min}\le E_{d}\le E_{d\,max},\nonumber
\end{eqnarray}
which are equivalent. It can be converted to the total LS momentum variable:
\begin{equation}
\frac{d\sigma_{pp}(p_p,p_d)}{dp_{d}}=\frac{d\sigma_{pp}}{dE_{d}}\frac{p_d}{E_d},  
\end{equation}
where the factor $p_d/E_d=dE_p/dp_p$ is the Jacobian.


\subsection{Heavier targets and/or projectiles}\label{cs:heavy}

For heavy targets $p+A$ $(A>4)$ and/or projectiles $^4$He $+A$ $(A\ge4)$ we follow a formalism described by \citet{2012A&A...539A..88C}.
The production of $^3$He, $^3$H, and $^2$H in $p+A$ interactions is scaled using $\sigma_{^3\rm He}$ defined in Eq.~(\ref{4He-p_pn-3He}):
\begin{equation}\label{cs:pA}
\sigma_{pA} = \Gamma_f(A) F(E_k,A) \sigma_{^3\rm He},
\end{equation}
where
\begin{eqnarray}
&&\Gamma_{^3\rm He}(A) = \Gamma_{^3\rm H}(A) =  1.3 \left[1+\left(\frac{A}{25}\right)^{1.5}\right],\nonumber \\
&&\Gamma_{^2\rm H}(A)=0.28 A^{1.2},\\
&&F(E_k,A)=
\begin{cases}
    \left(\frac{E_k}{1500\, \rm MeV/n}\right)^{0.8(A/26)^{0.5}},& E_k\le 1500 \, \rm MeV/n\\
    1,              & \text{otherwise}.
\end{cases}\nonumber
\end{eqnarray}

For the case of $^4$He $+A$ $(A\ge4)$ reactions, the $pA$ cross section $\sigma_{pA}$ (Eq.~\ref{cs:pA}) is scaled by a factor $A^{0.31}$.

The calculated cross sections are shown in Fig.~\ref{sigma_AA}.

\end{appendix}

\bibliography{gp_v57,imos,imos_icrc2015,ms_iso}

\begin{thebibliography}{}
\expandafter\ifx\csname natexlab\endcsname\relax\def\natexlab#1{#1}\fi
\providecommand{\url}[1]{\href{#1}{#1}}

\bibitem[{{Abdo} {et~al.}(2008){Abdo}, {Allen}, {Aune}, {Berley}, {Blaufuss},
  {Casanova}, {Chen}, {Dingus}, {Ellsworth}, {Fleysher}, {Fleysher},
  {Gonzalez}, {Goodman}, {Hoffman}, {H{\"u}ntemeyer}, {Kolterman}, {Lansdell},
  {Linnemann}, {McEnery}, {Mincer}, {Moskalenko}, {Nemethy}, {Noyes}, {Porter},
  {Pretz}, {Ryan}, {Parkinson}, {Shoup}, {Sinnis}, {Smith}, {Strong},
  {Sullivan}, {Vasileiou}, {Walker}, {Williams}, \&
  {Yodh}}]{2008ApJ...688.1078A}
{Abdo}, A.~A., {Allen}, B., {Aune}, T., {et~al.} 2008, \apj, 688, 1078

\bibitem[{{Abdullin} {et~al.}(1993){Abdullin}, {Blinov}, {Vanyushin},
  {Grechko}, {Ergakov}, {Zombkovskii}, {Kiselevich}, {Korolev}, \&
  {Selektor}}]{1993PAN....56..536A}
{Abdullin}, S.~K., {Blinov}, A.~V., {Vanyushin}, I.~A., {et~al.} 1993, Physics
  of Atomic Nuclei, 56, 536

\bibitem[{{Abdullin} {et~al.}(1994){Abdullin}, {Blinov}, {Chadeeva}, {Chuvilo},
  {Ergakov}, {Grechko}, {Kiselevich}, {Korolev}, {Selektor}, {Turov},
  {Vanyushin}, \& {Zombkovsky}}]{1994NuPhA.569..753A}
{Abdullin}, S.~K., {Blinov}, A.~V., {Chadeeva}, M.~V., {et~al.} 1994, \nphysa,
  569, 753

\bibitem[{{Abeysekara} {et~al.}(2017){Abeysekara}, {Albert}, {Alfaro},
  {Alvarez}, {{\'A}lvarez}, {Arceo}, {Arteaga-Vel{\'a}zquez}, {Avila Rojas},
  {Ayala Solares}, {Barber}, {Bautista-Elivar}, {Becerril}, {Belmont-Moreno},
  {BenZvi}, {Berley}, {Bernal}, {Braun}, {Brisbois}, {Caballero-Mora},
  {Capistr{\'a}n}, {Carrami{\~n}ana}, {Casanova}, {Castillo}, {Cotti},
  {Cotzomi}, {Couti{\~n}o de Le{\'o}n}, {De Le{\'o}n}, {De la Fuente},
  {Dingus}, {DuVernois}, {D{\'{\i}}az-V{\'e}lez}, {Ellsworth}, {Engel},
  {Enr{\'{\i}}quez-Rivera}, {Fiorino}, {Fraija}, {Garc{\'{\i}}a-Gonz{\'a}lez},
  {Garfias}, {Gerhardt}, {Gonz{\'a}lez Mu{\~n}oz}, {Gonz{\'a}lez}, {Goodman},
  {Hampel-Arias}, {Harding}, {Hern{\'a}ndez}, {Hern{\'a}ndez-Almada}, {Hinton},
  {Hona}, {Hui}, {H{\"u}ntemeyer}, {Iriarte}, {Jardin-Blicq}, {Joshi},
  {Kaufmann}, {Kieda}, {Lara}, {Lauer}, {Lee}, {Lennarz}, {Vargas},
  {Linnemann}, {Longinotti}, {Luis Raya}, {Luna-Garc{\'{\i}}a},
  {L{\'o}pez-Coto}, {Malone}, {Marinelli}, {Martinez}, {Martinez-Castellanos},
  {Mart{\'{\i}}nez-Castro}, {Mart{\'{\i}}nez-Huerta}, {Matthews},
  {Miranda-Romagnoli}, {Moreno}, {Mostaf{\'a}}, {Nellen}, {Newbold}, {Nisa},
  {Noriega-Papaqui}, {Pelayo}, {Pretz}, {P{\'e}rez-P{\'e}rez}, {Ren}, {Rho},
  {Rivi{\`e}re}, {Rosa-Gonz{\'a}lez}, {Rosenberg}, {Ruiz-Velasco}, {Salazar},
  {Salesa Greus}, {Sandoval}, {Schneider}, {Schoorlemmer}, {Sinnis}, {Smith},
  {Springer}, {Surajbali}, {Taboada}, {Tibolla}, {Tollefson}, {Torres},
  {Ukwatta}, {Vianello}, {Weisgarber}, {Westerhoff}, {Wisher}, {Wood},
  {Yapici}, {Yodh}, {Younk}, {Zepeda}, {Zhou}, {Guo}, {Hahn}, {Li}, \&
  {Zhang}}]{2017Sci...358..911A}
{Abeysekara}, A.~U., {Albert}, A., {Alfaro}, R., {et~al.} 2017, Science, 358,
  911

\bibitem[{{Abraham} {et~al.}(1966){Abraham}, {Brunstein}, \&
  {Cline}}]{1966PhRv..150.1088A}
{Abraham}, P.~B., {Brunstein}, K.~A., \& {Cline}, T.~L. 1966, Physical Review,
  150, 1088

\bibitem[{{Ackermann} {et~al.}(2012){Ackermann}, {Ajello}, {Atwood}, {Baldini},
  {Ballet}, {Barbiellini}, {Bastieri}, {Bechtol}, {Bellazzini}, {Berenji},
  {Blandford}, {Bloom}, {Bonamente}, {Borgland}, {Brandt}, {Bregeon},
  {Brigida}, {Bruel}, {Buehler}, {Buson}, {Caliandro}, {Cameron}, {Caraveo},
  {Cavazzuti}, {Cecchi}, {Charles}, {Chekhtman}, {Chiang}, {Ciprini}, {Claus},
  {Cohen-Tanugi}, {Conrad}, {Cutini}, {de Angelis}, {de Palma}, {Dermer},
  {Digel}, {Silva}, {Drell}, {Drlica-Wagner}, {Falletti}, {Favuzzi}, {Fegan},
  {Ferrara}, {Focke}, {Fortin}, {Fukazawa}, {Funk}, {Fusco}, {Gaggero},
  {Gargano}, {Germani}, {Giglietto}, {Giordano}, {Giroletti}, {Glanzman},
  {Godfrey}, {Grove}, {Guiriec}, {Gustafsson}, {Hadasch}, {Hanabata},
  {Harding}, {Hayashida}, {Hays}, {Horan}, {Hou}, {Hughes}, {J{\'o}hannesson},
  {Johnson}, {Johnson}, {Kamae}, {Katagiri}, {Kataoka}, {Kn{\"o}dlseder},
  {Kuss}, {Lande}, {Latronico}, {Lee}, {Lemoine-Goumard}, {Longo}, {Loparco},
  {Lott}, {Lovellette}, {Lubrano}, {Mazziotta}, {McEnery}, {Michelson},
  {Mitthumsiri}, {Mizuno}, {Monte}, {Monzani}, {Morselli}, {Moskalenko},
  {Murgia}, {Naumann-Godo}, {Norris}, {Nuss}, {Ohsugi}, {Okumura}, {Omodei},
  {Orlando}, {Ormes}, {Paneque}, {Panetta}, {Parent}, {Pesce-Rollins},
  {Pierbattista}, {Piron}, {Pivato}, {Porter}, {Rain{\`o}}, {Rando}, {Razzano},
  {Razzaque}, {Reimer}, {Reimer}, {Sadrozinski}, {Sgr{\`o}}, {Siskind},
  {Spandre}, {Spinelli}, {Strong}, {Suson}, {Takahashi}, {Tanaka}, {Thayer},
  {Thayer}, {Thompson}, {Tibaldo}, {Tinivella}, {Torres}, {Tosti}, {Troja},
  {Usher}, {Vandenbroucke}, {Vasileiou}, {Vianello}, {Vitale}, {Waite}, {Wang},
  {Winer}, {Wood}, {Wood}, {Yang}, {Ziegler}, \&
  {Zimmer}}]{2012ApJ...750....3A}
{Ackermann}, M., {Ajello}, M., {Atwood}, W.~B., {et~al.} 2012, \apj, 750, 3

\bibitem[{{Ackermann} {et~al.}(2015){Ackermann}, {Ajello}, {Albert}, {Atwood},
  {Baldini}, {Ballet}, {Barbiellini}, {Bastieri}, {Bechtol}, {Bellazzini},
  {Bissaldi}, {Blandford}, {Bloom}, {Bottacini}, {Brandt}, {Bregeon}, {Bruel},
  {Buehler}, {Buson}, {Caliandro}, {Cameron}, {Caragiulo}, {Caraveo},
  {Cavazzuti}, {Cecchi}, {Charles}, {Chekhtman}, {Chiang}, {Chiaro}, {Ciprini},
  {Claus}, {Cohen-Tanugi}, {Conrad}, {Cuoco}, {Cutini}, {D'Ammando}, {de
  Angelis}, {de Palma}, {Dermer}, {Digel}, {Silva}, {Drell}, {Favuzzi},
  {Ferrara}, {Focke}, {Franckowiak}, {Fukazawa}, {Funk}, {Fusco}, {Gargano},
  {Gasparrini}, {Germani}, {Giglietto}, {Giommi}, {Giordano}, {Giroletti},
  {Godfrey}, {Gomez-Vargas}, {Grenier}, {Guiriec}, {Gustafsson}, {Hadasch},
  {Hayashi}, {Hays}, {Hewitt}, {Ippoliti}, {Jogler}, {J{\'o}hannesson},
  {Johnson}, {Johnson}, {Kamae}, {Kataoka}, {Kn{\"o}dlseder}, {Kuss},
  {Larsson}, {Latronico}, {Li}, {Li}, {Longo}, {Loparco}, {Lott}, {Lovellette},
  {Lubrano}, {Madejski}, {Manfreda}, {Massaro}, {Mayer}, {Mazziotta},
  {McEnery}, {Michelson}, {Mitthumsiri}, {Mizuno}, {Moiseev}, {Monzani},
  {Morselli}, {Moskalenko}, {Murgia}, {Nemmen}, {Nuss}, {Ohsugi}, {Omodei},
  {Orlando}, {Ormes}, {Paneque}, {Panetta}, {Perkins}, {Pesce-Rollins},
  {Piron}, {Pivato}, {Porter}, {Rain{\`o}}, {Rando}, {Razzano}, {Razzaque},
  {Reimer}, {Reimer}, {Reposeur}, {Ritz}, {Romani}, {S{\'a}nchez-Conde},
  {Schaal}, {Schulz}, {Sgr{\`o}}, {Siskind}, {Spandre}, {Spinelli}, {Strong},
  {Suson}, {Takahashi}, {Thayer}, {Thayer}, {Tibaldo}, {Tinivella}, {Torres},
  {Tosti}, {Troja}, {Uchiyama}, {Vianello}, {Werner}, {Winer}, {Wood}, {Wood},
  {Zaharijas}, \& {Zimmer}}]{2015ApJ...799...86A}
{Ackermann}, M., {Ajello}, M., {Albert}, A., {et~al.} 2015, \apj, 799, 86

\bibitem[{{Ajello} {et~al.}(2016){Ajello}, {Albert}, {Atwood}, {Barbiellini},
  {Bastieri}, {Bechtol}, {Bellazzini}, {Bissaldi}, {Blandford}, {Bloom},
  {Bonino}, {Bottacini}, {Brandt}, {Bregeon}, {Bruel}, {Buehler}, {Buson},
  {Caliandro}, {Cameron}, {Caputo}, {Caragiulo}, {Caraveo}, {Cecchi},
  {Chekhtman}, {Chiang}, {Chiaro}, {Ciprini}, {Cohen-Tanugi}, {Cominsky},
  {Conrad}, {Cutini}, {D'Ammando}, {de Angelis}, {de Palma}, {Desiante}, {Di
  Venere}, {Drell}, {Favuzzi}, {Ferrara}, {Fusco}, {Gargano}, {Gasparrini},
  {Giglietto}, {Giommi}, {Giordano}, {Giroletti}, {Glanzman}, {Godfrey},
  {Gomez-Vargas}, {Grenier}, {Guiriec}, {Gustafsson}, {Harding}, {Hewitt},
  {Hill}, {Horan}, {Jogler}, {J{\'o}hannesson}, {Johnson}, {Kamae}, {Karwin},
  {Kn{\"o}dlseder}, {Kuss}, {Larsson}, {Latronico}, {Li}, {Li}, {Longo},
  {Loparco}, {Lovellette}, {Lubrano}, {Magill}, {Maldera}, {Malyshev},
  {Manfreda}, {Mayer}, {Mazziotta}, {Michelson}, {Mitthumsiri}, {Mizuno},
  {Moiseev}, {Monzani}, {Morselli}, {Moskalenko}, {Murgia}, {Nuss}, {Ohno},
  {Ohsugi}, {Omodei}, {Orlando}, {Ormes}, {Paneque}, {Pesce-Rollins}, {Piron},
  {Pivato}, {Porter}, {Rain{\`o}}, {Rando}, {Razzano}, {Reimer}, {Reimer},
  {Ritz}, {S{\'a}nchez-Conde}, {Saz Parkinson}, {Sgr{\`o}}, {Siskind}, {Smith},
  {Spada}, {Spandre}, {Spinelli}, {Suson}, {Tajima}, {Takahashi}, {Thayer},
  {Torres}, {Tosti}, {Troja}, {Uchiyama}, {Vianello}, {Winer}, {Wood},
  {Zaharijas}, \& {Zimmer}}]{2016ApJ...819...44A}
{Ajello}, M., {Albert}, A., {Atwood}, W.~B., {et~al.} 2016, \apj, 819, 44

\bibitem[{{Amenomori} {et~al.}(2021){Amenomori}, {Bao}, {Bi}, {Chen}, {Chen},
  {Chen}, {Chen}, {Chen}, {Cirennima}, {Danzengluobu}, {Fang}, {Fang}, {Feng},
  {Feng}, {Feng}, {Gao}, {Gou}, {Guo}, {Guo}, {He}, {He}, {Hibino}, {Hotta},
  {Hu}, {Hu}, {Huang}, {Jia}, {Jiang}, {Jin}, {Kasahara}, {Katayose}, {Kato},
  {Kato}, {Kawata}, {Kihara}, {Ko}, {Kozai}, {Labaciren}, {Li}, {Li}, {Li},
  {Lin}, {Liu}, {Liu}, {Liu}, {Liu}, {Liu}, {Lou}, {Lu}, {Meng}, {Munakata},
  {Nakada}, {Nakamura}, {Nanjo}, {Nishizawa}, {Ohnishi}, {Ohura}, {Ozawa},
  {Qian}, {Qu}, {Saito}, {Sakata}, {Sako}, {Shao}, {Shibata}, {Shiomi},
  {Sugimoto}, {Takano}, {Takita}, {Tan}, {Tateyama}, {Torii}, {Tsuchiya},
  {Udo}, {Wang}, {Wu}, {Xue}, {Yamamoto}, {Yang}, {Yokoe}, {Yuan}, {Zhai},
  {Zhang}, {Zhang}, {Zhang}, {Zhang}, {Zhang}, {Zhang}, {Zhang}, {Zhao},
  {Zhaxisangzhu}, \& {Tibet AS<SUB>{\ensuremath{\gamma}}</SUB>
  Collaboration}}]{2021PhRvL.126n1101A}
{Amenomori}, M., {Bao}, Y.~W., {Bi}, X.~J., {et~al.} 2021, \prl, 126, 141101

\bibitem[{{Atoyan} {et~al.}(1995){Atoyan}, {Aharonian}, \&
  {V{\"o}lk}}]{1995PhRvD..52.3265A}
{Atoyan}, A.~M., {Aharonian}, F.~A., \& {V{\"o}lk}, H.~J. 1995, \prd, 52, 3265

\bibitem[{{Axen} {et~al.}(1976){Axen}, {Duesdieker}, {Felawka}, {Ingram},
  {Johnson}, {Jones}, {Lepatourel}, {Salomon}, {Westlund}, \&
  {Robertson}}]{1976NuPhA.256..387A}
{Axen}, D., {Duesdieker}, G., {Felawka}, L., {et~al.} 1976, \nphysa, 256, 387

\bibitem[{{Barashenkov} \& {Polanski}(1994)}]{BarPol1994}
{Barashenkov}, V.~S., \& {Polanski}, A. 1994, Electronic Guide for Nuclear
  Cross Sections, Tech. Rep. E2-94-417, Comm.\ JINR, Dubna

\bibitem[{{Beck}(2001)}]{2001SSRv...99..243B}
{Beck}, R. 2001, \ssr, 99, 243

\bibitem[{{Bernard} {et~al.}(2012){Bernard}, {Delahaye}, {Salati}, \&
  {Taillet}}]{2012A&A...544A..92B}
{Bernard}, G., {Delahaye}, T., {Salati}, P., \& {Taillet}, R. 2012, \aap, 544,
  A92

\bibitem[{{Berrington} \& {Dermer}(2003)}]{2003ApJ...594..709B}
{Berrington}, R.~C., \& {Dermer}, C.~D. 2003, \apj, 594, 709

\bibitem[{{Beuermann} {et~al.}(1985){Beuermann}, {Kanbach}, \&
  {Berkhuijsen}}]{1985A&A...153...17B}
{Beuermann}, K., {Kanbach}, G., \& {Berkhuijsen}, E.~M. 1985, \aap, 153, 17

\bibitem[{{Bizard} {et~al.}(1977){Bizard}, {Le Brun}, {Berger}, {Duflo},
  {Goldzahl}, {Plouin}, {Oostens}, {Van Den Bossche}, {Vu Hai}, {Fabbri},
  {Picozza}, \& {Satta}}]{1977NuPhA.285..461B}
{Bizard}, G., {Le Brun}, C., {Berger}, J., {et~al.} 1977, \nphysa, 285, 461

\bibitem[{{Blinov} {et~al.}(2001){Blinov}, {Chadeyeva}, {Grechko}, \&
  {Turov}}]{2001PAN....64..907B}
{Blinov}, A.~V., {Chadeyeva}, M.~V., {Grechko}, V.~E., \& {Turov}, V.~F. 2001,
  Physics of Atomic Nuclei, 64, 907

\bibitem[{{Blinov} {et~al.}(1986){Blinov}, {Chuvilo}, {Drobot}, {Ergakov},
  {Grechko}, {Korolev}, {Selector}, {Soloviev}, {Shulyachenko}, {Trebukhovsky},
  {Turov}, {Vanyushin}, \& {Zombkovsky}}]{1986NuPhA.451..701B}
{Blinov}, A.~V., {Chuvilo}, I.~V., {Drobot}, V.~V., {et~al.} 1986, \nphysa,
  451, 701

\bibitem[{{Bolatto} {et~al.}(2013){Bolatto}, {Wolfire}, \&
  {Leroy}}]{2013ARA&A..51..207B}
{Bolatto}, A.~D., {Wolfire}, M., \& {Leroy}, A.~K. 2013, \araa, 51, 207

\bibitem[{{Boschini} {et~al.}(2018{\natexlab{a}}){Boschini}, {Della Torre},
  {Gervasi}, {La Vacca}, \& {Rancoita}}]{2018AdSpR..62.2859B}
{Boschini}, M.~J., {Della Torre}, S., {Gervasi}, M., {La Vacca}, G., \&
  {Rancoita}, P.~G. 2018{\natexlab{a}}, \asr, 62, 2859

\bibitem[{{Boschini} {et~al.}(2019){Boschini}, {Della Torre}, {Gervasi}, {La
  Vacca}, \& {Rancoita}}]{2019AdSpR..64.2459B}
---. 2019, Advances in Space Research, 64, 2459

\bibitem[{{Boschini} {et~al.}(2017){Boschini}, {Della Torre}, {Gervasi},
  {Grandi}, {J{\'o}hannesson}, {Kachelriess}, {La Vacca}, {Masi}, {Moskalenko},
  {Orlando}, {Ostapchenko}, {Pensotti}, {Porter}, {Quadrani}, {Rancoita},
  {Rozza}, \& {Tacconi}}]{2017ApJ...840..115B}
{Boschini}, M.~J., {Della Torre}, S., {Gervasi}, M., {et~al.} 2017, \apj, 840,
  115

\bibitem[{{Boschini} {et~al.}(2018{\natexlab{b}}){Boschini}, {Della Torre},
  {Gervasi}, {Grandi}, {J{\'o}hannesson}, {La Vacca}, {Masi}, {Moskalenko},
  {Pensotti}, {Porter}, {Quadrani}, {Rancoita}, {Rozza}, \&
  {Tacconi}}]{2018ApJ...854...94B}
---. 2018{\natexlab{b}}, \apj, 854, 94

\bibitem[{{Boschini} {et~al.}(2018{\natexlab{c}}){Boschini}, {Della Torre},
  {Gervasi}, {Grandi}, {J{\'o}hannesson}, {La Vacca}, {Masi}, {Moskalenko},
  {Pensotti}, {Porter}, {Quadrani}, {Rancoita}, {Rozza}, \&
  {Tacconi}}]{2018ApJ...858...61B}
---. 2018{\natexlab{c}}, \apj, 858, 61

\bibitem[{{Boschini} {et~al.}(2020){Boschini}, {Della Torre}, {Gervasi}, {Grand
  i}, {J{\'o}hannesson}, {La Vacca}, {Masi}, {Moskalenko}, {Pensotti},
  {Porter}, {Quadrani}, {Rancoita}, {Rozza}, \&
  {Tacconi}}]{2020ApJS..250...27B}
---. 2020, \apjs, 250, 27

\bibitem[{{Boschini} {et~al.}(2021{\natexlab{a}}){Boschini}, {Della Torre},
  {Gervasi}, {Grandi}, {J{\'o}hannesson}, {La Vacca}, {Masi}, {Moskalenko},
  {Pensotti}, {Porter}, {Quadrani}, {Rancoita}, {Rozza}, \&
  {Tacconi}}]{2021ApJ...913....5B}
---. 2021{\natexlab{a}}, \apj, 913, 5

\bibitem[{{Boschini} {et~al.}(2021{\natexlab{b}}){Boschini}, {Della Torre},
  {Gervasi}, {Grandi}, {Johannesson}, {La Vacca}, {Masi}, {Moskalenko},
  {Pensotti}, {Porter}, {Quadrani}, {Rancoita}, {Rozza}, \&
  {Tacconi}}]{2021arXiv210601626B}
---. 2021{\natexlab{b}}, arXiv e-prints, arXiv:2106.01626

\bibitem[{{Boswell} {et~al.}(1982){Boswell}, {Altemus}, {Minehart}, {Orphanos},
  {Ziock}, \& {Wadlinger}}]{1982PhRvC..25.2540B}
{Boswell}, J., {Altemus}, R., {Minehart}, R., {et~al.} 1982, \prc, 25, 2540

\bibitem[{{Brandenburg} \& {Dobler}(2002)}]{2002CoPhC.147..471B}
{Brandenburg}, A., \& {Dobler}, W. 2002, Computer Physics Communications, 147,
  471

\bibitem[{{Bringmann} {et~al.}(2018){Bringmann}, {Edsj{\"o}}, {Gondolo},
  {Ullio}, \& {Bergstr{\"o}m}}]{2018JCAP...07..033B}
{Bringmann}, T., {Edsj{\"o}}, J., {Gondolo}, P., {Ullio}, P., \&
  {Bergstr{\"o}m}, L. 2018, \jcap, 2018, 033

\bibitem[{{Bronfman} {et~al.}(1988){Bronfman}, {Cohen}, {Alvarez}, {May}, \&
  {Thaddeus}}]{1988ApJ...324..248B}
{Bronfman}, L., {Cohen}, R.~S., {Alvarez}, H., {May}, J., \& {Thaddeus}, P.
  1988, \apj, 324, 248

\bibitem[{{Bugg} {et~al.}(1964){Bugg}, {Oxley}, {Zoll}, {Rushbrooke}, {Barnes},
  {Kinson}, {Dodd}, {Doran}, \& {Riddiford}}]{1964PhRv..133.1017B}
{Bugg}, D.~V., {Oxley}, A.~J., {Zoll}, J.~A., {et~al.} 1964, Physical Review,
  133, 1017

\bibitem[{{Bulanov} \& {Dogel}(1974)}]{1974Ap&SS..29..305B}
{Bulanov}, S.~V., \& {Dogel}, V.~A. 1974, \apss, 29, 305

\bibitem[{{Carlson} {et~al.}(1973){Carlson}, {Doherty}, {Margaziotis},
  {et~al.}}]{1973LNC_8_319C}
{Carlson}, R.~F., {Doherty}, P., {Margaziotis}, D.~J., {et~al.} 1973, Lett.
  Nuovo Cimento, 8, 319?323

\bibitem[{{Cartwright} {et~al.}(1953){Cartwright}, {Richman}, {Whitehead}, \&
  {Wilcox}}]{1953PhRv...91..677C}
{Cartwright}, W.~F., {Richman}, C., {Whitehead}, M.~N., \& {Wilcox}, H.~A.
  1953, Physical Review, 91, 677

\bibitem[{{Case} \& {Bhattacharya}(1998)}]{1998ApJ...504..761C}
{Case}, G.~L., \& {Bhattacharya}, D. 1998, \apj, 504, 761

\bibitem[{{Cataldo} {et~al.}(2019){Cataldo}, {Pagliaroli}, {Vecchiotti}, \&
  {Villante}}]{2019JCAP...12..050C}
{Cataldo}, M., {Pagliaroli}, G., {Vecchiotti}, V., \& {Villante}, F.~L. 2019,
  \jcap, 2019, 050

\bibitem[{{Chapman} {et~al.}(1964){Chapman}, {Jones}, {Khan}, {McKee}, {Van Der
  Raay}, \& {Tanimura}}]{1964PhL....11..253C}
{Chapman}, K.~R., {Jones}, T.~W., {Khan}, Q.~H., {et~al.} 1964, Physics
  Letters, 11, 253

\bibitem[{{Cordes}(2004)}]{2004ASPC..317..211C}
{Cordes}, J.~M. 2004, in Astronomical Society of the Pacific Conference Series,
  Vol. 317, Milky Way Surveys: The Structure and Evolution of our Galaxy, ed.
  D.~{Clemens}, R.~{Shah}, \& T.~{Brainerd}, 211

\bibitem[{{Cordes} \& {Lazio}(2002)}]{2002astro.ph..7156C}
{Cordes}, J.~M., \& {Lazio}, T.~J.~W. 2002, arXiv e-prints, astro

\bibitem[{{Cordes} \& {Lazio}(2003)}]{2003astro.ph..1598C}
---. 2003, arXiv e-prints, astro

\bibitem[{{Coste} {et~al.}(2012){Coste}, {Derome}, {Maurin}, \&
  {Putze}}]{2012A&A...539A..88C}
{Coste}, B., {Derome}, L., {Maurin}, D., \& {Putze}, A. 2012, \aap, 539, A88

\bibitem[{{Cox} {et~al.}(1986){Cox}, {Kruegel}, \&
  {Mezger}}]{1986A&A...155..380C}
{Cox}, P., {Kruegel}, E., \& {Mezger}, P.~G. 1986, \aap, 155, 380

\bibitem[{{Crawford}(1979)}]{1979PhDT........67C}
{Crawford}, H.~J. 1979, PhD thesis, Single Electron Attachment and Stripping
  Cross Sections for Relativistic Heavy Ions, University of California at
  Berkeley

\bibitem[{{Cucinotta}(1993)}]{1993STIN...9417666C}
{Cucinotta}, F.~A. 1993, {Calculations of cosmic-ray helium transport in
  shielding materials}, NASA Technical Paper 3354, ,

\bibitem[{{Dame} {et~al.}(2001){Dame}, {Hartmann}, \&
  {Thaddeus}}]{2001ApJ...547..792D}
{Dame}, T.~M., {Hartmann}, D., \& {Thaddeus}, P. 2001, \apj, 547, 792

\bibitem[{{Dame} \& {Thaddeus}(2004)}]{2004ASPC..317...66D}
{Dame}, T.~M., \& {Thaddeus}, P. 2004, in Astronomical Society of the Pacific
  Conference Series, Vol. 317, Milky Way Surveys: The Structure and Evolution
  of our Galaxy, ed. D.~{Clemens}, R.~{Shah}, \& T.~{Brainerd}, 66

\bibitem[{{de Marco} {et~al.}(2008){de Marco}, {Blasi}, \&
  {Stanev}}]{2008ICRC....2..195D}
{de Marco}, D., {Blasi}, P., \& {Stanev}, T. 2008, in International Cosmic Ray
  Conference, Vol.~2, International Cosmic Ray Conference, 195--198

\bibitem[{{de Vries} {et~al.}(1987){de Vries}, {de Jager}, \& {de
  Vries}}]{1987ADNDT..36..495D}
{de Vries}, H., {de Jager}, C.~W., \& {de Vries}, C. 1987, Atomic Data and
  Nuclear Data Tables, 36, 495

\bibitem[{{DeMarco} {et~al.}(2007){DeMarco}, {Blasi}, \&
  {Stanev}}]{2007JCAP...06..027D}
{DeMarco}, D., {Blasi}, P., \& {Stanev}, T. 2007, \jcap, 2007, 027

\bibitem[{{Dermer}(1986{\natexlab{a}})}]{1986A&A...157..223D}
{Dermer}, C.~D. 1986{\natexlab{a}}, \aap, 157, 223

\bibitem[{{Dermer}(1986{\natexlab{b}})}]{1986ApJ...307...47D}
---. 1986{\natexlab{b}}, \apj, 307, 47

\bibitem[{{Derome} {et~al.}(2019){Derome}, {Maurin}, {Salati}, {Boudaud},
  {G{\'e}nolini}, \& {Kunz{\'e}}}]{2019A&A...627A.158D}
{Derome}, L., {Maurin}, D., {Salati}, P., {et~al.} 2019, \aap, 627, A158

\bibitem[{{Di Mauro} {et~al.}(2019){Di Mauro}, {Manconi}, \&
  {Donato}}]{2019PhRvD.100l3015D}
{Di Mauro}, M., {Manconi}, S., \& {Donato}, F. 2019, \prd, 100, 123015

\bibitem[{{Di Mauro} {et~al.}(2021){Di Mauro}, {Manconi}, \&
  {Donato}}]{2021PhRvD.104h9903D}
---. 2021, \prd, 104, 089903

\bibitem[{{Dickey} \& {Lockman}(1990)}]{1990ARA&A..28..215D}
{Dickey}, J.~M., \& {Lockman}, F.~J. 1990, \araa, 28, 215

\bibitem[{{Durbin} {et~al.}(1951){Durbin}, {Loar}, \&
  {Steinberger}}]{1951PhRv...84..581D}
{Durbin}, R., {Loar}, H., \& {Steinberger}, J. 1951, Physical Review, 84, 581

\bibitem[{{Dzhatdoev}(2021)}]{2021arXiv210402838D}
{Dzhatdoev}, T. 2021, arXiv e-prints, arXiv:2104.02838

\bibitem[{{Evoli} {et~al.}(2019){Evoli}, {Aloisio}, \&
  {Blasi}}]{2019PhRvD..99j3023E}
{Evoli}, C., {Aloisio}, R., \& {Blasi}, P. 2019, \prd, 99, 103023

\bibitem[{{Evoli} {et~al.}(2008){Evoli}, {Gaggero}, {Grasso}, \&
  {Maccione}}]{2008JCAP...10..018E}
{Evoli}, C., {Gaggero}, D., {Grasso}, D., \& {Maccione}, L. 2008, \jcap, 2008,
  018

\bibitem[{{Evoli} {et~al.}(2016){Evoli}, {Gaggero}, {Grasso}, \&
  {Maccione}}]{2016JCAP...04E.001E}
---. 2016, \jcap, 2016, E01

\bibitem[{{Evoli} {et~al.}(2017){Evoli}, {Gaggero}, {Vittino}, {Di Bernardo},
  {Di Mauro}, {Ligorini}, {Ullio}, \& {Grasso}}]{2017JCAP...02..015E}
{Evoli}, C., {Gaggero}, D., {Vittino}, A., {et~al.} 2017, \jcap, 2017, 015

\bibitem[{{Faherty} {et~al.}(2007){Faherty}, {Walter}, \&
  {Anderson}}]{FahertyEtAl:2007}
{Faherty}, J., {Walter}, F.~M., \& {Anderson}, J. 2007, \apss, 308, 225

\bibitem[{{Fang} {et~al.}(2018){Fang}, {Bi}, {Yin}, \& {Yuan}}]{FangEtAl:2018}
{Fang}, K., {Bi}, X.-J., {Yin}, P.-F., \& {Yuan}, Q. 2018, \apj, 863, 30

\bibitem[{{Fang} \& {Murase}(2021)}]{2021ApJ...919...93F}
{Fang}, K., \& {Murase}, K. 2021, \apj, 919, 93

\bibitem[{{Ferrando} {et~al.}(1988){Ferrando}, {Webber}, {Goret}, {Kish},
  {Schrier}, {Soutoul}, \& {Testard}}]{1988PhRvC..37.1490F}
{Ferrando}, P., {Webber}, W.~R., {Goret}, P., {et~al.} 1988, \prc, 37, 1490

\bibitem[{{Ferri{\`e}re} {et~al.}(2007){Ferri{\`e}re}, {Gillard}, \&
  {Jean}}]{2007A&A...467..611F}
{Ferri{\`e}re}, K., {Gillard}, W., \& {Jean}, P. 2007, \aap, 467, 611

\bibitem[{{Ferri{\`e}re}(2001)}]{2001RvMP...73.1031F}
{Ferri{\`e}re}, K.~M. 2001, Reviews of Modern Physics, 73, 1031

\bibitem[{{Freudenreich}(1998)}]{1998ApJ...492..495F}
{Freudenreich}, H.~T. 1998, \apj, 492, 495

\bibitem[{{Gaensler} \& {Johnston}(1995)}]{1995MNRAS.277.1243G}
{Gaensler}, B.~M., \& {Johnston}, S. 1995, \mnras, 277, 1243

\bibitem[{{Gaensler} {et~al.}(2008){Gaensler}, {Madsen}, {Chatterjee}, \&
  {Mao}}]{2008PASA...25..184G}
{Gaensler}, B.~M., {Madsen}, G.~J., {Chatterjee}, S., \& {Mao}, S.~A. 2008,
  \pasa, 25, 184

\bibitem[{{G{\'e}nolini} {et~al.}(2018){G{\'e}nolini}, {Maurin}, {Moskalenko},
  \& {Unger}}]{2018PhRvC..98c4611G}
{G{\'e}nolini}, Y., {Maurin}, D., {Moskalenko}, I.~V., \& {Unger}, M. 2018,
  \prc, 98, 034611

\bibitem[{{Genolini} {et~al.}(2017){Genolini}, {Salati}, {Serpico}, \&
  {Taillet}}]{2017A&A...600A..68G}
{Genolini}, Y., {Salati}, P., {Serpico}, P.~D., \& {Taillet}, R. 2017, \aap,
  600, A68

\bibitem[{{G{\'e}nolini} {et~al.}(2019){G{\'e}nolini}, {Boudaud}, {Batista},
  {Caroff}, {Derome}, {Lavalle}, {Marcowith}, {Maurin}, {Poireau}, {Poulin},
  {Rosier}, {Salati}, {Serpico}, \& {Vecchi}}]{2019PhRvD..99l3028G}
{G{\'e}nolini}, Y., {Boudaud}, M., {Batista}, P.~I., {et~al.} 2019, \prd, 99,
  123028

\bibitem[{{Glagolev} {et~al.}(1993){Glagolev}, {Lebedev}, {Pestova},
  {Shimansky}, {Krav{\v{c}}{\'\i}kov{\'a}}, {Seman}, {{\v{S}}{\'a}ndor},
  {Dirner}, {Hlav{\'a}{\v{c}}ov{\'a}}, {Martinsk{\'a}}, {Urb{\'a}n},
  {Khairetdinov}, {Braun}, {Gerber}, {Juillot}, {Michalon}, {Kacharava},
  {Menteshashvili}, {Nioradze}, {Salukvadze}, {Sobczak}, \&
  {Stepaniak}}]{1993ZPhyC..60..421G}
{Glagolev}, V.~V., {Lebedev}, R.~M., {Pestova}, G.~D., {et~al.} 1993,
  Zeitschrift fur Physik C Particles and Fields, 60, 421

\bibitem[{{Gleeson} \& {Axford}(1968)}]{1968ApJ...154.1011G}
{Gleeson}, L.~J., \& {Axford}, W.~I. 1968, \apj, 154, 1011

\bibitem[{{Gondolo} {et~al.}(2004){Gondolo}, {Edsj{\"o}}, {Ullio},
  {Bergstr{\"o}m}, {Schelke}, \& {Baltz}}]{2004JCAP...07..008G}
{Gondolo}, P., {Edsj{\"o}}, J., {Ullio}, P., {et~al.} 2004, \jcap, 2004, 008

\bibitem[{{Gordon} \& {Burton}(1976)}]{1976ApJ...208..346G}
{Gordon}, M.~A., \& {Burton}, W.~B. 1976, \apj, 208, 346

\bibitem[{{G{\'o}rski} {et~al.}(2005){G{\'o}rski}, {Hivon}, {Banday},
  {Wandelt}, {Hansen}, {Reinecke}, \& {Bartelmann}}]{2005ApJ...622..759G}
{G{\'o}rski}, K.~M., {Hivon}, E., {Banday}, A.~J., {et~al.} 2005, \apj, 622,
  759

\bibitem[{{Green} {et~al.}(2019){Green}, {Schlafly}, {Zucker}, {Speagle}, \&
  {Finkbeiner}}]{2019ApJ...887...93G}
{Green}, G.~M., {Schlafly}, E., {Zucker}, C., {Speagle}, J.~S., \&
  {Finkbeiner}, D. 2019, \apj, 887, 93

\bibitem[{{Griffiths} \& {Harbison}(1969)}]{1969ApJ...158..711G}
{Griffiths}, R.~J., \& {Harbison}, S.~A. 1969, \apj, 158, 711

\bibitem[{{Haverkorn} {et~al.}(2008){Haverkorn}, {Brown}, {Gaensler}, \&
  {McClure-Griffiths}}]{2008ApJ...680..362H}
{Haverkorn}, M., {Brown}, J.~C., {Gaensler}, B.~M., \& {McClure-Griffiths},
  N.~M. 2008, \apj, 680, 362

\bibitem[{{Heinz} {et~al.}(1968){Heinz}, {Overseth}, {Pellett}, \&
  {Perl}}]{1968PhRv..167.1232H}
{Heinz}, R.~M., {Overseth}, O.~E., {Pellett}, D.~E., \& {Perl}, M.~L. 1968,
  Physical Review, 167, 1232

\bibitem[{{Higdon} \& {Lingenfelter}(2003)}]{2003ApJ...582..330H}
{Higdon}, J.~C., \& {Lingenfelter}, R.~E. 2003, \apj, 582, 330

\bibitem[{{Hooper} \& {Linden}(2021)}]{2021arXiv210400014H}
{Hooper}, D., \& {Linden}, T. 2021, arXiv e-prints, arXiv:2104.00014

\bibitem[{{Horst} {et~al.}(2019){Horst}, {Aric{\`o}}, {Brinkmann}, {Brons},
  {Ferrari}, {Haberer}, {Mairani}, {Parodi}, {Reidel}, {Weber}, {Zink}, \&
  {Schuy}}]{2019PhRvC..99a4603H}
{Horst}, F., {Aric{\`o}}, G., {Brinkmann}, K.-T., {et~al.} 2019, \prc, 99,
  014603

\bibitem[{{Iroshnikov}(1964)}]{1964SvA.....7..566I}
{Iroshnikov}, P.~S. 1964, \sovast, 7, 566

\bibitem[{{Jaffe} {et~al.}(2011){Jaffe}, {Banday}, {Leahy}, {Leach}, \&
  {Strong}}]{2011MNRAS.416.1152J}
{Jaffe}, T.~R., {Banday}, A.~J., {Leahy}, J.~P., {Leach}, S., \& {Strong},
  A.~W. 2011, \mnras, 416, 1152

\bibitem[{{Jaffe} {et~al.}(2010){Jaffe}, {Leahy}, {Banday}, {Leach}, {Lowe}, \&
  {Wilkinson}}]{2010MNRAS.401.1013J}
{Jaffe}, T.~R., {Leahy}, J.~P., {Banday}, A.~J., {et~al.} 2010, \mnras, 401,
  1013

\bibitem[{{Jansson} \& {Farrar}(2012)}]{2012ApJ...757...14J}
{Jansson}, R., \& {Farrar}, G.~R. 2012, \apj, 757, 14

\bibitem[{{J{\'o}hannesson} {et~al.}(2018){J{\'o}hannesson}, {Porter}, \&
  {Moskalenko}}]{2018ApJ...856...45J}
{J{\'o}hannesson}, G., {Porter}, T.~A., \& {Moskalenko}, I.~V. 2018, \apj, 856,
  45

\bibitem[{{J{\'o}hannesson} {et~al.}(2019){J{\'o}hannesson}, {Porter}, \&
  {Moskalenko}}]{2019ApJ...879...91J}
---. 2019, \apj, 879, 91

\bibitem[{{J{\'o}hannesson} {et~al.}(2016){J{\'o}hannesson}, {Ruiz de Austri},
  {Vincent}, {Moskalenko}, {Orlando}, {Porter}, {Strong}, {Trotta}, {Feroz},
  {Graff}, \& {Hobson}}]{2016ApJ...824...16J}
{J{\'o}hannesson}, G., {Ruiz de Austri}, R., {Vincent}, A.~C., {et~al.} 2016,
  \apj, 824, 16

\bibitem[{{Jung} {et~al.}(1973){Jung}, {Sakamoto}, {Suren}, {Jacquot},
  {Girardin}, \& {Schmitt}}]{1973PhRvC...7.2209J}
{Jung}, M., {Sakamoto}, Y., {Suren}, J.~N., {et~al.} 1973, \prc, 7, 2209

\bibitem[{{Kachelrie{\ss}} {et~al.}(2019){Kachelrie{\ss}}, {Moskalenko}, \&
  {Ostapchenko}}]{2019CoPhC.24506846K}
{Kachelrie{\ss}}, M., {Moskalenko}, I.~V., \& {Ostapchenko}, S. 2019, Computer
  Physics Communications, 245, 106846

\bibitem[{{Kachelriess} {et~al.}(2014){Kachelriess}, {Moskalenko}, \&
  {Ostapchenko}}]{2014ApJ...789..136K}
{Kachelriess}, M., {Moskalenko}, I.~V., \& {Ostapchenko}, S.~S. 2014, \apj,
  789, 136

\bibitem[{{Kachelriess} {et~al.}(2015){Kachelriess}, {Moskalenko}, \&
  {Ostapchenko}}]{2015ApJ...803...54K}
---. 2015, \apj, 803, 54

\bibitem[{{Kachelrie{\ss}} \& {Ostapchenko}(2012)}]{2012PhRvD..86d3004K}
{Kachelrie{\ss}}, M., \& {Ostapchenko}, S. 2012, \prd, 86, 043004

\bibitem[{{Kalberla} {et~al.}(2005){Kalberla}, {Burton}, {Hartmann}, {Arnal},
  {Bajaja}, {Morras}, \& {P{\"o}ppel}}]{2005A&A...440..775K}
{Kalberla}, P.~M.~W., {Burton}, W.~B., {Hartmann}, D., {et~al.} 2005, \aap,
  440, 775

\bibitem[{{Kamae} {et~al.}(2006){Kamae}, {Karlsson}, {Mizuno}, {Abe}, \&
  {Koi}}]{2006ApJ...647..692K}
{Kamae}, T., {Karlsson}, N., {Mizuno}, T., {Abe}, T., \& {Koi}, T. 2006, \apj,
  647, 692

\bibitem[{{Kerr} \& {Lynden-Bell}(1986)}]{1986MNRAS.221.1023K}
{Kerr}, F.~J., \& {Lynden-Bell}, D. 1986, \mnras, 221, 1023

\bibitem[{{Kolmogorov}(1941)}]{1941DoSSR..30..301K}
{Kolmogorov}, A. 1941, Akademiia Nauk SSSR Doklady, 30, 301

\bibitem[{{Kraichnan}(1965)}]{1965PhFl....8.1385K}
{Kraichnan}, R.~H. 1965, Physics of Fluids, 8, 1385

\bibitem[{{Lallement} {et~al.}(2019){Lallement}, {Babusiaux}, {Vergely},
  {Katz}, {Arenou}, {Valette}, {Hottier}, \& {Capitanio}}]{2019A&A...625A.135L}
{Lallement}, R., {Babusiaux}, C., {Vergely}, J.~L., {et~al.} 2019, \aap, 625,
  A135

\bibitem[{{Linden} \& {Buckman}(2018)}]{2018PhRvL.120l1101L}
{Linden}, T., \& {Buckman}, B.~J. 2018, \prl, 120, 121101

\bibitem[{{Lipari} \& {Vernetto}(2018)}]{2018PhRvD..98d3003L}
{Lipari}, P., \& {Vernetto}, S. 2018, \prd, 98, 043003

\bibitem[{{Liu} \& {Wang}(2021)}]{2021ApJ...914L...7L}
{Liu}, R.-Y., \& {Wang}, X.-Y. 2021, \apjl, 914, L7

\bibitem[{{Liu} {et~al.}(2015){Liu}, {Salati}, \& {Chen}}]{2015RAA....15...15L}
{Liu}, W., {Salati}, P., \& {Chen}, X. 2015, Research in Astronomy and
  Astrophysics, 15, 15

\bibitem[{{Lock} \& {Measday}(1970)}]{LockMeasday1970}
{Lock}, W.~O., \& {Measday}, D.~F. 1970, Intermediate Energy Nuclear Physics
  (Methuen \& Company, Limited)

\bibitem[{{Luoni} {et~al.}(2021){Luoni}, {Horst}, {Reidel}, {Quarz}, {Bagnale},
  {Sihver}, {Weber}, {Norman}, {de Wet}, {Giraudo}, {Santin}, {Norbury}, \&
  {Durante}}]{2021NJPh...23j1201L}
{Luoni}, F., {Horst}, F., {Reidel}, C.~A., {et~al.} 2021, New Journal of
  Physics, 23, 101201

\bibitem[{{Malkov} {et~al.}(2013){Malkov}, {Diamond}, {Sagdeev}, {Aharonian},
  \& {Moskalenko}}]{2013ApJ...768...73M}
{Malkov}, M.~A., {Diamond}, P.~H., {Sagdeev}, R.~Z., {Aharonian}, F.~A., \&
  {Moskalenko}, I.~V. 2013, \apj, 768, 73

\bibitem[{{Mertsch}(2011)}]{2011JCAP...02..031M}
{Mertsch}, P. 2011, \jcap, 2, 031

\bibitem[{{Mertsch}(2018)}]{2018JCAP...11..045M}
---. 2018, \jcap, 11, 045

\bibitem[{{Meshcheryakov} \& {Neganov}(1955)}]{1955Doklady_100_677M}
{Meshcheryakov}, M.~G., \& {Neganov}, B.~S. 1955, Doklady Akademii Nauk SSSR,
  100, 677

\bibitem[{{Meshcheryakov} {et~al.}(1955){Meshcheryakov}, {Neganov}, {Bogachev},
  \& {Sidorov}}]{1955Doklady_100_673M}
{Meshcheryakov}, M.~G., {Neganov}, B.~S., {Bogachev}, N.~P., \& {Sidorov},
  B.~M. 1955, Doklady Akademii Nauk SSSR, 100, 673

\bibitem[{{Meyer}(1972)}]{1972A&AS....7..417M}
{Meyer}, J.~P. 1972, \aaps, 7, 417

\bibitem[{{Miyake} {et~al.}(2015){Miyake}, {Muraishi}, \&
  {Yanagita}}]{2015A&A...573A.134M}
{Miyake}, S., {Muraishi}, H., \& {Yanagita}, S. 2015, \aap, 573, A134

\bibitem[{{Moskalenko} {et~al.}(2016){Moskalenko}, {J\'{o}hannesson},
  {Orlando}, {Porter}, {Strong}, \& {Vladimirov}}]{Moskalenko:2015ptr}
{Moskalenko}, I.~V., {J\'{o}hannesson}, G., {Orlando}, E., {et~al.} 2016, PoS,
  ICRC2015, 492

\bibitem[{{Moskalenko} {et~al.}(2006){Moskalenko}, {Porter}, \&
  {Strong}}]{2006ApJ...640L.155M}
{Moskalenko}, I.~V., {Porter}, T.~A., \& {Strong}, A.~W. 2006, \apjl, 640, L155

\bibitem[{{Moskalenko} \& {Strong}(1998)}]{1998ApJ...493..694M}
{Moskalenko}, I.~V., \& {Strong}, A.~W. 1998, \apj, 493, 694

\bibitem[{{Moskalenko} \& {Strong}(2000)}]{2000ApJ...528..357M}
---. 2000, \apj, 528, 357

\bibitem[{{Moskalenko} {et~al.}(2007){Moskalenko}, {Strong}, {Digel}, \&
  {Porter}}]{2007AIPC..921..490M}
{Moskalenko}, I.~V., {Strong}, A.~W., {Digel}, S.~W., \& {Porter}, T.~A. 2007,
  in American Institute of Physics Conference Series, Vol. 921, The First GLAST
  Symposium, ed. S.~{Ritz}, P.~{Michelson}, \& C.~A. {Meegan}, 490--491

\bibitem[{{Moskalenko} {et~al.}(2003){Moskalenko}, {Strong}, {Mashnik}, \&
  {Ormes}}]{2003ApJ...586.1050M}
{Moskalenko}, I.~V., {Strong}, A.~W., {Mashnik}, S.~G., \& {Ormes}, J.~F. 2003,
  \apj, 586, 1050

\bibitem[{{Moskalenko} {et~al.}(2002){Moskalenko}, {Strong}, {Ormes}, \&
  {Potgieter}}]{2002ApJ...565..280M}
{Moskalenko}, I.~V., {Strong}, A.~W., {Ormes}, J.~F., \& {Potgieter}, M.~S.
  2002, \apj, 565, 280

\bibitem[{{Moskalenko} {et~al.}(1998){Moskalenko}, {Strong}, \&
  {Reimer}}]{1998A&A...338L..75M}
{Moskalenko}, I.~V., {Strong}, A.~W., \& {Reimer}, O. 1998, \aap, 338, L75

\bibitem[{{Murphy} {et~al.}(2012){Murphy}, {Porter}, {Moskalenko}, {Helou}, \&
  {Strong}}]{2012ApJ...750..126M}
{Murphy}, E.~J., {Porter}, T.~A., {Moskalenko}, I.~V., {Helou}, G., \&
  {Strong}, A.~W. 2012, \apj, 750, 126

\bibitem[{{Nava} {et~al.}(2016){Nava}, {Gabici}, {Marcowith}, {Morlino}, \&
  {Ptuskin}}]{2016MNRAS.461.3552N}
{Nava}, L., {Gabici}, S., {Marcowith}, A., {Morlino}, G., \& {Ptuskin}, V.~S.
  2016, \mnras, 461, 3552

\bibitem[{{Neganov} \& {Parfenov}(1958)}]{1958JETP...34..767N}
{Neganov}, B.~S., \& {Parfenov}, L.~B. 1958, JETP, 34, 767

\bibitem[{{Nicholls} {et~al.}(1972){Nicholls}, {Craig}, {Griffith}, {Imrie},
  {Lush}, \& {Metheringham}}]{1972NuPhA.181..329N}
{Nicholls}, J.~E., {Craig}, A., {Griffith}, T.~C., {et~al.} 1972, \nphysa, 181,
  329

\bibitem[{{Overseth} {et~al.}(1964){Overseth}, {Heinz}, {Jones}, {Longo},
  {Pellett}, {Perl}, \& {Martin}}]{1964PhRvL..13...59O}
{Overseth}, O.~E., {Heinz}, R.~M., {Jones}, L.~W., {et~al.} 1964, \prl, 13, 59

\bibitem[{{Parker}(1965)}]{1965P&SS...13....9P}
{Parker}, E.~N. 1965, \planss, 13, 9

\bibitem[{{Porter} {et~al.}(2015){Porter}, {Johannesson}, \&
  {Moskalenko}}]{2015ICRC...34..908P}
{Porter}, T., {Johannesson}, G., \& {Moskalenko}, I.~V. 2015, in International
  Cosmic Ray Conference, Vol.~34, 34th International Cosmic Ray Conference
  (ICRC2015), 908

\bibitem[{{Porter} {et~al.}(2017){Porter}, {J{\'o}hannesson}, \&
  {Moskalenko}}]{2017ApJ...846...67P}
{Porter}, T.~A., {J{\'o}hannesson}, G., \& {Moskalenko}, I.~V. 2017, \apj, 846,
  67

\bibitem[{{Porter} {et~al.}(2019){Porter}, {J{\'o}hannesson}, \&
  {Moskalenko}}]{2019ApJ...887..250P}
---. 2019, \apj, 887, 250

\bibitem[{{Porter} {et~al.}(2008){Porter}, {Moskalenko}, {Strong}, {Orlando},
  \& {Bouchet}}]{2008ApJ...682..400P}
{Porter}, T.~A., {Moskalenko}, I.~V., {Strong}, A.~W., {Orlando}, E., \&
  {Bouchet}, L. 2008, \apj, 682, 400

\bibitem[{{Porter} {et~al.}(2018){Porter}, {Rowell}, {J{\'o}hannesson}, \&
  {Moskalenko}}]{2018PhRvD..98d1302P}
{Porter}, T.~A., {Rowell}, G.~P., {J{\'o}hannesson}, G., \& {Moskalenko}, I.~V.
  2018, \prd, 98, 041302

\bibitem[{{Porter} \& {Strong}(2005)}]{2005ICRC....4...77P}
{Porter}, T.~A., \& {Strong}, A.~W. 2005, Proc.\ 29th \icrc\ (Pune), 4, 77

\bibitem[{{Pratt} {et~al.}(1973){Pratt}, {Ron}, \&
  {Tseng}}]{1973RvMP...45..273P}
{Pratt}, R.~H., {Ron}, A., \& {Tseng}, H.~K. 1973, Reviews of Modern Physics,
  45, 273

\bibitem[{{Profumo} {et~al.}(2018){Profumo}, {Reynoso-Cordova}, {Kaaz}, \&
  {Silverman}}]{2018PhRvD..97l3008P}
{Profumo}, S., {Reynoso-Cordova}, J., {Kaaz}, N., \& {Silverman}, M. 2018,
  \prd, 97, 123008

\bibitem[{{Pshirkov} {et~al.}(2011){Pshirkov}, {Tinyakov}, {Kronberg}, \&
  {Newton-McGee}}]{2011ApJ...738..192P}
{Pshirkov}, M.~S., {Tinyakov}, P.~G., {Kronberg}, P.~P., \& {Newton-McGee},
  K.~J. 2011, \apj, 738, 192

\bibitem[{{Ptuskin} {et~al.}(2006){Ptuskin}, {Jones}, {Seo}, \&
  {Sina}}]{2006AdSpR..37.1909P}
{Ptuskin}, V.~S., {Jones}, F.~C., {Seo}, E.~S., \& {Sina}, R. 2006, \adv, 37,
  1909

\bibitem[{{Ptuskin} {et~al.}(2008){Ptuskin}, {Zirakashvili}, \&
  {Plesser}}]{2008AdSpR..42..486P}
{Ptuskin}, V.~S., {Zirakashvili}, V.~N., \& {Plesser}, A.~A. 2008, Advances in
  Space Research, 42, 486

\bibitem[{{Qiao} {et~al.}(2021){Qiao}, {Liu}, {Zhao}, {Bi}, \&
  {Guo}}]{2021arXiv210403729Q}
{Qiao}, B.-Q., {Liu}, W., {Zhao}, M.-J., {Bi}, X.-J., \& {Guo}, Y.-Q. 2021,
  arXiv e-prints, arXiv:2104.03729

\bibitem[{{Richard-Serre} {et~al.}(1970){Richard-Serre}, {Hirt}, {Measday},
  {Michaelis}, {Saltmarsh}, \& {Skarek}}]{1970NuPhB..20..413R}
{Richard-Serre}, C., {Hirt}, W., {Measday}, D.~F., {et~al.} 1970, Nuclear
  Physics B, 20, 413

\bibitem[{{Ritchie}(1983)}]{1983PhRvC..28..926R}
{Ritchie}, B.~G. 1983, \prc, 28, 926

\bibitem[{{Ritchie} {et~al.}(1981){Ritchie}, {Edge}, {Malbrough}, {Preedom},
  {Bertrand}, {Gross}, {Obenshain}, {Wu}, {Blecher}, {Gotow}, {Burman},
  {Carlini}, {Hamm}, {Leitch}, \& {Moinester}}]{1981PhRvC..24..552R}
{Ritchie}, B.~G., {Edge}, R.~D., {Malbrough}, D.~J., {et~al.} 1981, \prc, 24,
  552

\bibitem[{{Ritchie} {et~al.}(1983){Ritchie}, {Blanpied}, {Moore}, {Preedom},
  {Gotow}, {Minehart}, {Boswell}, {Das}, {Ziock}, {Chant}, {Roos}, {Burger},
  {Gilad}, \& {Redwine}}]{1983PhRvC..27.1685R}
{Ritchie}, B.~G., {Blanpied}, G.~S., {Moore}, R.~S., {et~al.} 1983, \prc, 27,
  1685

\bibitem[{{Robitaille} {et~al.}(2012){Robitaille}, {Churchwell}, {Benjamin},
  {Whitney}, {Wood}, {Babler}, \& {Meade}}]{2012A&A...545A..39R}
{Robitaille}, T.~P., {Churchwell}, E., {Benjamin}, R.~A., {et~al.} 2012, \aap,
  545, A39

\bibitem[{{Rogers} \& {Lederman}(1957)}]{1957PhRv..105..247R}
{Rogers}, K.~C., \& {Lederman}, L.~M. 1957, Physical Review, 105, 247

\bibitem[{{Rose}(1967)}]{1967PhRv..154.1305R}
{Rose}, C.~M. 1967, Physical Review, 154, 1305

\bibitem[{{Sachs} {et~al.}(1958){Sachs}, {Winick}, \&
  {Wooten}}]{1958PhRv..109.1733S}
{Sachs}, A.~M., {Winick}, H., \& {Wooten}, B.~A. 1958, Physical Review, 109,
  1733

\bibitem[{{Seo} \& {Ptuskin}(1994)}]{1994ApJ...431..705S}
{Seo}, E.~S., \& {Ptuskin}, V.~S. 1994, \apj, 431, 705

\bibitem[{{Strong} \& {Moskalenko}(1998)}]{1998ApJ...509..212S}
{Strong}, A.~W., \& {Moskalenko}, I.~V. 1998, \apj, 509, 212

\bibitem[{{Strong} \& {Moskalenko}(2001{\natexlab{a}})}]{2001AIPC..587..533S}
{Strong}, A.~W., \& {Moskalenko}, I.~V. 2001{\natexlab{a}}, in American
  Institute of Physics Conference Series, Vol. 587, Gamma 2001: Gamma-Ray
  Astrophysics, ed. S.~{Ritz}, N.~{Gehrels}, \& C.~R. {Shrader}, 533--537

\bibitem[{{Strong} \& {Moskalenko}(2001{\natexlab{b}})}]{2001ICRC....5.1964S}
---. 2001{\natexlab{b}}, International Cosmic Ray Conference, 5, 1964

\bibitem[{{Strong} {et~al.}(2009){Strong}, {Moskalenko}, {Porter},
  {J{\'o}hannesson}, {Orlando}, \& {Digel}}]{2009arXiv0907.0559S}
{Strong}, A.~W., {Moskalenko}, I.~V., {Porter}, T.~A., {et~al.} 2009, arXiv:
  0907.0559, arXiv:0907.0559

\bibitem[{{Strong} {et~al.}(2007){Strong}, {Moskalenko}, \&
  {Ptuskin}}]{2007ARNPS..57..285S}
{Strong}, A.~W., {Moskalenko}, I.~V., \& {Ptuskin}, V.~S. 2007, \arnps, 57, 285

\bibitem[{{Strong} {et~al.}(2000){Strong}, {Moskalenko}, \&
  {Reimer}}]{2000ApJ...537..763S}
{Strong}, A.~W., {Moskalenko}, I.~V., \& {Reimer}, O. 2000, \apj, 537, 763

\bibitem[{{Strong} {et~al.}(2004{\natexlab{a}}){Strong}, {Moskalenko}, \&
  {Reimer}}]{2004ApJ...613..962S}
---. 2004{\natexlab{a}}, \apj, 613, 962

\bibitem[{{Strong} {et~al.}(2004{\natexlab{b}}){Strong}, {Moskalenko},
  {Reimer}, {Digel}, \& {Diehl}}]{2004A&A...422L..47S}
{Strong}, A.~W., {Moskalenko}, I.~V., {Reimer}, O., {Digel}, S., \& {Diehl}, R.
  2004{\natexlab{b}}, \aap, 422, L47

\bibitem[{{Sudoh} {et~al.}(2019){Sudoh}, {Linden}, \&
  {Beacom}}]{2019PhRvD.100d3016S}
{Sudoh}, T., {Linden}, T., \& {Beacom}, J.~F. 2019, \prd, 100, 043016

\bibitem[{{Sudoh} {et~al.}(2021){Sudoh}, {Linden}, \&
  {Hooper}}]{2021arXiv210111026S}
{Sudoh}, T., {Linden}, T., \& {Hooper}, D. 2021, arXiv e-prints,
  arXiv:2101.11026

\bibitem[{{Sun} \& {Reich}(2010)}]{2010RAA....10.1287S}
{Sun}, X.-H., \& {Reich}, W. 2010, Research in Astronomy and Astrophysics, 10,
  1287

\bibitem[{{Sun} {et~al.}(2008{\natexlab{a}}){Sun}, {Reich}, {Waelkens}, \&
  {En{\ss}lin}}]{2008A&A...477..573S}
{Sun}, X.~H., {Reich}, W., {Waelkens}, A., \& {En{\ss}lin}, T.~A.
  2008{\natexlab{a}}, \aap, 477, 573

\bibitem[{{Sun} {et~al.}(2008{\natexlab{b}}){Sun}, {Reich}, {Waelkens}, \&
  {En{\ss}lin}}]{SunEtAl:2008}
---. 2008{\natexlab{b}}, \aap, 477, 573

\bibitem[{{Swordy}(2003)}]{2003ICRC....4.1989S}
{Swordy}, S.~P. 2003, International Cosmic Ray Conference, 4, 1989

\bibitem[{{Taillet} {et~al.}(2004){Taillet}, {Salati}, {Maurin},
  {Vangioni-Flam}, \& {Cass{\'e}}}]{2004ApJ...609..173T}
{Taillet}, R., {Salati}, P., {Maurin}, D., {Vangioni-Flam}, E., \& {Cass{\'e}},
  M. 2004, \apj, 609, 173

\bibitem[{{Tang} \& {Piran}(2018)}]{TangPiran:2018}
{Tang}, X., \& {Piran}, T. 2018, ArXiv e-prints, arXiv:1808.02445

\bibitem[{{Tomassetti}(2015)}]{2015arXiv151009212T}
{Tomassetti}, N. 2015, arXiv e-prints, arXiv:1510.09212

\bibitem[{{Tripathi} {et~al.}(1996){Tripathi}, {Cucinotta}, \&
  {Wilson}}]{1996NIMPB.117..347T}
{Tripathi}, R.~K., {Cucinotta}, F.~A., \& {Wilson}, J.~W. 1996, Nuclear
  Instruments and Methods in Physics Research B, 117, 347

\bibitem[{{Tripathi} {et~al.}(1997){Tripathi}, {Cucinotta}, \&
  {Wilson}}]{1997lrc..reptQ....T}
---. 1997, {Universal Parameterization of Absorption Cross Sections}, NASA
  Technical Paper 3621, ,

\bibitem[{{Tripathi} {et~al.}(1999{\natexlab{a}}){Tripathi}, {Cucinotta}, \&
  {Wilson}}]{1999NIMPB.155..349T}
---. 1999{\natexlab{a}}, Nuclear Instruments and Methods in Physics Research B,
  155, 349

\bibitem[{{Tripathi} {et~al.}(1999{\natexlab{b}}){Tripathi}, {Cucinotta}, \&
  {Wilson}}]{1999STIN...0004259T}
---. 1999{\natexlab{b}}, {Universal Parameterization of Absorption Cross
  Sections. Light Systems}, NASA Technical Paper TP-1999-209726, ,

\bibitem[{{Trotta} {et~al.}(2011){Trotta}, {J{\'o}hannesson}, {Moskalenko},
  {Porter}, {Ruiz de Austri}, \& {Strong}}]{2011ApJ...729..106T}
{Trotta}, R., {J{\'o}hannesson}, G., {Moskalenko}, I.~V., {et~al.} 2011, \apj,
  729, 106

\bibitem[{{Turkot} {et~al.}(1963){Turkot}, {Collins}, \&
  {Fujii}}]{1963PhRvL..11..474T}
{Turkot}, F., {Collins}, G.~B., \& {Fujii}, T. 1963, \prl, 11, 474

\bibitem[{{Webber}(1990)}]{1990AIPC..203..294W}
{Webber}, W.~R. 1990, in American Institute of Physics Conference Series, Vol.
  203, Particle Astrophysics - The NASA Cosmic Ray Program for the 1990s and
  Beyond, ed. W.~V. {Jones}, F.~J. {Kerr}, \& J.~F. {Ormes}, 294--298

\bibitem[{{Wellisch} \& {Axen}(1996)}]{1996PhRvC..54.1329W}
{Wellisch}, H.~P., \& {Axen}, D. 1996, \prc, 54, 1329

\bibitem[{{Wilson}(1978)}]{1978PhDT........12W}
{Wilson}, L.~W. 1978, PhD thesis, The Nuclear and Atomic Physics Governing
  Changes in the Composition of Relativistic Cosmic Rays, University of
  California at Berkeley

\bibitem[{{Wouterloot} {et~al.}(1990){Wouterloot}, {Brand}, {Burton}, \&
  {Kwee}}]{1990A&A...230...21W}
{Wouterloot}, J.~G.~A., {Brand}, J., {Burton}, W.~B., \& {Kwee}, K.~K. 1990,
  \aap, 230, 21

\bibitem[{{Yusifov} \& {K{\"u}{\c c}{\"u}k}(2004)}]{2004A&A...422..545Y}
{Yusifov}, I., \& {K{\"u}{\c c}{\"u}k}, I. 2004, \aap, 422, 545

\bibitem[{{Zhang} {et~al.}(2020){Zhang}, {Liu}, {Chen}, \&
  {Wang}}]{2020arXiv201015731Z}
{Zhang}, Y., {Liu}, R.-Y., {Chen}, S.~Z., \& {Wang}, X.-Y. 2020, arXiv
  e-prints, arXiv:2010.15731

\end{thebibliography}
  
\end{document}